\documentclass[12pt,aps]{revtex4}
\usepackage{graphicx}
\usepackage{amsmath}
\usepackage{amsfonts}
\usepackage{amssymb}
\usepackage{wasysym}
\usepackage{float}
\usepackage{mathrsfs}

\begin{document}
\title{Integrable turbulence generated from modulational instability of cnoidal waves}
\author{D.S. Agafontsev$^{(a)}$, V.E. Zakharov$^{(a),(b),(c),(d)}$}

\affiliation{\small \textit{ $^{(a)}$ P.P. Shirshov Institute of Oceanology, 36 Nakhimovsky prosp., Moscow 117218, Russia.\\
$^{(b)}$ Department of Mathematics, University of Arizona, Tucson, AZ, 857201, USA.\\
$^{(c)}$ P.N. Lebedev Physical Institute, 53 Leninsky ave., 119991 Moscow, Russia.\\
$^{(d)}$ Novosibirsk State University, 2 Pirogova, 630090 Novosibirsk, Russia.}}

\begin{abstract}
We study numerically the nonlinear stage of modulational instability (MI) of cnoidal waves, in the framework of the focusing one-dimensional Nonlinear Schr{\"o}dinger (NLS) equation. 
Cnoidal waves are exact periodic solutions of the NLS equation which can be represented as lattices of overlapping solitons. 
MI of these lattices leads to development of ``integrable turbulence'' \textit{[Zakharov V.\,E., Stud. Appl. Math. 122, 219-234 (2009)]}. 
We study the major characteristics of the turbulence for dn-branch of cnoidal waves and demonstrate how these characteristics depend on the degree of ``overlapping'' between the solitons within the cnoidal wave.

Integrable turbulence, that develops from MI of dn-branch of cnoidal waves, asymptotically approaches to its stationary state in oscillatory way. During this process kinetic and potential energies oscillate around their asymptotic values. The amplitudes of these oscillations decay with time as $t^{-\alpha}$, $1<\alpha< 1.5$, the phases contain nonlinear phase shift decaying as $t^{-1/2}$, and the frequency of the oscillations is equal to the double maximal growth rate of the MI, $s=2\gamma_{\max}$. In the asymptotic stationary state the ratio of potential to kinetic energy is equal to $-2$. The asymptotic PDF of wave intensity is close to exponential distribution for cnoidal waves with strong overlapping, and is significantly non-exponential one for cnoidal waves with weak overlapping of solitons. In the latter case the dynamics of the system reduces to two-soliton collisions, which occur with exponentially small rate and provide up to two-fold increase in amplitude compared with the original cnoidal wave. 
For all cnoidal waves of dn-branch, rogue waves at the time of their maximal elevation have quasi-rational profile similar to that of the Peregrine solution. 
\end{abstract}

\maketitle


\section{Introduction}
\label{Sec:Intro}

The statistics of waves for different nonlinear systems was intensively studied in the recent years~\cite{mussot2009observation, genty2010collisions, taki2010third, hammani2010emergence, chung2011strong, randoux2014intermittency, agafontsev2014integrable, walczak2015optical, agafontsev2014intermittency,suret2016direct}, especially since the first experimental observation of optical rogue waves~\cite{solli2007optical}. Known previously for hydrodynamics~\cite{kharif2003physical, dysthe2008oceanic, onorato2013rogue}, rogue waves are short very large pulses that may endanger marine navigation and optical communications. These pulses appear randomly from initially smooth waves and their statistics may significantly exceed that predicted by the approximation of random wave field governed by linear equations. 

Let us suppose, that wave field $\psi$ is a random superposition of a multitude of uncorrelated linear waves, 
$$
\psi(x) = \sum_{k}|\psi_{k}|e^{i(kx+\phi_{k})}.
$$
If phases $\phi_{k}$ are random and uncorrelated, the number of waves $\{k\}$ is large, and amplitudes $|\psi_{k}|$ fall under the conditions of central limit theorem, then real $\mathrm{Re}\,\psi(x)$ and imaginary $\mathrm{Im}\,\psi(x)$ parts are Gaussian-distributed, and the probability density function (PDF) for wave amplitude is Rayleigh distribution~\cite{nazarenko2011wave},
\begin{equation}\label{Rayleigh0}
\mathcal{P}_{R}(|\psi|) = \frac{2|\psi|}{\sigma^{2}}e^{-|\psi|^{2}/\sigma^{2}}.
\end{equation}
Here $\sigma^{2}=\langle|\psi|^{2}\rangle$ is the average square amplitude, and we use normalization for the PDF as $\int\mathcal{P}(|\psi|)\,d |\psi|=1$. For convenience, below we study PDFs for normalized square amplitude $I=|\psi|^{2}/\langle|\psi|^{2}\rangle$, which has the meaning of relative intensity: small waves correspond to $I\ll 1$, moderate - to $I\sim 1$, and large -- to $I\gg 1$. Then, Rayleigh PDF~(\ref{Rayleigh0}) takes the simple form 
\begin{equation}\label{Rayleigh}
\mathcal{P}_{R}(I) = e^{-I},
\end{equation}
which we will call as exponential PDF (note that we called the same PDF as Rayleigh one in our previous paper~\cite{agafontsev2014integrable}). If evolution is governed by linear equations, then superposition of linear waves stays uncorrelated, and its PDF remains exponential~(\ref{Rayleigh}). Nonlinear evolution may introduce correlation, which in turn may lead to enhanced appearance of large waves.

With a certain degree of accuracy, many physical systems can be described by completely integrable (nonlinear) mathematical models. In comparison with non-integrable models, the corresponding integrable equations demonstrate significantly different statistical properties \cite{zakharov2009turbulence, suret2011wave, picozzi2014optical}. The new emerging field of nonlinear science which studies these properties was introduced in 2009 by V.E.\,Zakharov~\cite{zakharov2009turbulence} as ``integrable turbulence''. One-dimensional Nonlinear Schr{\"o}dinger (NLS) equation of focusing type, 
\begin{equation}
i\psi_{t}+\psi_{xx}+|\psi|^2 \psi = 0, \label{nlse}
\end{equation}
is paid a special attention in these studies, since it is a simple mathematical model suitable for the description of rogue waves in optics and hydrodynamics~\cite{kharif2003physical, dysthe2008oceanic, onorato2013rogue}. The simplest ``condensate'' solution $\psi=e^{it}$ of Eq.~(\ref{nlse}) is modulationally unstable, and development of this instability from initially small perturbation may lead to appearance of rogue waves~\cite{kharif2003physical, dysthe2008oceanic}.

However, as we demonstrated in our previous paper~\cite{agafontsev2014integrable}, for the NLS equation and in the scenario of modulational instability (MI) of the condensate, the PDF of wave intensity, averaged over realizations of initial perturbation, does not exceed significantly exponential PDF~(\ref{Rayleigh}). Development of MI leads to integrable turbulence, which asymptotically approaches to its stationary state in oscillatory way. The PDF in this state is exponential~(\ref{Rayleigh}). During the evolution toward the stationary state, the PDF significantly deviates from~(\ref{Rayleigh}), however it does not exceed the exponential PDF by more than a few times. 

Another physically relevant scenario of rogue waves emergence in the framework of the NLS equation was studied in~\cite{walczak2015optical,suret2016direct} for incoherent waves as initial conditions (see also the earlier studies~\cite{onorato2000occurrence,onorato2004observation,onorato2006extreme} with the similar results for long crested water waves of Jonswap spectrum, and also the similar study for defocusing NLS equation~\cite{randoux2014intermittency}). 
For incoherent waves, integrable turbulence quickly reaches its stationary state, in which the tail of the PDF at large intensities exceeds exponential distribution~(\ref{Rayleigh}) by orders of magnitude. 

The fact that different initial conditions of these studies lead to entirely different results is not surprising: integrable systems ``remember'' their initial condition through conservation of infinite series of invariants (integrals of motion). These invariants are different for different types of initial conditions, and thus the stationary states and the evolution toward them are different too. However, so far there is no explanation why in one case integrable turbulence approaches to its stationary state for a very long time and the probability of rogue waves appearance is small, while in the other case the stationary state is reached very quickly and rogue waves appear much more frequently.

\begin{figure}[t] \centering
\includegraphics[width=8.0cm]{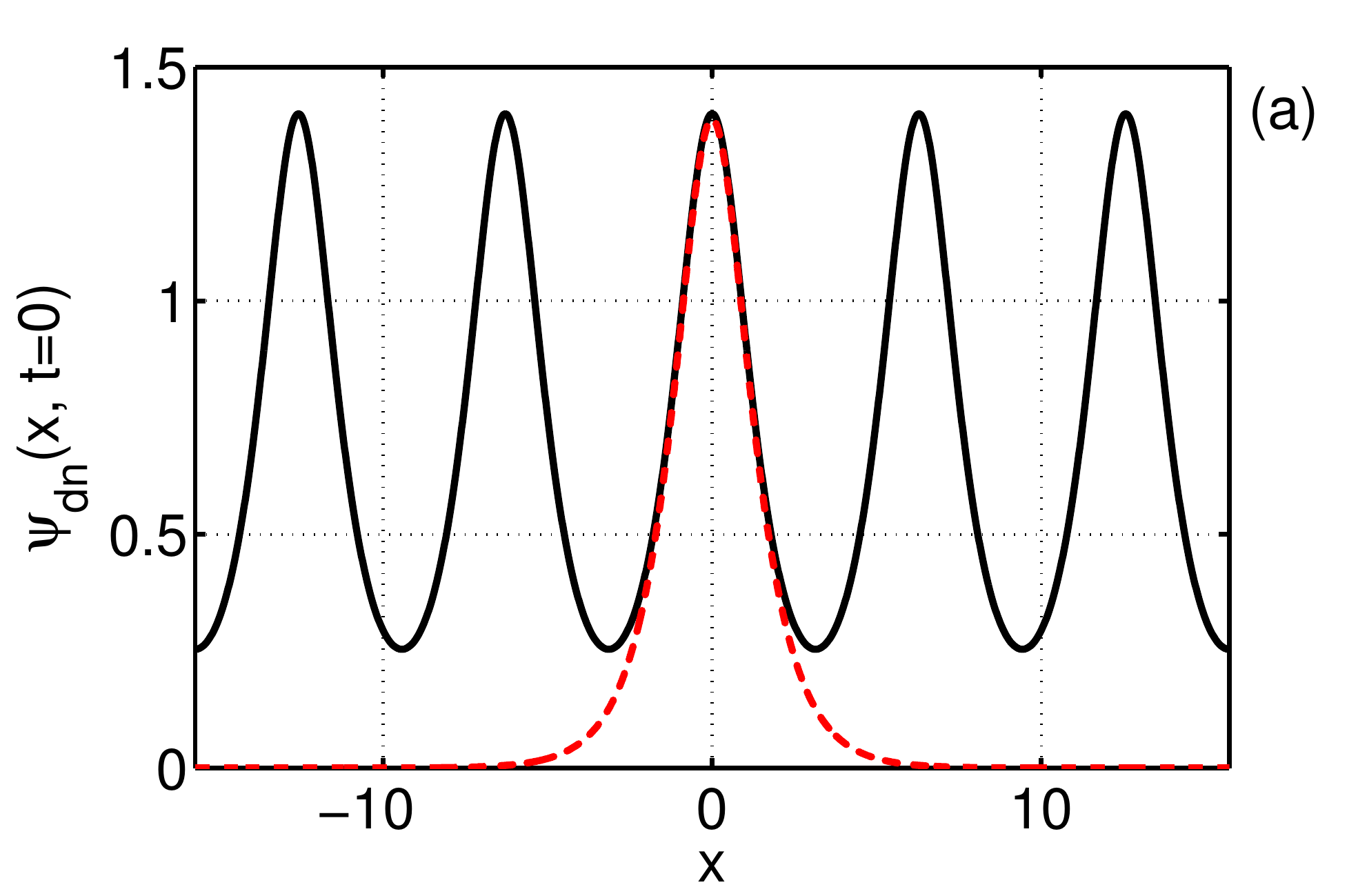}
\includegraphics[width=8.0cm]{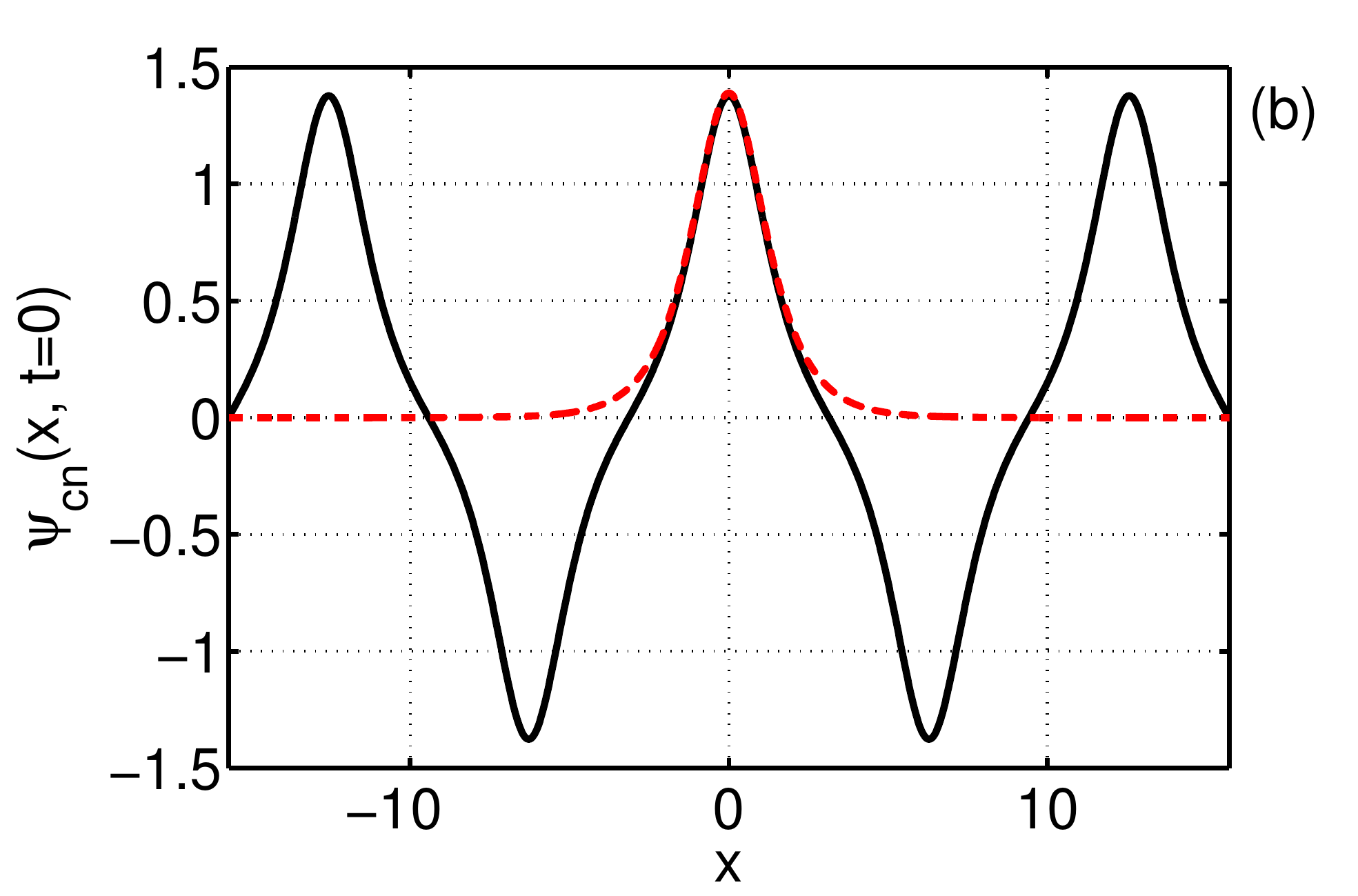}

\caption{\small {\it (Color on-line)} 
Cnoidal waves of (a) dn-branch~(\ref{cnoidal1}) and (b) cn-branch~(\ref{cnoidal2}) with $\omega_{0}=\pi$, $\omega_{1}=1.6$, at $t=0$ when they are purely real, $\mathrm{Im}\,\psi_{dn}(x,0)=\mathrm{Im}\,\psi_{cn}(x,0)=0$. 
Dashed red lines show soliton~(\ref{nlse_soliton}) with $\lambda=\pi/2\omega_{1}$. }
\label{fig:cnoidal_wave_B1}
\end{figure}

In this publication we study one more scenario, when integrable turbulence develops from MI of cnoidal waves. Cnoidal waves are exact periodic solutions of the NLS equation~(\ref{nlse}) essentially depending on two parameters $\omega_{0}$ and $\omega_{1}$, which we will call as real and imaginary half-periods respectively. 
There are dn- and cn-branches of such solutions. 
The dn-branch of cnoidal waves can be written as
\begin{eqnarray}\label{cnoidal1}
\psi_{dn}(x,t) = e^{i\Omega t} \sqrt{2}\,\nu\,\,\mathrm{dn}(\nu x; s^{2}),
\end{eqnarray}
where $\mathrm{dn}(x; s^{2})$ is the corresponding Jacobi elliptic function and $\Omega$, $\nu$ and $s$ are specific values defined by half-periods $\omega_{0}$ and $\omega_{1}$ (see Appendix~\ref{Sec:Annex-A} for more details).
Solutions~(\ref{cnoidal1}) are periodic with period $2\omega_{0}$, at $t=0$ they are purely real and positive, $\psi_{dn}(x,0)>0$; the example of one such solution with half-periods $\omega_{0}=\pi$ and $\omega_{1}=1.6$ is shown in Fig.~\ref{fig:cnoidal_wave_B1}(a). The cn-branch of cnoidal waves can be written as
\begin{eqnarray}\label{cnoidal2}
\psi_{cn}(x,t) = e^{i\Omega t} \sqrt{2}\,s\nu\,\,\mathrm{cn}(\nu x; s^{2}),
\end{eqnarray}
where $\mathrm{cn}(x; s^{2})$ is the corresponding Jacobi elliptic function. These solutions are periodic with period $4\omega_{0}$, at $t=0$ they are purely real and change their sign periodically with $x$. The example of such solution for the same half-periods $\omega_{0}=\pi$ and $\omega_{1}=1.6$ is shown in Fig.~\ref{fig:cnoidal_wave_B1}(b).

As described in Appendix~\ref{Sec:Annex-A}, both types of cnoidal waves can be viewed as infinite lattices of overlapping NLS solitons, 
\begin{eqnarray}\label{nlse_soliton}
\psi_{s}(x,t) = e^{i\lambda^{2}t}\frac{\sqrt{2}\,\lambda}{\cosh\,\lambda\,x}, \quad \lambda=\pi/2\omega_{1},
\end{eqnarray}
with the width of the solitons proportional to $\omega_{1}$ and the distance between them equal to $2\omega_{0}$.
For weak ``overlapping'' between the solitons $\omega_{1}/\omega_{0}\ll 1$, cnoidal waves transform into arithmetic sum of NLS solitons,
\begin{eqnarray}\label{cnoidal_limit_w0}
\psi(x,t) \to e^{i\lambda^{2}t}\sum_{m=-\infty}^{+\infty}\frac{(-1)^{\varrho\, m}\,\sqrt{2}\,\lambda}{\cosh\,\lambda (x-2m\omega_{0})},
\end{eqnarray}
where $\varrho=0$ for dn-branch and $\varrho=1$ for cn-branch. For strong ``overlapping'' $\omega_{1}/\omega_{0}\gg 1$, cnoidal waves of dn-branch transform into condensate, 
\begin{eqnarray}\label{cnoidal_limit_w1}
\psi_{dn}(x,t) \to \sqrt{2}\,\kappa\,e^{2i\kappa^{2}t},
\end{eqnarray}
while cnoidal waves of cn-branch -- into sinusoidal wave with exponentially small amplitude,
\begin{eqnarray}\label{cnoidal2_limit_w1}
\psi_{cn}(x,t) \to \bigg[4\sqrt{2}\,\kappa\,\exp(-\kappa\,\omega_{1})\bigg]\,e^{-i\kappa^{2}t}\,\cos(\kappa\, x),
\end{eqnarray} 
where $\kappa = \pi/2\omega_{0}$. Both branches of cnoidal waves are modulationally unstable. 
For dn-branch, the maximal growth rate of MI was found in~\cite{kuznetsov1999modulation},
\begin{eqnarray}\label{cnoidal1_increment}
\gamma_{\max} = 2\nu\,(e_{1}-e_{2})^{1/2}. 
\end{eqnarray}
For fixed $\omega_{0}$, this relation is exponentially small for small $\omega_{1}$, monotonically increases with $\omega_{1}$, and approaches to $2\kappa^{2}$ as $\omega_{1}\to +\infty$. 
To the authors knowledge, $\gamma_{\max}$ was not found for cn-branch so far. 

In this paper we study statistical properties of MI for dn-branch of cnoidal waves only. It will be more convenient to work with the stationary variants of cnoidal waves,
\begin{eqnarray}\label{cnoidal1_stationary}
\Psi_{dn}(x) = \sqrt{2}\,\nu\,\,\mathrm{dn}(\nu x; s^{2}),
\end{eqnarray}
which are solutions of slightly modified NLS equation
\begin{eqnarray}\label{nlse_stationary}
i\Psi_{t}-\Omega\,\Psi+\Psi_{xx}+|\Psi|^{2}\Psi = 0,
\end{eqnarray}
obtained from Eq.~(\ref{nlse}) after gauge transformation $\psi = e^{i\Omega t}\Psi$. To study statistical properties of MI of cnoidal waves~(\ref{cnoidal1_stationary}), one has to solve Eq.~(\ref{nlse_stationary}) with initial conditions
\begin{eqnarray}\label{cnoidal1_MI}
\Psi|_{t=0} = \Psi_{dn}(x) + \epsilon(x),\quad |\epsilon(x)|\ll |\Psi_{dn}(x)|,
\end{eqnarray}
and average the results over different realizations of initial noise $\epsilon(x)$. Without loss of generality we will consider cnoidal waves with real half-period $\omega_{0}=\pi$ only. Indeed, NLS equation can be scaled $t\to\chi\, t$, $x\to\eta\, x$ and $\Psi\to\mu\,\Psi$ using 3 independent parameters $\chi$, $\eta$ and $\mu$. Two of these parameters can be used to scale dispersion and nonlinearity coefficients to unity, while the last parameter -- to scale $\omega_{0}$ to $\pi$. Then, in the limits~(\ref{cnoidal_limit_w1}) and~(\ref{cnoidal2_limit_w1}) constant $\kappa=\pi/2\omega_{0}$ is equal to  $1/2$, and cnoidal waves from dn-branch transform into condensate with amplitude $1/\sqrt{2}$ as $\omega_{1}\to +\infty$.

Formally, the problem~(\ref{nlse_stationary}) with initial conditions~(\ref{cnoidal1_MI}) can be solved analytically, as any periodic solution of the NLS equation can be expressed explicitly in terms of the Jacobi $\theta$-functions over a certain hyperbolic curve, see e.g.~\cite{ma1981periodic,zakharov1984theory,belokolos1994algebro}. However, to study the general case of MI, we have to use initial noise $\epsilon(x)$ with a very large number of excited modes, which makes the genus of the curve also very large (for the present study we use at least $10^{4}$ harmonics). It is unrealistic so far to follow this evolution with the exact analytical methods.

Therefore, we rely completely on numerical experiments, solving NLS equation~(\ref{nlse_stationary}) in the box $x\in[-L/2, L/2]$ with periodic boundary. Integrability implies conservation of infinite series of integrals of motion. The first three of these invariants are wave action, 
\begin{equation}\label{wave_action}
N = \frac{1}{L}\int_{-L/2}^{L/2}|\Psi(x,t)|^{2}\,dx,
\end{equation}
momentum,
\begin{equation}\label{momentum}
P=\frac{i}{2L}\int_{-L/2}^{L/2}(\Psi_{x}^{*}\Psi-\Psi_{x}\Psi^{*})\,dx,
\end{equation}
and total energy,
\begin{equation}\label{energy}
E=H_{d}+H_{4},\quad\quad H_{d}=\frac{1}{L}\int_{-L/2}^{L/2} |\Psi_{x}|^2\,dx,\quad\quad H_{4} = -\frac{1}{2L}\int_{-L/2}^{L/2} |\Psi|^4\,dx.
\end{equation}
Here $H_{d}$ is kinetic and $H_{4}$ is potential energy. The other invariants, 
\begin{equation}\label{all_integrals1}
c_{n}[\Psi]=\frac{1}{L}\int_{-L/2}^{L/2}\phi_{n}\,dx,
\end{equation}
can be calculated using the following recurrent series of equations~\cite{zakharov1984theory}:
\begin{equation}\label{all_integrals2}
\phi_{n+1} = \Psi\frac{\partial}{\partial x}\bigg(\frac{\phi_{n}}{\Psi}\bigg)+\sum_{l+m=n}\phi_{l}\phi_{m},\quad \phi_{1} = |\Psi|^{2}/2.
\end{equation}
Our method of numerical simulations conserves very well the first 10 invariants. 

The general motivation of our study is to improve our understanding of integrable turbulence. In~\cite{agafontsev2014integrable, walczak2015optical,suret2016direct} two cases of integrable turbulence were studied, with the condensate and incoherent wave initial conditions respectively. With the help of the scaling transformations, the problem of MI of the condensate renormalizes to Eq.(\ref{nlse}) and condensate $\psi=e^{it}$ -- for all dispersion and nonlinearity coefficients and for all condensates~\cite{agafontsev2014integrable}. Thus, except for the noise, this problem does not depend on any free parameter. The evolution of incoherent wave in formulation~\cite{walczak2015optical,suret2016direct} essentially depends on one free parameter, and initial potential to kinetic energy ratio can be used as such. 
However, so far an extensive study of how integrable turbulence depends on this parameter was not performed. 
Cnoidal waves are the class of modulationally unstable solutions of the focusing NLS equation. For fixed $\omega_{0}=\pi$, the properties of these solutions essentially depend on one parameter $\omega_{1}$, which determines the degree of ``overlapping'' between the solitons within the cnoidal wave (see Appendix~\ref{Sec:Annex-A}). Thus, we can expect that integrable turbulence generated from MI of these waves will also significantly depend on $\omega_{1}$.

Cnoidal waves with small $\omega_{1}$ are very close to arithmetic sum of equally spaced very thin and high NLS solitons~(\ref{cnoidal_limit_w0}). Turbulence generated from MI of such waves should be close to soliton turbulence in integrable system (for soliton turbulence in nonintegrable systems see e.g.~\cite{zakharov1988soliton, d1989soliton}). For large $\omega_{1}$ we should obtain results similar to that for the condensate initial conditions~\cite{agafontsev2014integrable}, since such cnoidal waves are close to condensate~(\ref{cnoidal_limit_w1}). 
Changing $\omega_{1}$, we can study how properties of the integrable turbulence transform from that for integrable soliton turbulence to that for MI of the condensate.

As we demonstrate in this publication, many of the presented facts do not have theoretical explanation so far. We hope that results of our study, together with the studies of the condensate and incoherent wave initial conditions~\cite{agafontsev2014integrable,walczak2015optical,suret2016direct}, will help in development of a consistent theory of integrable turbulence. Our study has also a practical meaning. Solitons~(\ref{nlse_soliton}) are proposed as information bit in optical communications, which are generally very well described by the NLS equation. To increase communication bit-rate, it is necessary to pack these solitons sufficiently close to each other (see e.g.~\cite{mamyshev1991generation, haus1996solitons}). Our study shows, how often large waves may appear in some of the regimes of these communications.

In this publication we consider the following ensemble-averaged characteristics of the turbulence: (1) kinetic $\langle H_{d}(t)\rangle$ and potential $\langle H_{4}(t)\rangle$ energies, (2) wave-action spectrum $S_{k}(t)$ and spatial correlation function $g(x,t)$, and (3) moments of amplitude $M^{(n)}(t)$ and the PDF for relative intensity $\mathcal{P}(I,t)$. Here and below $\langle ... \rangle$ stands for arithmetic average across ensemble of initial conditions. We define wave-action spectrum as 
\begin{equation}\label{Ik}
S_{k}(t)=\langle|\Psi_{k}(t)|^{2}\rangle,
\end{equation}
where $\Psi_{k}(t)$ is Fourier transform of $\Psi(x,t)$,
\begin{eqnarray}
\Psi_{k}(t)&=&\mathscr{F}[\Psi(x,t)]=\frac{1}{L}\int_{-L/2}^{L/2}\Psi(x,t)e^{-ikx}\,dx, \label{Fourier1}\\
\Psi(x,t)&=&\mathscr{F}^{-1}[\Psi_{k}(t)]=\sum_{k}\Psi_{k}(t)e^{ikx}.\label{Fourier2}
\end{eqnarray}
Here $k=2\pi m/L$ is wavenumber and $m\in\mathbb{Z}$ is integer. Wave-action spectrum is the spectral density of wave action $\langle N\rangle$, 
\begin{eqnarray}\label{spectral_density}
\langle N\rangle = \langle |\Psi|^{2}\rangle = \sum_{k} S_{k}(t),
\end{eqnarray}
where $\langle |\Psi|^{2}\rangle$ is ensemble and space average of square amplitude. Spatial correlation function, 
\begin{equation}\label{gx}
g(x,t) = \bigg\langle\frac{1}{L}\int_{-L/2}^{L/2}\Psi(y,t)\Psi^{*}(y-x,t)\,dy\bigg\rangle \,/\langle N\rangle,
\end{equation}
is connected with wave-action spectrum as
\begin{equation}\label{gx2}
g(x,t) = \mathscr{F}^{-1}[S_{k}(t)]\,/\langle N\rangle.
\end{equation}
Due to this definition, at $x=0$ the correlation function is always fixed to unity, $g(0,t)=1$.

Moments of amplitude, 
\begin{eqnarray}\label{Mn}
M^{(n)}(t)=\bigg\langle\frac{1}{L}\int_{-L/2}^{+L/2} |\Psi(x,t)|^{n}\,dx\bigg\rangle,
\end{eqnarray}
are connected with the PDF $\mathcal{P}(|\Psi|,t)$ of wave amplitude $|\Psi|$ as 
\begin{equation}\label{MnPDF}
M^{(n)}(t)=\int_{0}^{+\infty} |\Psi|^{n} \mathcal{P}(|\Psi|,t)\,d|\Psi|.
\end{equation}
The second moment coincides with wave action, $M^{(2)}(t)=\langle |\Psi|^{2}\rangle=\langle N\rangle$, and thus does not change with time. Potential energy is connected with the 4th moment as $\langle H_{4}(t)\rangle=-M^{(4)}(t)/2$ (see e.g.~\cite{onorato2016origin} where this relation is extensively exploited). For exponential PDF~(\ref{Rayleigh}) the moments would be equal to
\begin{equation}\label{MnR}
M_{E}^{(n)}=\langle N\rangle^{n/2}\,\Gamma(n/2+1),
\end{equation}
where $\Gamma(m)$ is gamma-function. Below we will call moments~(\ref{MnR}) as exponential moments.

In the present study we demonstrate that, after development of the MI, all of the considered characteristics of the resulted integrable turbulence evolve with time in oscillatory way, approaching at late times to their asymptotics. 
Hence, one can say that MI of cnoidal waves lead to integrable turbulence, which asymptotically approaches in oscillatory way to its stationary state defined by infinite series of invariants~(\ref{all_integrals1}), (\ref{all_integrals2}). 
Numerical simulations presented below show that during the evolution toward the stationary state, kinetic $\langle H_{d}(t)\rangle$ and potential $\langle H_{4}(t)\rangle$ energies, as well as the moments $M^{(n)}(t)$, oscillate with time around their asymptotic values. 
The amplitudes of these oscillations decay with time as $t^{-\alpha}$,  $1<\alpha< 1.5$, the phases contain nonlinear phase shift decaying as $t^{-1/2}$, and the frequency of the oscillations is equal to the double maximal growth rate of the MI, $s=2\gamma_{\max}$.
Very similar oscillations are present in the condensate case~\cite{agafontsev2014integrable}. 
Remarkably, the asymptotic potential to kinetic energy ratio turns out to be equal to $Q_{A}=\langle H_{4}\rangle/\langle H_{d}\rangle=-2$ for all cnoidal waves of dn-branch. Wave-action spectrum, spatial correlation function and the PDF evolve coherently with oscillations of potential energy, so that at the local maximums and minimums of $|\langle H_{4}(t)\rangle|$ their evolution changes to roughly the opposite. 
We describe evolution of these functions in details in Sections~\ref{Sec:Evolution} and~\ref{Sec:Dependence} of this paper. 

For cnoidal waves with small $\omega_{1}$, we observe that wave field at all times remains close to a collection of solitons~(\ref{nlse_soliton}) with different positions and phases, even when the system is close to the asymptotic state. Thus, turbulence generated from MI of such cnoidal waves is indeed close to integrable soliton turbulence of very thin and high solitons~(\ref{nlse_soliton}). In its asymptotic stationary state, the PDF is significantly non-exponential and dynamics of the system reduces to two-soliton collisions. These collisions provide up to two-fold increase in amplitude and occur with exponentially small rate $\propto e^{-\pi\omega_{0}/\omega_{1}}$. The potential to kinetic energy ratio $Q(t)=\langle H_{4}(t)\rangle/\langle H_{d}(t)\rangle$ at all times remains very close to -2, the same as for a 
singular soliton~(\ref{nlse_soliton}). 
For cnoidal waves with large $\omega_{1}$, the asymptotic PDF coincides with exponential PDF~(\ref{Rayleigh}).

The properties of the integrable turbulence change gradually with $\omega_{1}$. MI of cnoidal waves with ``intermediate'' $\omega_{1}$ lead to turbulence with ``intermediate'' properties between the two limits $\omega_{1}\to 0$ and $\omega_{1}\to +\infty$. All rogue waves that we examined, for all cnoidal waves that we studied, at the time of their maximal elevation have quasi-rational profile similar to that of the Peregrine solution~\cite{peregrine1983water,kibler2010peregrine} of the NLS equation.

The paper is organized as follows. In Section~\ref{Sec:NumMeth} we describe numerical methods used in our study. In Section~\ref{Sec:Evolution} we consider the general properties of integrable turbulence, that develops from MI of dn-branch of cnoidal waves, on the example of one cnoidal wave with fixed imaginary half-period $\omega_{1}$. The dependence of these properties on $\omega_{1}$ is described in Section~\ref{Sec:Dependence}. The final Section~\ref{Sec:Conclusions} contains conclusions. 
In Appendix~\ref{Sec:Annex-A} we explain cnoidal wave solutions in more details, and in Appendix~\ref{Sec:Annex-B} we demonstrate how MI develops on the background of three cnoidal waves with $\omega_{1}=0.8$ (weak overlapping), $\omega_{1}=1.6$ (moderate overlapping) and $\omega_{1}=5$ (strong overlapping).


\section{Numerical methods}
\label{Sec:NumMeth}

We integrate Eq.~(\ref{nlse_stationary}) numerically in the box $x\in[-L/2, L/2]$ with periodic boundary. Typically, we use $L=256\pi$ and integrate Eq.~(\ref{nlse_stationary}) up to final time $t=200$, but in some cases we use larger boxes and/or integration times. Large integration times $t\geq 200$ are necessary, since for our initial conditions the evolution close to the asymptotic stationary state is a very long process. Large boxes $L$ are necessary, because starting from some time $T$ we encounter with recurrence, which might be connected with Fermi - Pasta - Ulam (FPU) phenomenon~\cite{fermi1955studies, infeld1981quantitive}. The time of this recurrence increases linearly with the box size~\cite{agafontsev2014integrable}, $T\propto L$. To avoid its influence on our results, we used sufficiently large boxes $L$ and additionally run our experiments on twice larger boxes $2L$ to ensure that our results quantitatively do not depend on $L$.

As in~\cite{agafontsev2014integrable}, we use Runge-Kutta 4th-order method with adaptive change of the spatial grid size $\Delta x$ and Fourier interpolation of the solution between the grids. To avoid appearance of numerical instabilities, time step $\Delta t$ changes with $\Delta x$ as $\Delta t = h\Delta x^{2}$, $h \le 0.1$. 
The simulations conserve the first 10 invariants~(\ref{all_integrals1}), (\ref{all_integrals2}) with accuracy better than $10^{-6}$. Note, that we measure relative errors  for integrals $c_{n}[\Psi]$ with odd orders $\mathrm{mod}(n,2)=1$ and absolute errors for integrals with even orders $\mathrm{mod}(n,2)=0$, since for unperturbed cnoidal waves~(\ref{cnoidal1_stationary}) the latter ones are zeroth. The first three invariants -- wave action~(\ref{wave_action}), momentum~(\ref{momentum}) and total energy~(\ref{energy}) -- are conserved with accuracy better than $10^{-10}$.

We start simulations in grid with $M=16\,384$ nodes (or proportionally larger when using larger boxes $L$) from initial conditions~(\ref{cnoidal1_MI}) where the real half-period is fixed to $\omega_{0}=\pi$. For each of the studied cnoidal wave characterized by imaginary half-period $\omega_{1}$, we average our results across ensemble of 1000 random realizations of initial noise $\epsilon(x)$. We use statistically homogeneous in space noise 
\begin{equation}\label{noise}
\epsilon(x)=A_{0}\bigg(\frac{\sqrt{8\pi}}{\theta L}\bigg)^{1/2} \sum_{k}e^{-k^{2}/\theta^{2}+i\xi_{k}+ikx},
\end{equation}
where $A_{0}$ is noise amplitude, $k=2\pi\,m/L$ is wavenumber, $m\in\mathbb{Z}$ is integer, $\theta$ is noise width in k-space and $\xi_{k}$ are arbitrary phases for each $k$ and each noise realization within the ensemble of initial conditions. As was shown in~\cite{agafontsev2014integrable}, in x-space the average square amplitude of such noise is $\langle|\epsilon|^{2}\rangle=A_{0}^{2}$. Below we will present our results for initial noise with parameters $A_{0}=10^{-5}$ and $\theta=5$. We performed experiments with other parameters $A_{0}$ and $\theta$ too, but didn't find significant difference. We checked our statistical results against the size of the ensembles and parameters of our numerical scheme, and found no difference. 


\section{Evolution toward the asymptotic stationary state}
\label{Sec:Evolution}

The results of this section are illustrated on the example of MI of cnoidal wave~(\ref{cnoidal1_stationary}) with $\omega_{1}=1.6$, which is an ``intermediate'' cnoidal wave between the two limits $\omega_{1}\to 0$ and $\omega_{1}\to +\infty$, see Fig.~\ref{fig:cnoidal_wave_B1}(a). The corresponding numerical simulations were carried out in the box $L=1024\pi$ up to final time $t=2000$. Cnoidal waves with other $\omega_{1}$ give qualitatively similar results; the dependence on $\omega_{1}$ will be considered in more details in Section~\ref{Sec:Dependence}.

\begin{figure}[t] \centering
\includegraphics[width=8.0cm]{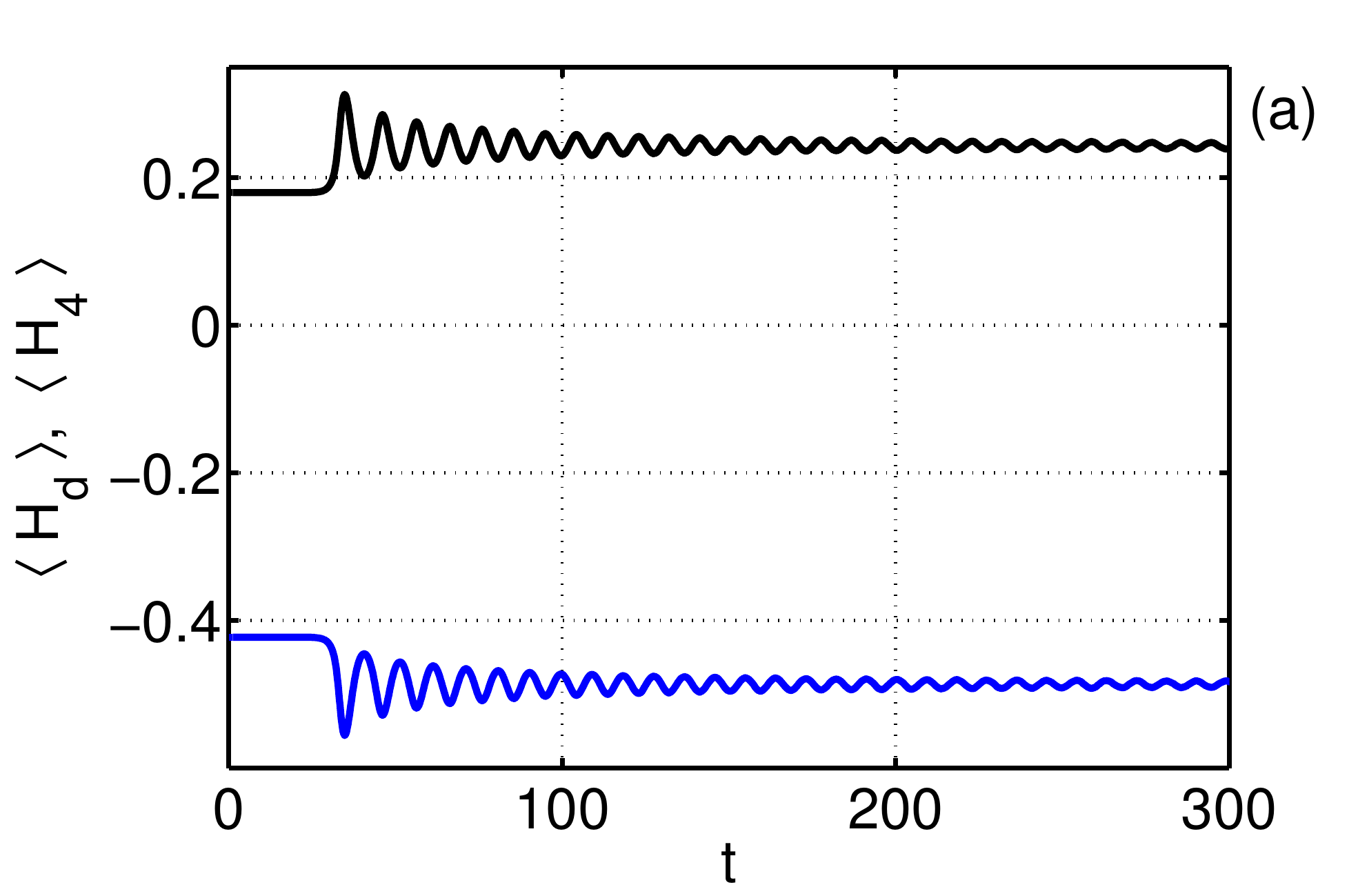}
\includegraphics[width=8.0cm]{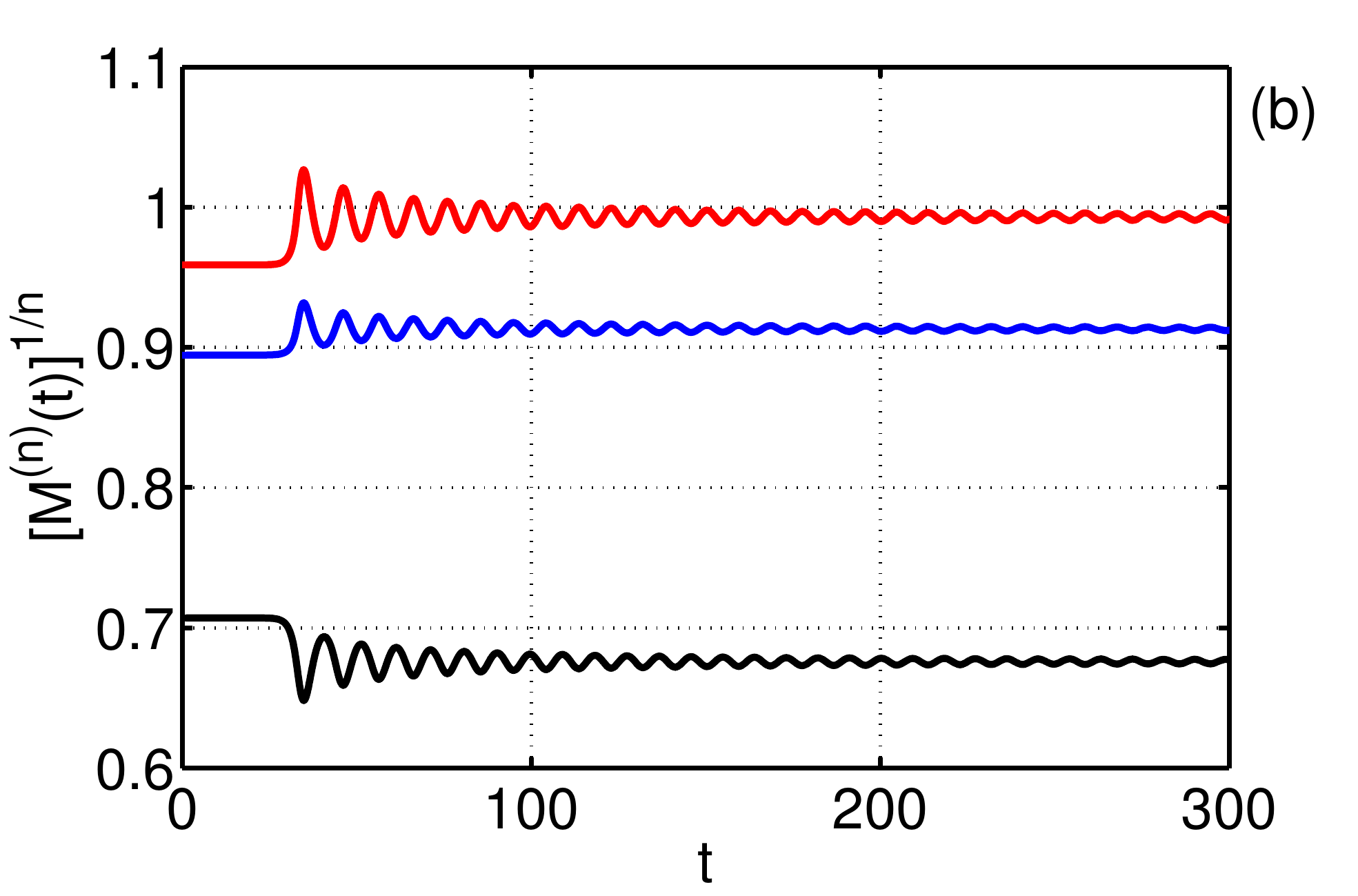}

\caption{\small {\it (Color on-line)} 
(a) Evolution of ensemble average kinetic $\langle H_{d}(t)\rangle$ (black) and potential $\langle H_{4}(t)\rangle$ (blue) energies. In the asymptotic stationary state their ratio is equal to $Q_{A} = \langle H_{4}\rangle/\langle H_{d}\rangle = -2$. 
(b) Evolution of moments $M^{(1)}(t)$ (black), $[M^{(3)}(t)]^{1/3}$ (blue) and $[M^{(4)}(t)]^{1/4}$ (red).}
\label{fig:evolutionEM}
\end{figure}

MI of cnoidal wave~(\ref{cnoidal1_stationary}) with $\omega_{1}=1.6$ has maximal increment~(\ref{cnoidal1_increment}) $\gamma_{\max}=0.356$ and reaches its nonlinear stage at about $t\sim 30$ (see Fig.~\ref{fig:CW2_MI} in Appendix~\ref{Sec:Annex-B}). Then, all statistical characteristics that we studied start to evolve in oscillatory way, approaching at late times to their asymptotics. The example of such evolution for kinetic $\langle H_{d}(t)\rangle$ and potential $\langle H_{4}(t)\rangle$ energies, and also moments $M^{(n)}(t)$, is shown in Fig.~\ref{fig:evolutionEM}(a),(b). 
Thus, one can conclude that after development of the MI, the system asymptotically approaches in oscillatory way to its stationary state, which in its turn is defined by infinite series of invariants~(\ref{all_integrals1}),~(\ref{all_integrals2}).
In order to determine characteristics of this asymptotic state (e.g. kinetic and potential energies, the moments, the PDF, etc.), we average the corresponding functions both across the ensemble of initial conditions and over time close to the asymptotic state $t\in[1800, 2000]$.

\begin{figure}[t] \centering
\includegraphics[width=16cm]{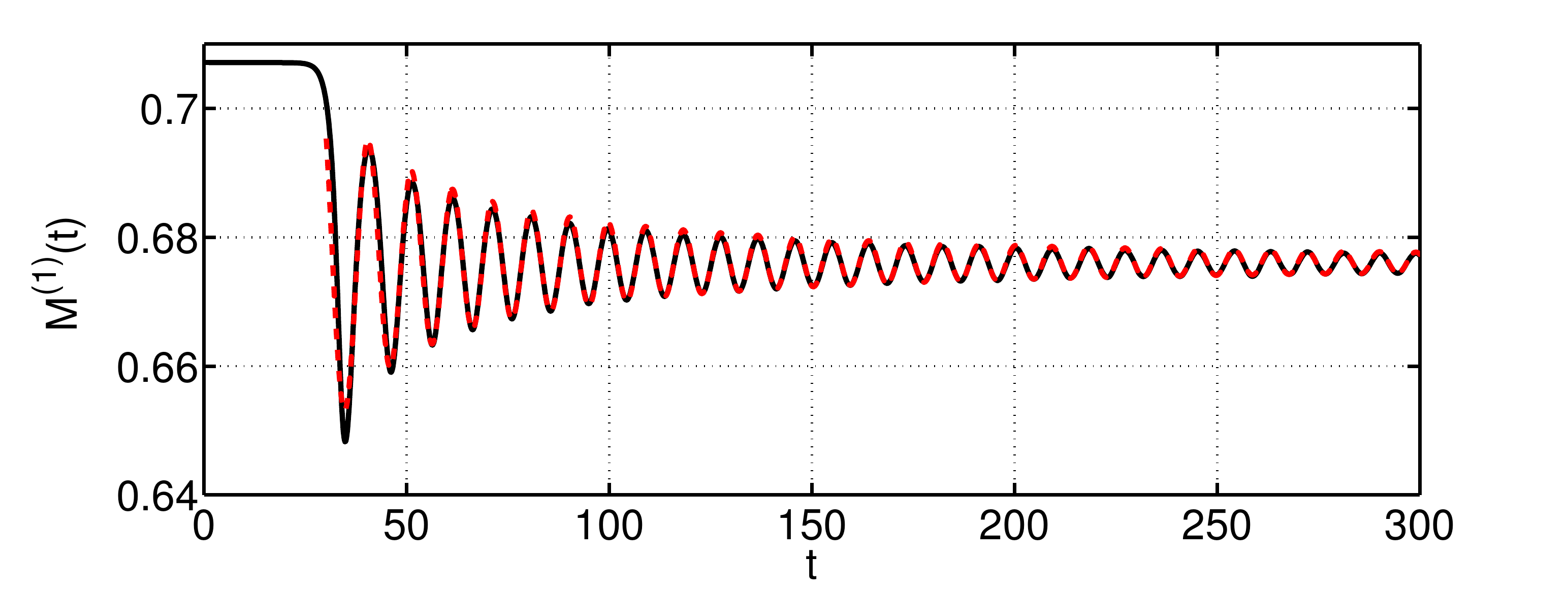}

\caption{\small {\it (Color on-line)} Evolution of moment $M^{(1)}(t)$ (solid black) and its fit by function $f(t)=M_{A}^{(1)}+[p/t^{\alpha}]\,\sin(st+q/\sqrt{t}+\Phi_{0})$ with parameters $M_{A}^{(1)}=0.676$, $\alpha=1.23$, $p=1.82$, $s=0.71$, $q=74.6$, $\Phi_{0}=-1.23$ (dashed red).}
\label{fig:evolutionM1fit}
\end{figure}

\begin{figure}[t] \centering
\includegraphics[width=8.0cm]{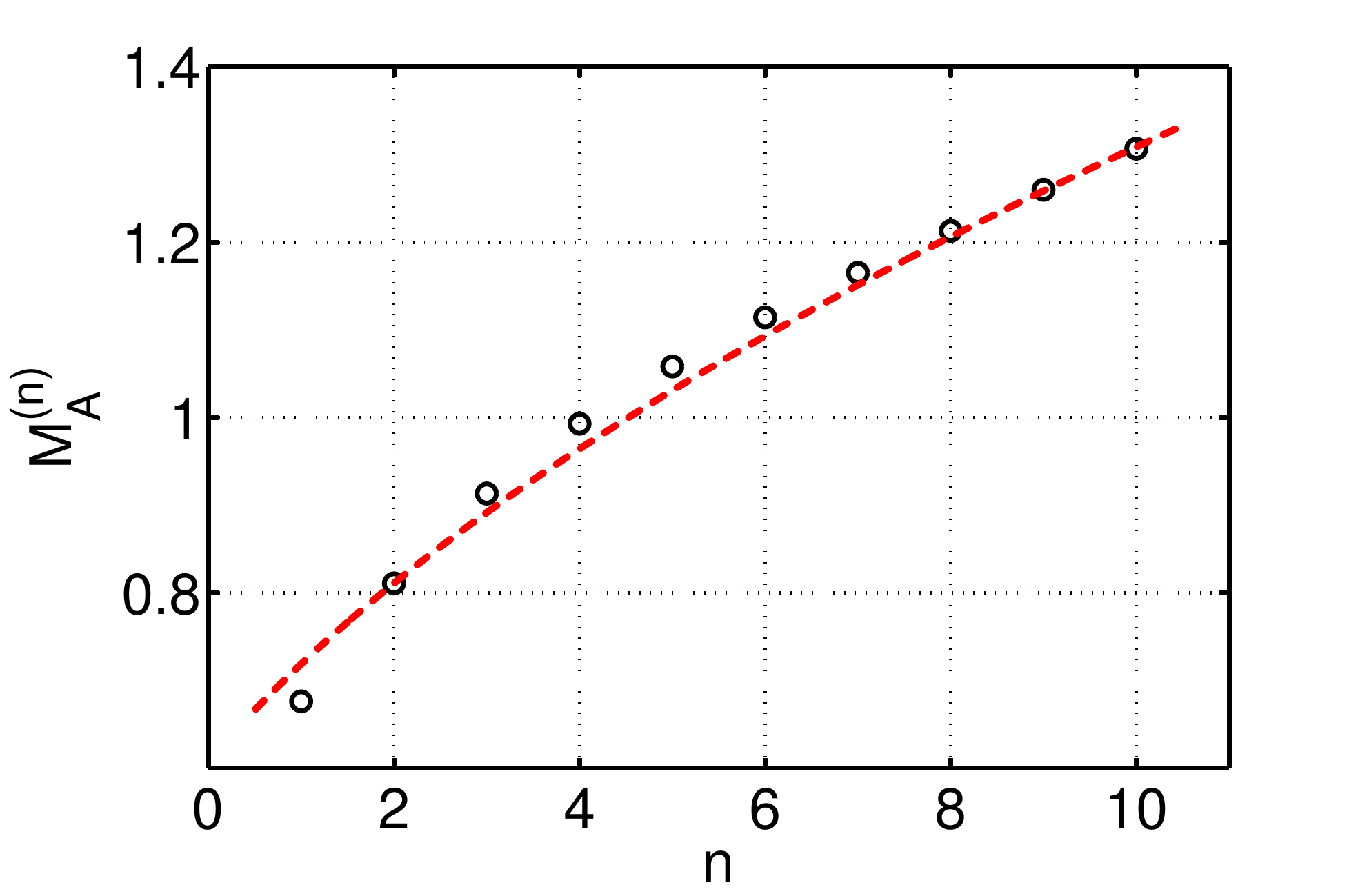}

\caption{\small {\it (Color on-line)} Asymptotic moments $[M^{(n)}_{A}]^{1/n}$, $n=1,...,10$ (black circles) and exponential moments $[M^{(n)}_{E}]^{1/n}$~(\ref{MnR}) (dashed red line).}
\label{fig:moments_asymptotic}
\end{figure}

The potential to kinetic energy ratio $Q(t)=\langle H_{4}(t)\rangle/\langle H_{d}(t)\rangle$ changes from $Q(t)=-2.3$ at $t=0$ to $Q_{A}=-2$ in the asymptotic stationary state. The same asymptotic energy ratio $Q_{A}=-2$ is observed for MI of the condensate. Following~\cite{agafontsev2014integrable}, we determine that in the nonlinear stage of the MI the evolution of moments $M^{(n)}(t)$ is very well approximated by functions
\begin{equation}\label{oscillations}
M^{(n)}(t)\approx M^{(n)}_{A} + [p/t^{\alpha}]\sin(st+q/\sqrt{t}+\Phi_{0}),
\end{equation}
where $M^{(n)}_{A}$ are asymptotic moments and $\alpha$, $p$, $s$, $q$, $\Phi_{0}$ are constants (different for different moments). The example of such approximation for the first moment $M^{(1)}(t)$ is shown in Fig.~\ref{fig:evolutionM1fit}. Kinetic $\langle H_{d}(t)\rangle$ and potential $\langle H_{4}(t)\rangle$ energies also oscillate according to~(\ref{oscillations}) due to relation $\langle H_{4}(t)\rangle=-M^{(4)}(t)/2$ and conservation of total energy; the second moment does not oscillate since $M^{(2)}(t)=\langle N\rangle$. Potential energy $\langle H_{4}(t)\rangle$ and the first moment $M^{(1)}(t)$ oscillate in-phase, with parameters $s=0.71$, $q=74.6$ and $\Phi_{0}=-1.23$. Kinetic energy $\langle H_{d}(t)\rangle$ and higher moments $M^{(n)}(t)$, $n\ge 3$, oscillate with parameters $s=0.71$, $q=74.6$ and $\Phi_{0}=1.91$, that is in-phase with each other and exactly anti-phase with potential energy and the first moment. 
Amplitudes of these oscillations decay with time by power law $\propto t^{-\alpha}$, with the exponent decreasing from $\alpha=1.23$ for the first moment to $\alpha=1.08$ for the 10th moment. 
Asymptotic moments $M^{(n)}_{A}$ slightly differ from exponential moments~(\ref{MnR}), as shown in Fig.~\ref{fig:moments_asymptotic}.

Wave-action spectrum of the original cnoidal wave represents a collection of peaks at integer wavenumbers $k_{0}\in\mathbb{Z}$. Since for all cnoidal waves~(\ref{cnoidal1_stationary}) the following equality is valid,
$$
\Psi_{k=0}=\frac{1}{L}\int_{-L/2}^{L/2}\Psi_{dn}(x)\,dx = \pi/\sqrt{2}\,\omega_{0},
$$
the peak at zeroth harmonic is equal to $S_{0}=|\Psi_{k=0}|^{2}=0.5$. Peaks at nonzero wavenumbers $|k_{0}|>0$ nearly exponentially decay with $|k_{0}|$. The spatial correlation function for unperturbed cnoidal wave is periodic with the same period $2\pi$ as the original cnoidal wave, everywhere positive $g(x)>0$, takes maximal values $\max g(x)=g(0)=1$ at $x=2\pi m$, $m\in\mathbb{Z}$, and minimal values at $x=2\pi (m+1/2)$.

\begin{figure}[t] \centering
\includegraphics[width=8.0cm]{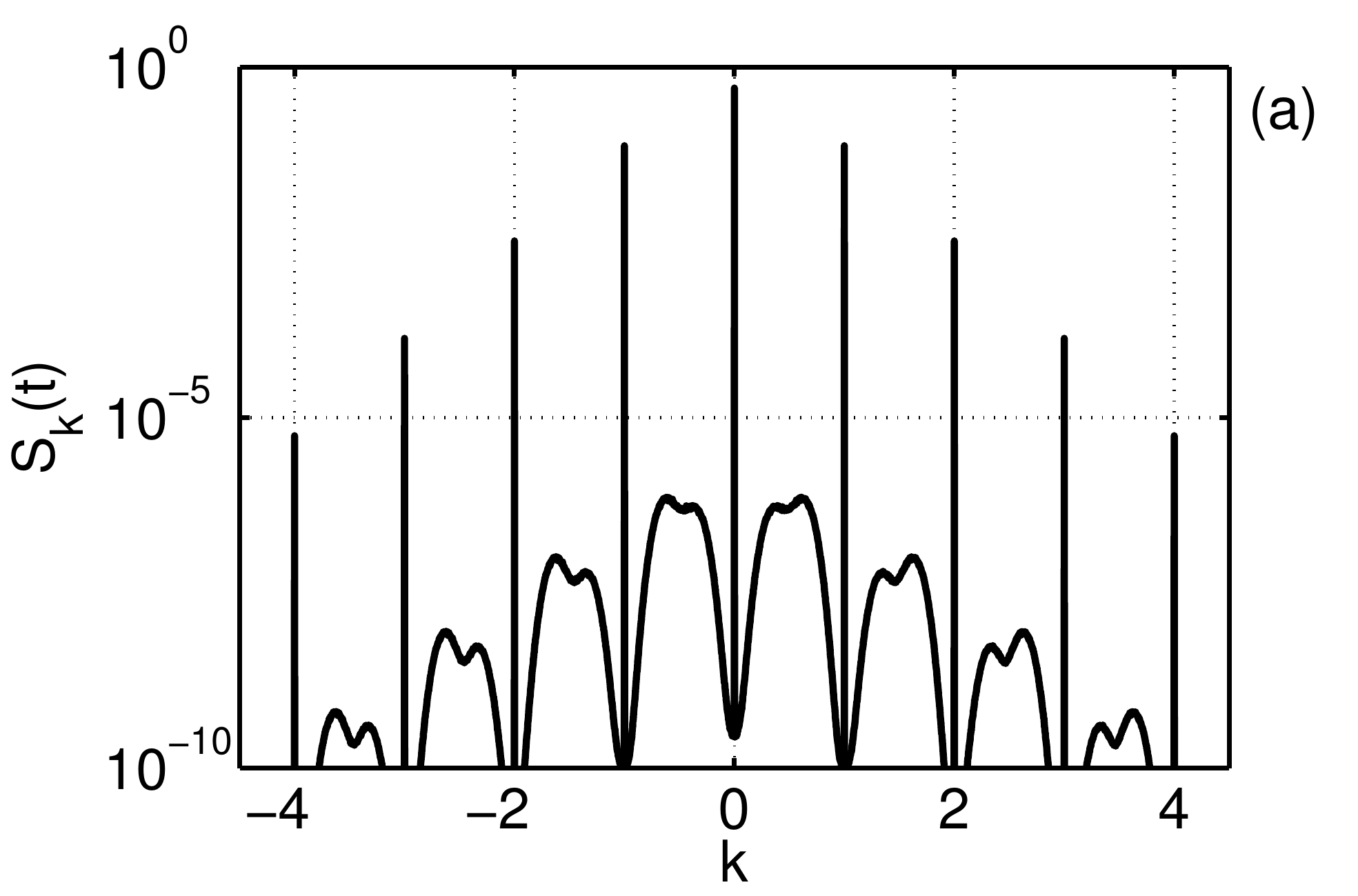}
\includegraphics[width=8.0cm]{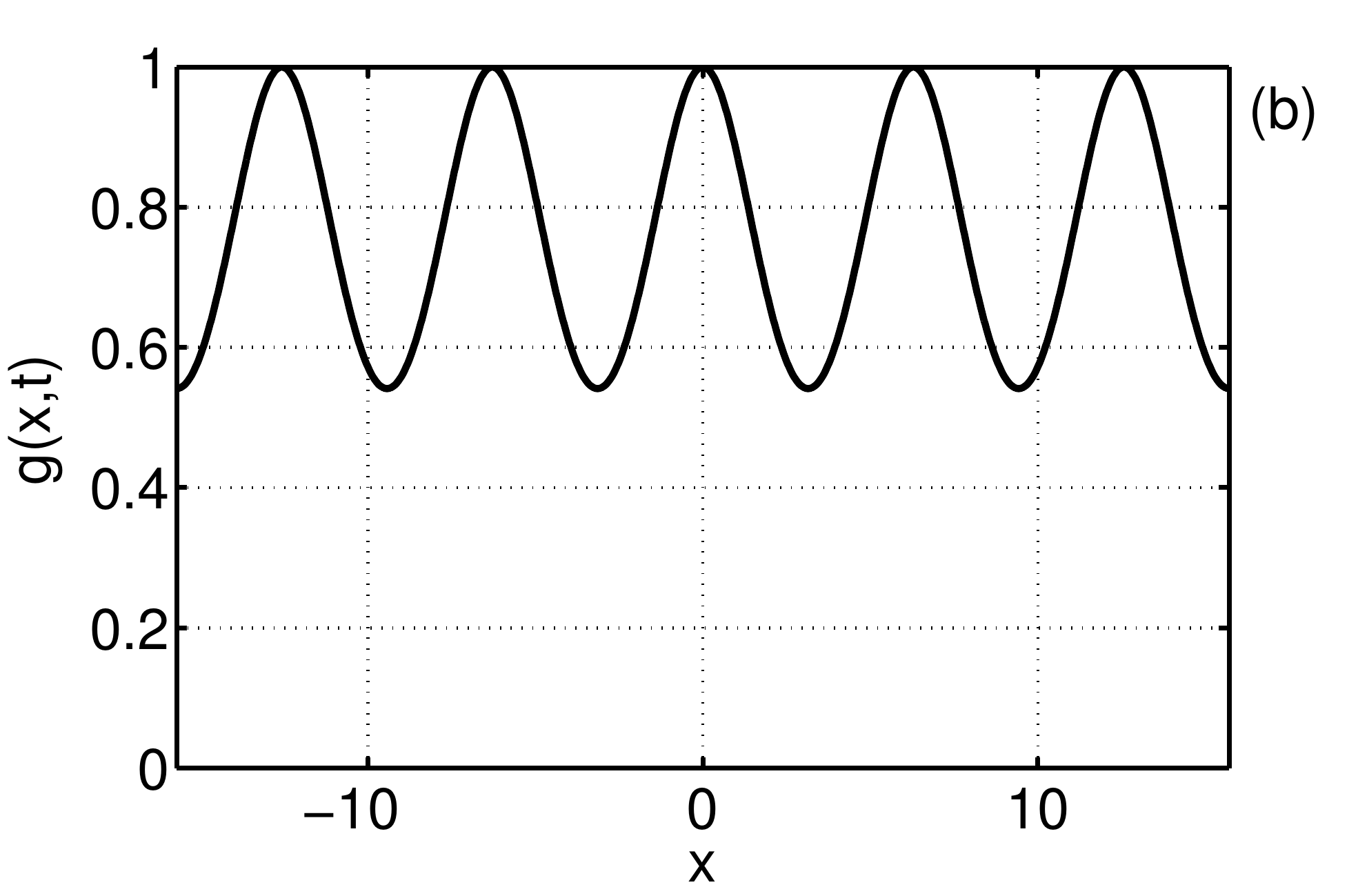}

\caption{\small Wave-action spectrum $S_{k}(t)$ (a) and spatial correlation function $g(x,t)$ (b) in the linear stage of MI at $t=20$. }
\label{fig:MI_linear}
\end{figure}

\begin{figure}[t] \centering
\includegraphics[width=8.0cm]{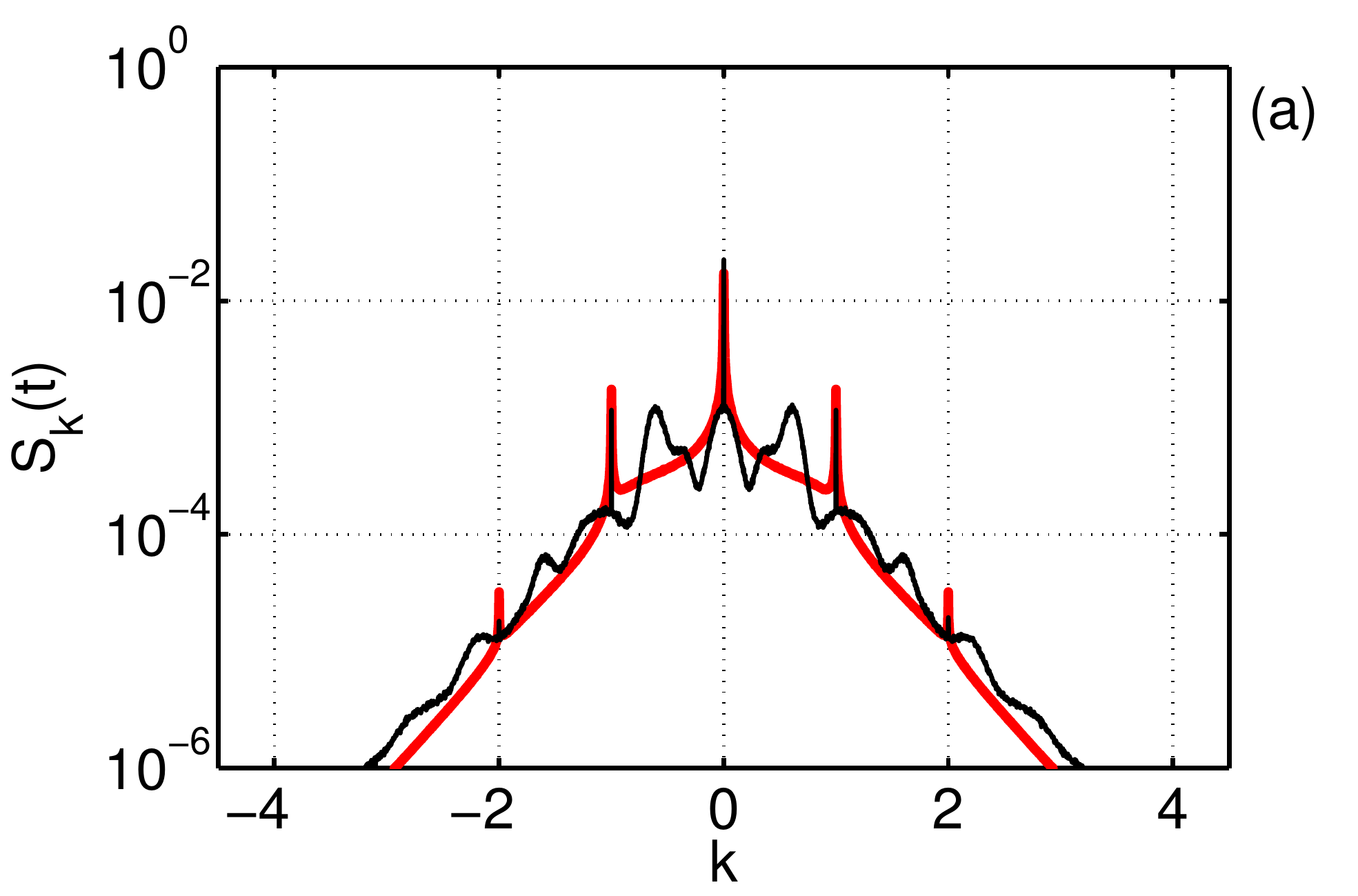}
\includegraphics[width=8.0cm]{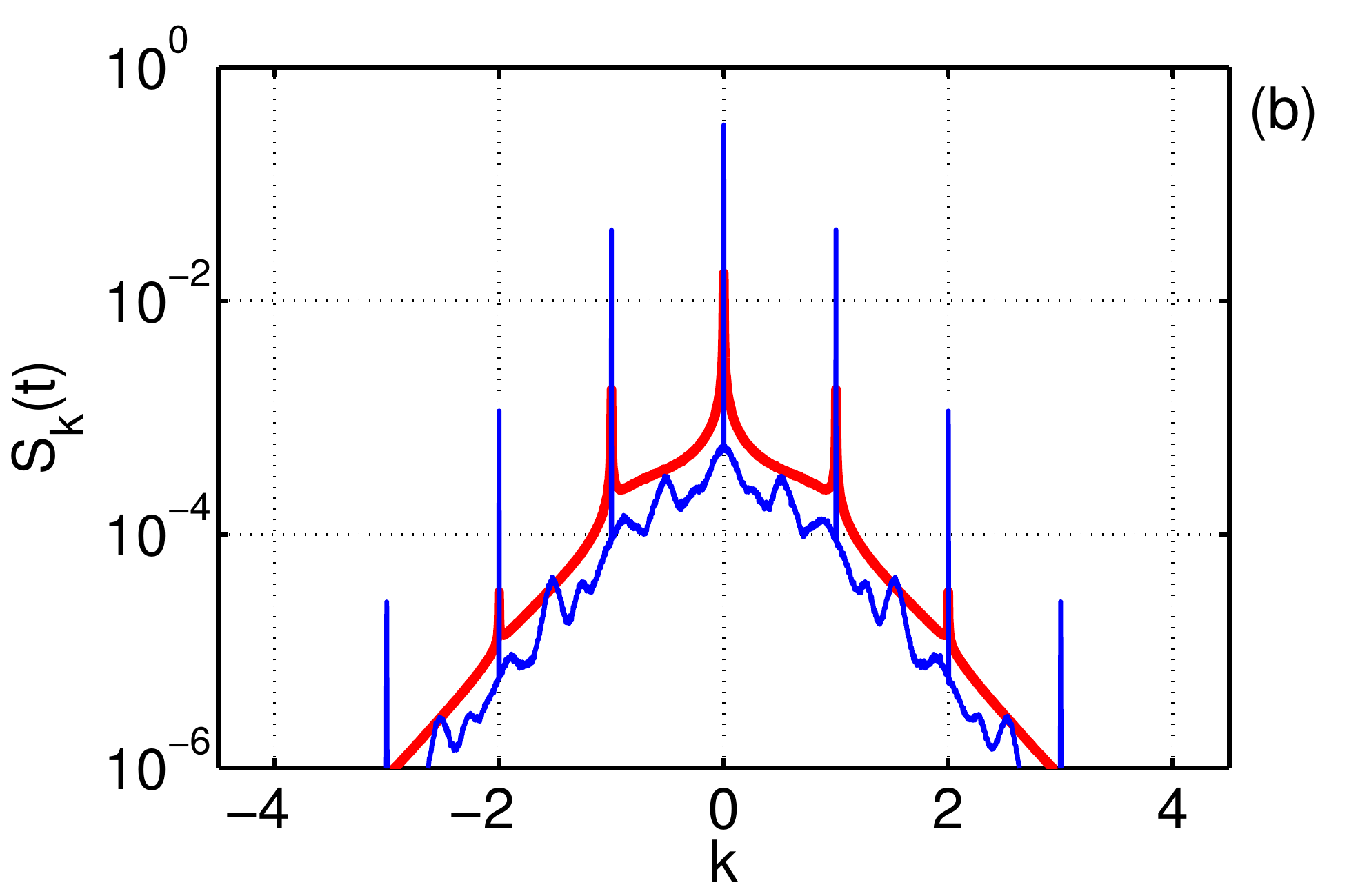}

\caption{\small {\it (Color on-line)} Wave-action spectrum $S_{k}(t)$: (a) at the first local maximum of potential energy modulus $|\langle H_{4}(t)\rangle|$ at $t=34.8$ (black) and (b) at the first local minimum of $|\langle H_{4}(t)\rangle|$ at $t=40.6$ (blue). The asymptotic wave-action spectrum is shown in thick red.}
\label{fig:spectra12}
\end{figure}

\begin{figure}[t] \centering
\includegraphics[width=8.0cm]{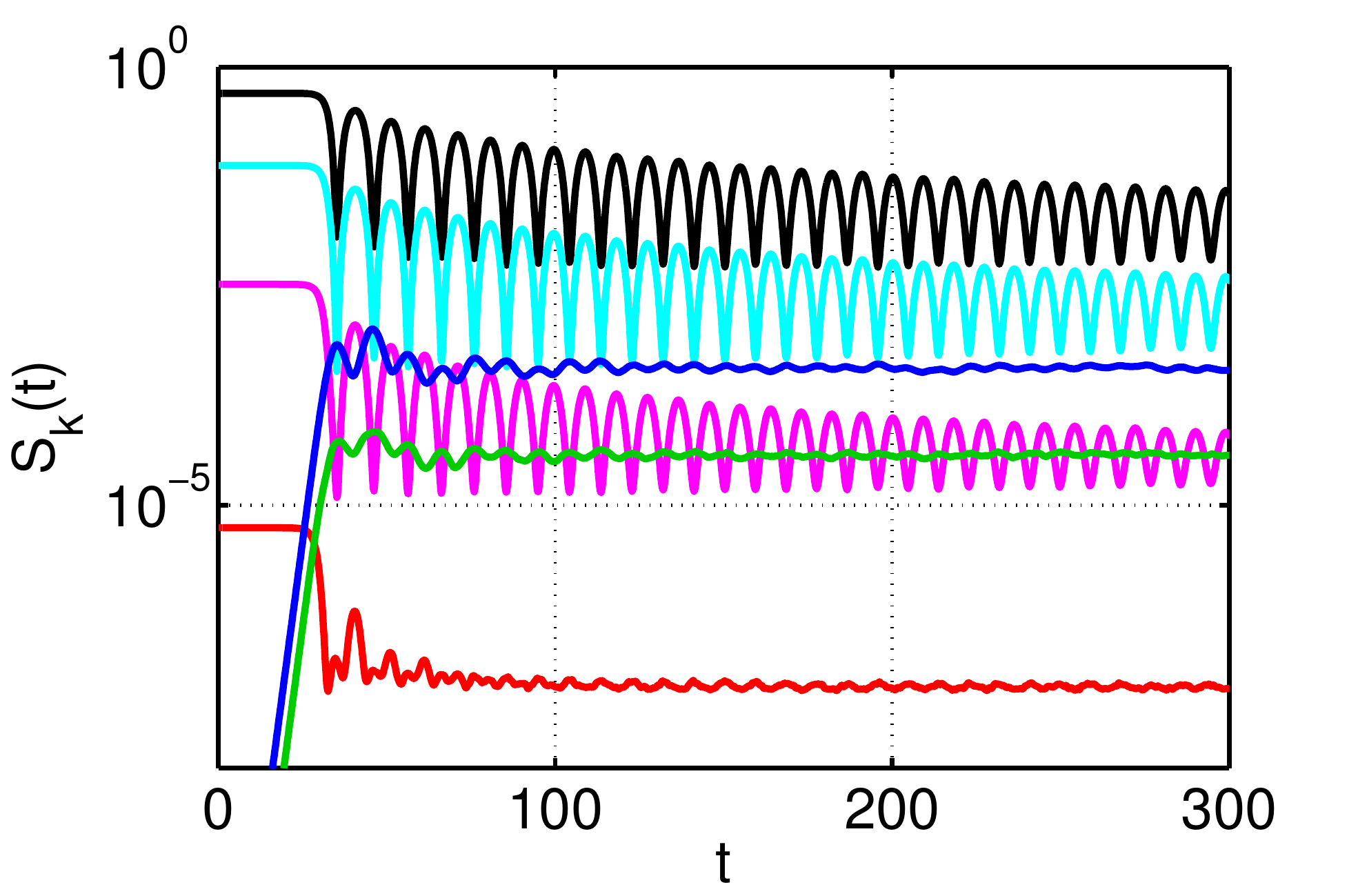}

\caption{\small {\it (Color on-line)} Time dependence of wave-action spectrum $S_{k}(t)$ at $k=0$ (black), $k=0.5$ (blue), $k=1$ (cyan), $k=1.5$ (green), $k=2$ (pink) and $k=4$ (red).}
\label{fig:evolution_spectra}
\end{figure}

\begin{figure}[t] \centering
\includegraphics[width=8.0cm]{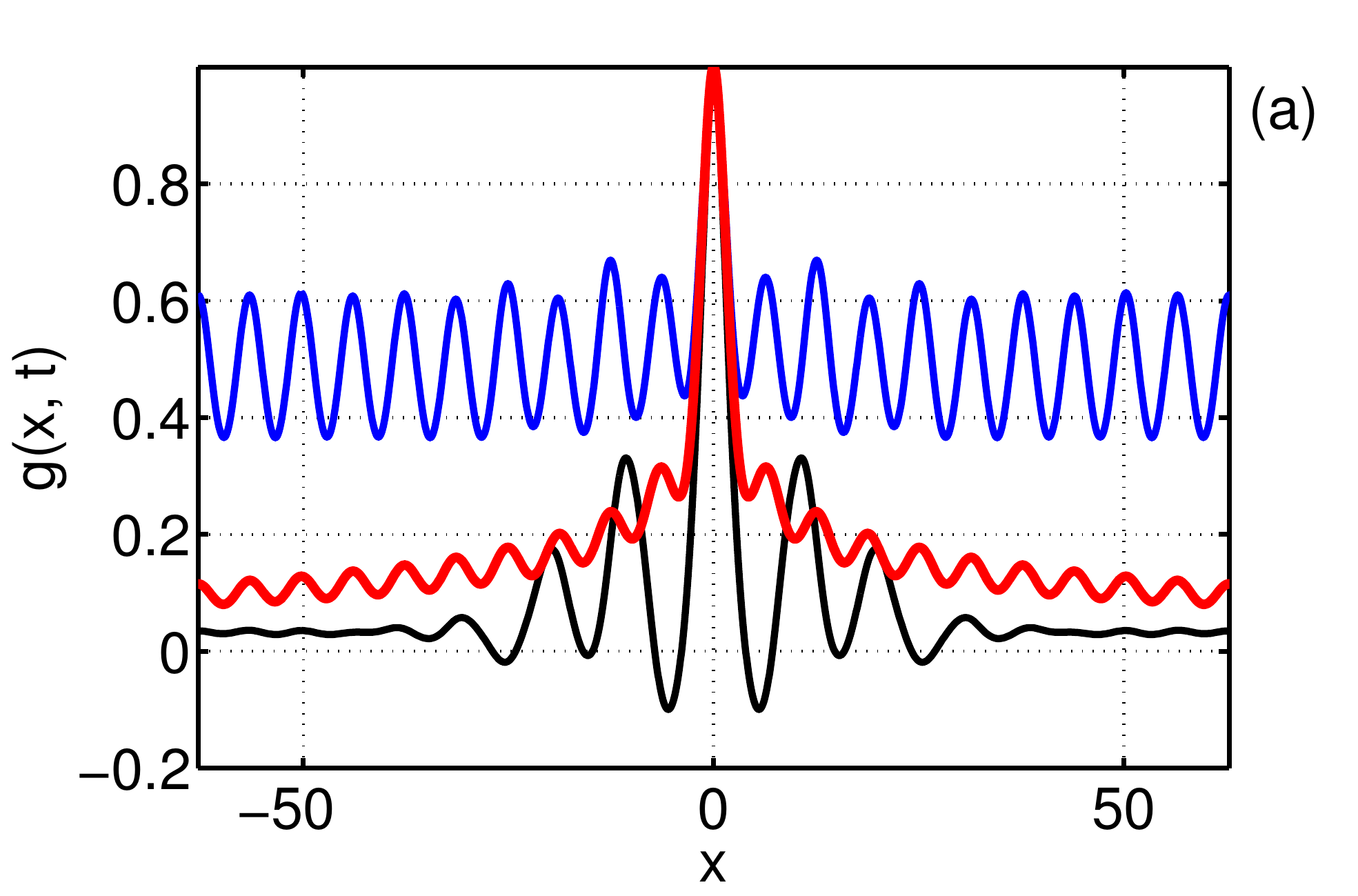}
\includegraphics[width=8.0cm]{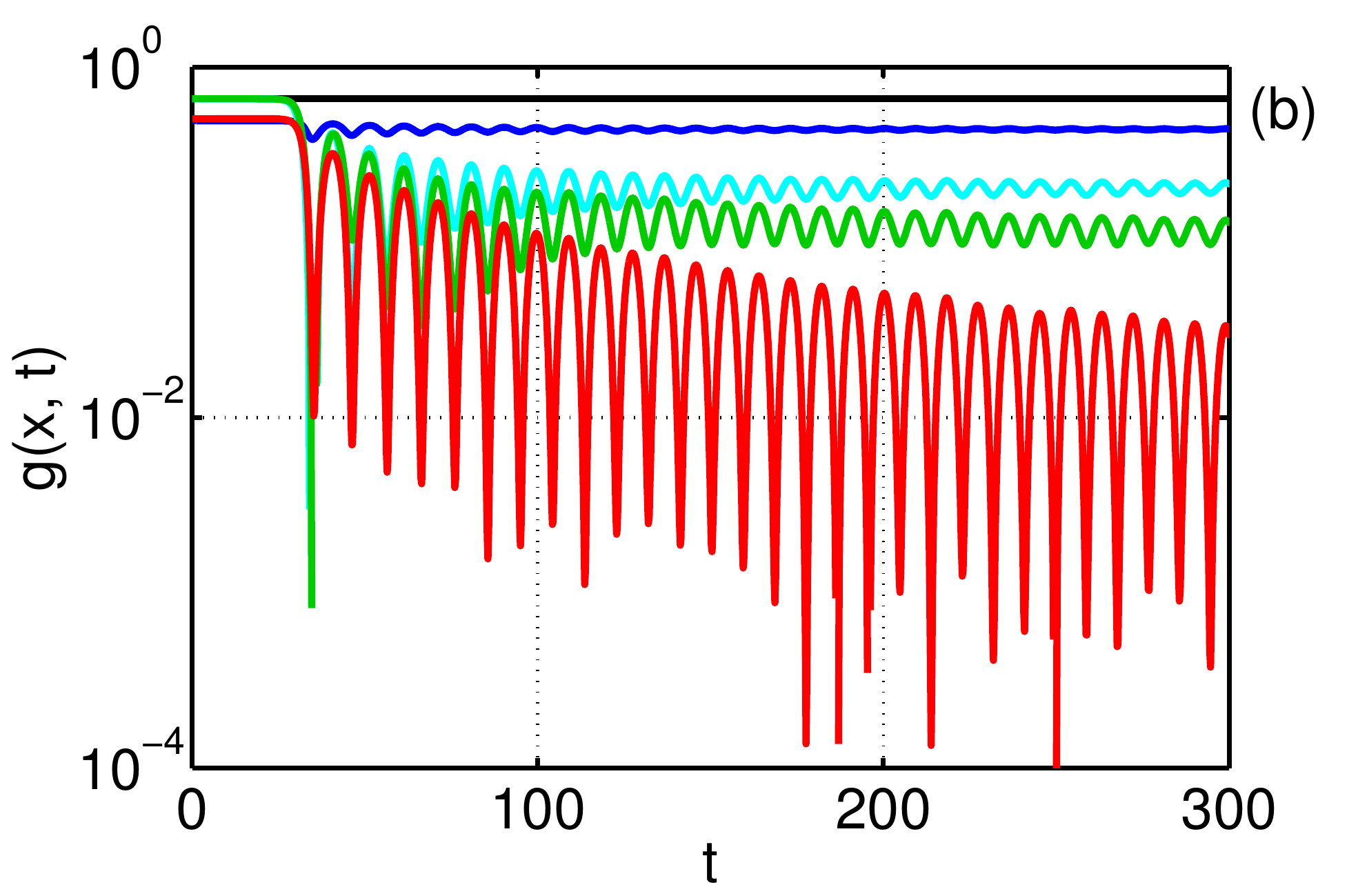}

\caption{\small {\it (Color on-line)} 
(a) Spatial correlation function $g(x,t)$ at the first local maximum of potential energy modulus $|\langle H_{4}(t)\rangle|$ at $t=34.8$ (black) and the first local minimum of $|\langle H_{4}(t)\rangle|$ at $t=40.6$ (blue). The asymptotic correlation function is shown in thick red.
(b) Time dependence of spatial correlation function $g(x,t)$ at $x=0$ (black), $x=\pi/2$ (blue), $x=2\pi$ (cyan), $x=8\pi$ (green), and at the border of the computational box $x=L/2$ (red).
}
\label{fig:corr_x12}
\end{figure}

In the linear stage of MI, wave-action spectrum $S_{k}(t)$ at non-integer wavenumbers starts to rise; the fastest increase is observed close to half-integer wavenumbers, Fig.~\ref{fig:MI_linear}(a). 
At this time the spatial correlation function $g(x,t)$ does not change visibly, Fig.~\ref{fig:MI_linear}(b). 
In the nonlinear stage, the spectrum and the correlation function evolve with time in oscillatory way approaching at late times to their asymptotics. The ``turning points'' for this oscillatory evolution -- i.e. the points in time when their evolution changes to roughly the opposite -- coincide with time when moments $M^{(n)}(t)$, and also kinetic $\langle H_{d}(t)\rangle$ and potential $\langle H_{4}(t)\rangle$ energies, take their maximal or minimal values. For definiteness, below we will refer to such points in time on the example of local maximums and minimums of potential energy modulus $|\langle H_{4}(t)\rangle|$. 
At the local maximums of $|\langle H_{4}(t)\rangle|$, the peaks at integer wavenumbers $k_{0}$ in the spectrum $S_{k}(t)$ are the smallest and the rest of the spectrum is maximally excited, Fig.~\ref{fig:spectra12}(a),~\ref{fig:evolution_spectra}, while the correlation function $g(x,t)$ takes (locally in time) minimal values at $|x|>0$, Fig.~\ref{fig:corr_x12}(a),(b). At the local minimums of $|\langle H_{4}(t)\rangle|$, the peaks at $k_{0}$ are the largest and the rest of the spectrum is minimally excited, Fig.~\ref{fig:spectra12}(b),~\ref{fig:evolution_spectra}, while the correlation function takes (locally in time) maximal values at $|x|>0$, Fig.~\ref{fig:corr_x12}(a),(b). 
Thus, one can say that during the evolution toward the asymptotic state, wave action is being ``pumped'' in oscillatory way between the peaks at integer wavenumbers and the rest of the spectrum, while the correlation function ``forms'' its tails at large lengths.

\begin{figure}[t] \centering
\includegraphics[width=8.0cm]{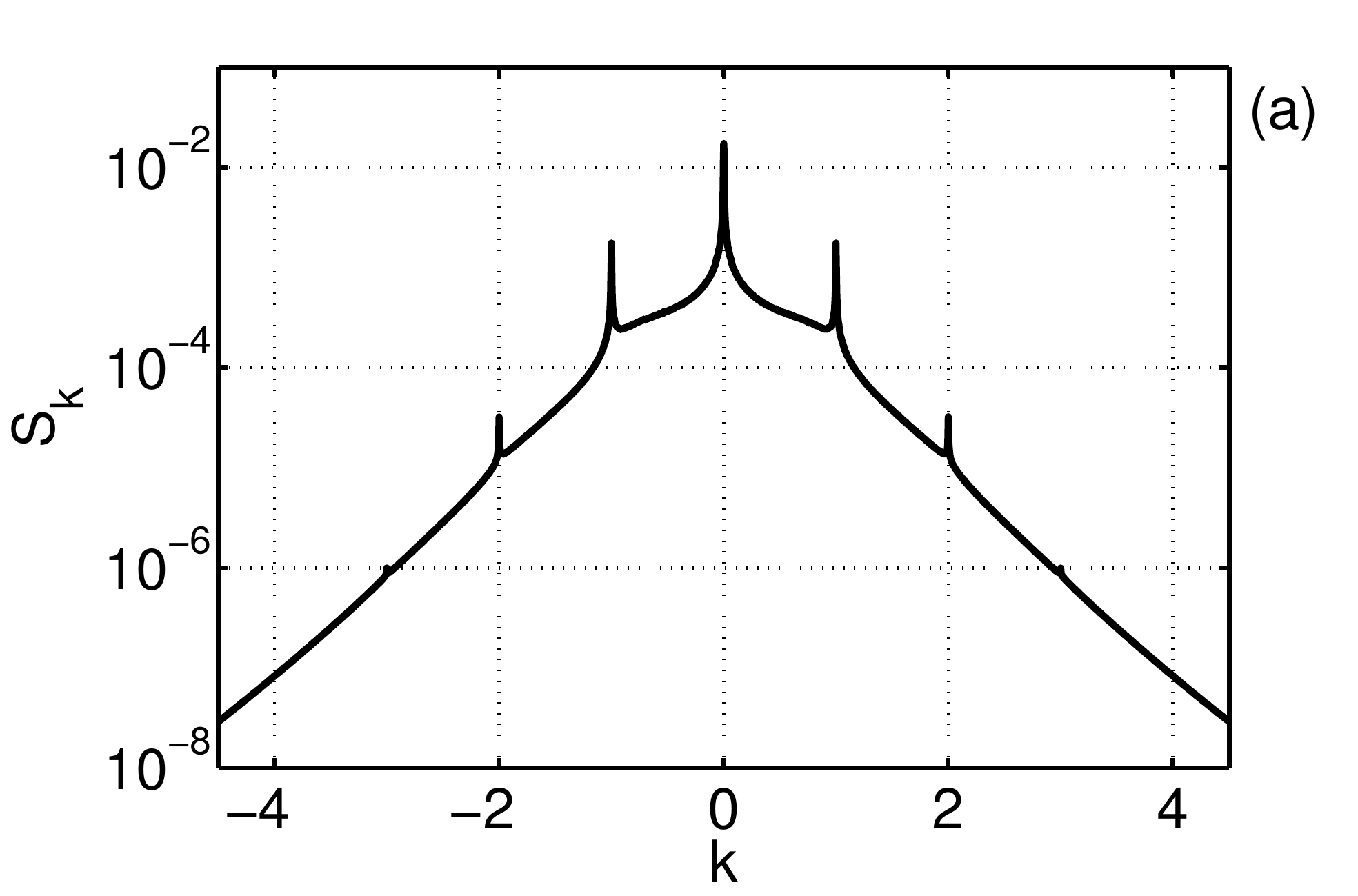}
\includegraphics[width=8.0cm]{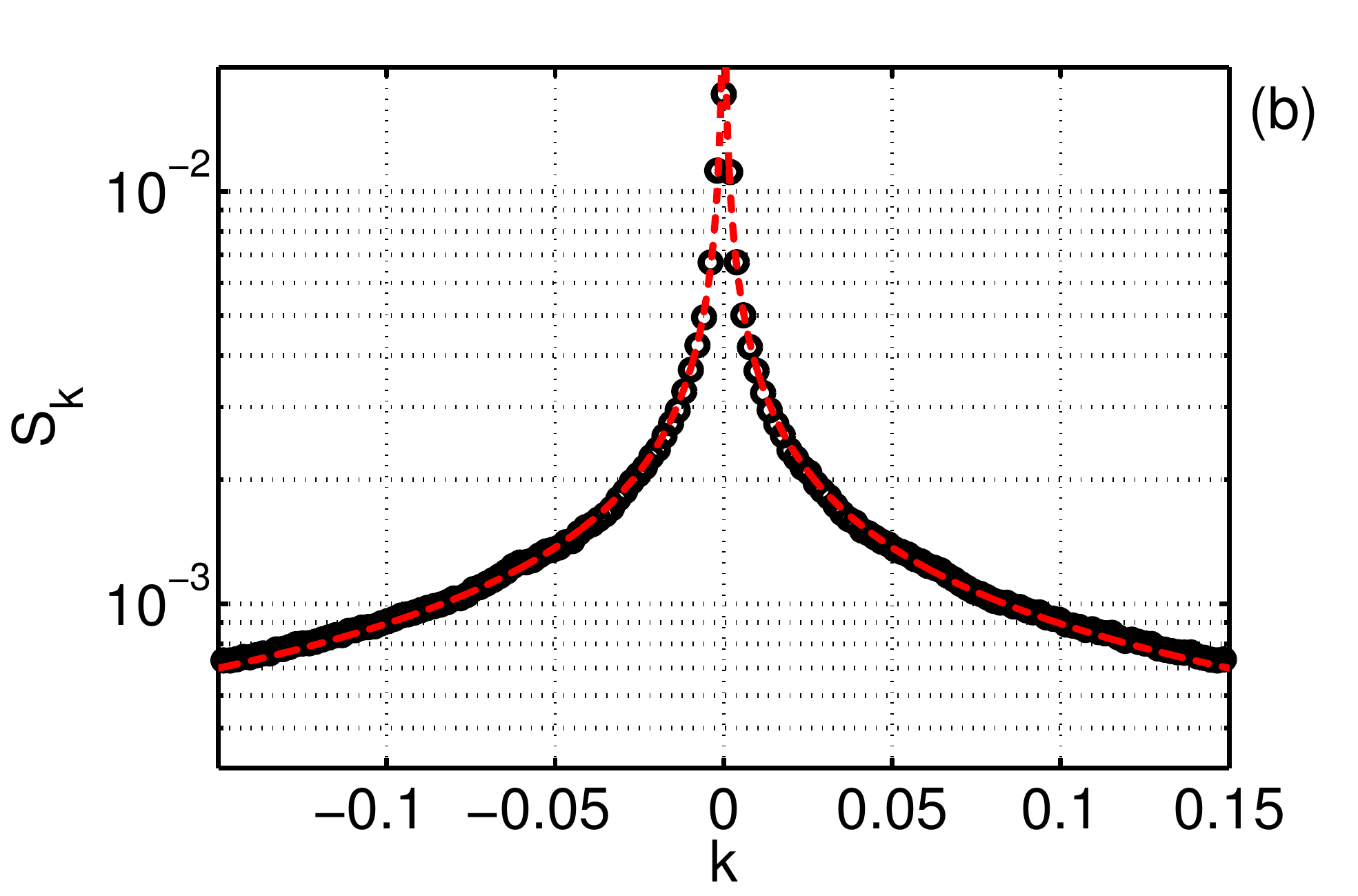}

\caption{\small {\it (Color on-line)} 
(a) Asymptotic wave-action spectrum $S_{k}$.
(b) Asymptotic wave-action spectrum $S_{k}$ in the vicinity of $k=0$ (black circles) and its fit by function $f(k)=b|k|^{-\beta}$, $b=2.2\times 10^{-4}$, $\beta=0.61$ (dashed red). At $k=0$ the asymptotic spectrum is finite, $S_{0}=1.72\times 10^{-2}$.
Graphs (a) and (b) contain about 4600 and 150 harmonics respectively, with the distance between them $\Delta k=2\pi/L=1/512$.}
\label{fig:spectra0}
\end{figure}

The asymptotic wave-action spectrum decays exponentially at large $k$ as $\propto e^{-\rho |k|}$, $\rho= 1.15$, and contains peaks at $k_{0}=0,\pm 1, \pm 2, \pm 3$, Fig.~\ref{fig:spectra0}(a). Contrary to the original cnoidal wave, these peaks now occupy not only integer wavenumbers $k_{0}$, but also small regions around these wavenumbers. One such region around the zeroth harmonic is shown in Fig.~\ref{fig:spectra0}(b). Similarly to the condensate case~\cite{agafontsev2014integrable}, the spectrum in this region behaves by power law $S_{k}\propto |k|^{-\beta}$, with almost the same exponent $\beta= 0.61$. At $k=0$ the asymptotic spectrum is finite, $S_{0}=1.72\times 10^{-2}$. 
The other peaks in the spectrum are also power-law, $S_{k}\propto |k-k_{0}|^{-\beta}$, with different exponents $\beta$ for different peaks ($\beta=0.56$ for $k_{0}=\pm 1$, $\beta=0.25$ for $k_{0}=\pm 2$, and peaks at $k_{0}=\pm 3$ are too small for analysis); at $k_{0}$ the spectrum is finite. 
The power-law behavior of the peaks means that wave action is concentrated in the corresponding modes. 
The peak at zeroth harmonic is sufficiently wide, with the power-law expanding in modes $|k|\le \delta k$, $\delta k=0.15$. The other peaks are much narrower, with $\delta k=0.02$ for $k_{0}=\pm 1$ and $\delta k=0.01$ for $k_{0}=\pm 2$. Modes $|k-k_{0}|\le \delta k$ contain about 39\% of all wave action $\langle N\rangle$ for the zeroth harmonic $k_{0}=0$, about 4\% -- for $k_{0}=\pm 1$, and less than 0.1\% -- for $k_{0}=\pm 2$; in total, the seven peaks contain about 43\% of all wave action $\langle N\rangle$. Modes $|k|\le 0.15$, which contain most of this wave action, have extremely large 
scales $\ell\gg 2\pi$ in the physical space, and can be called as quasi-condensate~\cite{agafontsev2014integrable}.

\begin{figure}[t] \centering
\includegraphics[width=8.0cm]{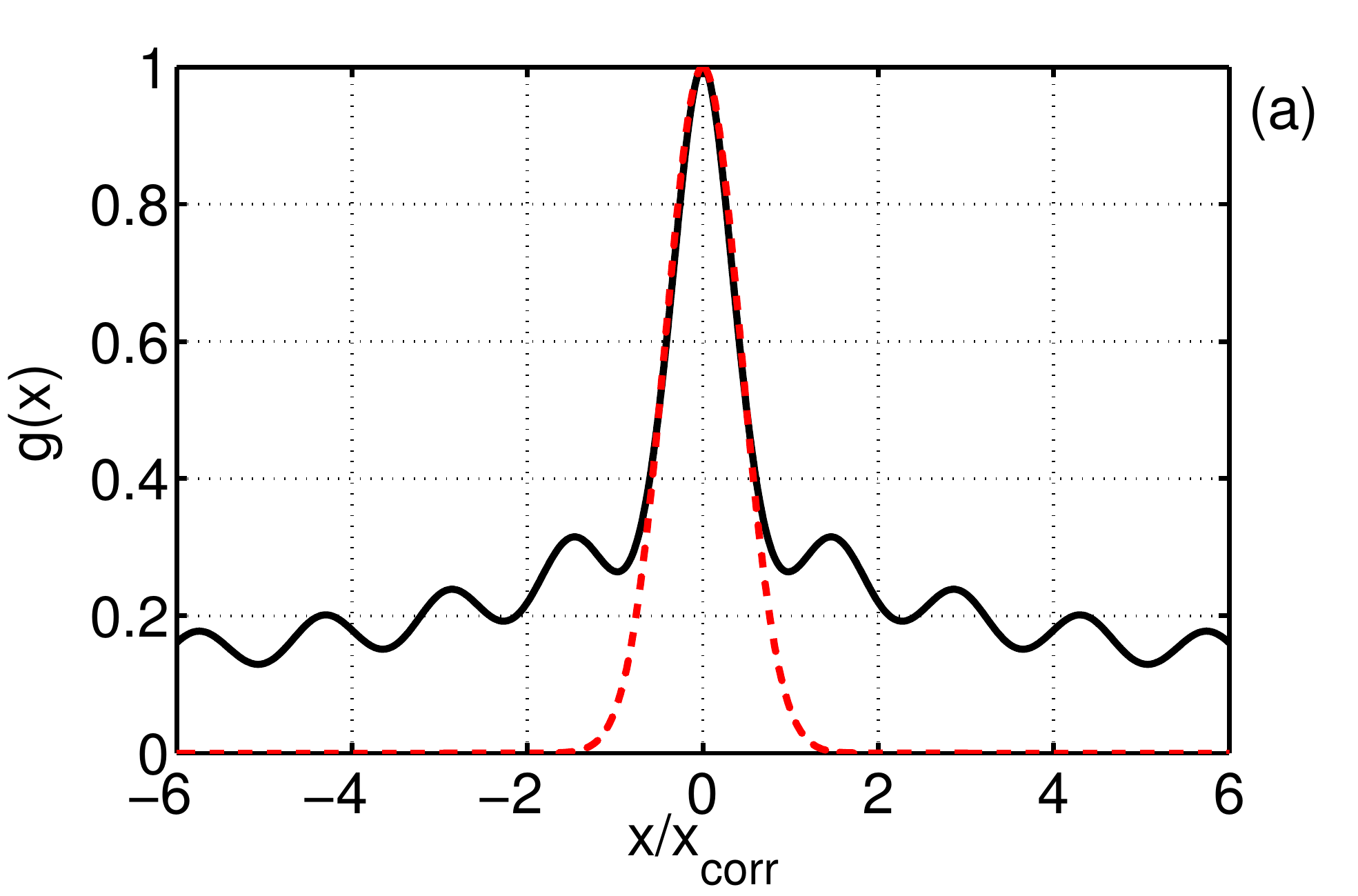}
\includegraphics[width=8.0cm]{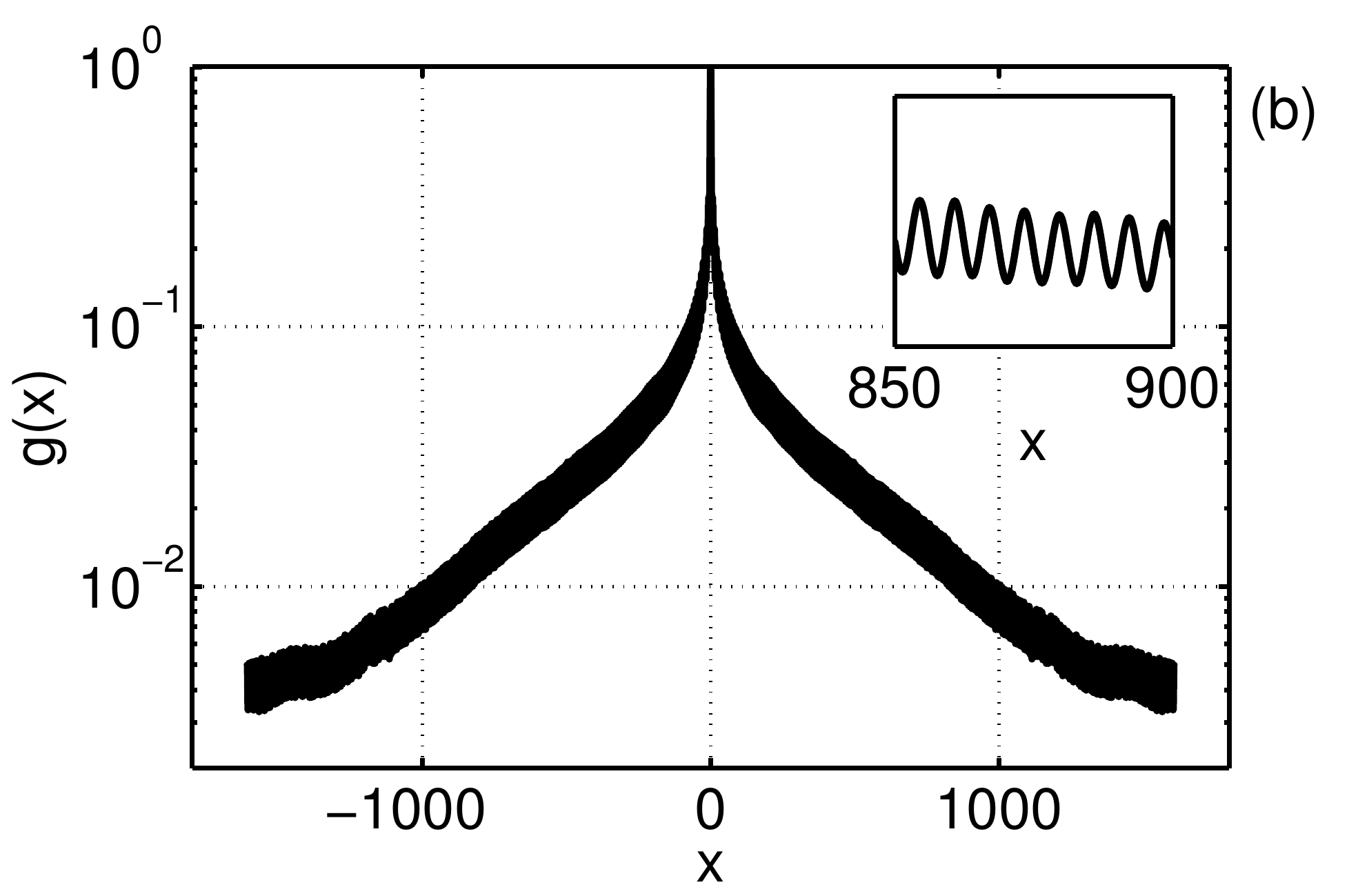}

\caption{\small {\it (Color on-line)} Asymptotic spatial correlation function $g(x)$ versus $x/x_{corr}$, $x_{corr}= 4.4$ (a) and $x$ (b). Dashed red line in graph (a) shows Gaussian distribution~(\ref{corr_Gaussian}), inset in graph (b) shows oscillations of $g(x)$ with period $2\pi$.}
\label{fig:corr_x0}
\end{figure}

The asymptotic spatial correlation function is shown in Fig.~\ref{fig:corr_x0}(a),(b). It's characteristic scale, defined as full width at half maximum, is $x_{corr}= 4.4$. At small lengths $|x|<x_{corr}/2$ the correlation function is well-approximated by Gaussian
\begin{equation}\label{corr_Gaussian}
g(x)\approx \exp\bigg[-\ln 2\, \bigg(\frac{2 x}{x_{corr}}\bigg)^{2}\bigg].
\end{equation}
At large lengths $|x|\gg x_{corr}$ it decays close to exponentially and in oscillatory way, as shown in Fig.~\ref{fig:corr_x0}(b); the period of these oscillations is equal to $2\pi$.

\begin{figure}[t] \centering
\includegraphics[width=8.0cm]{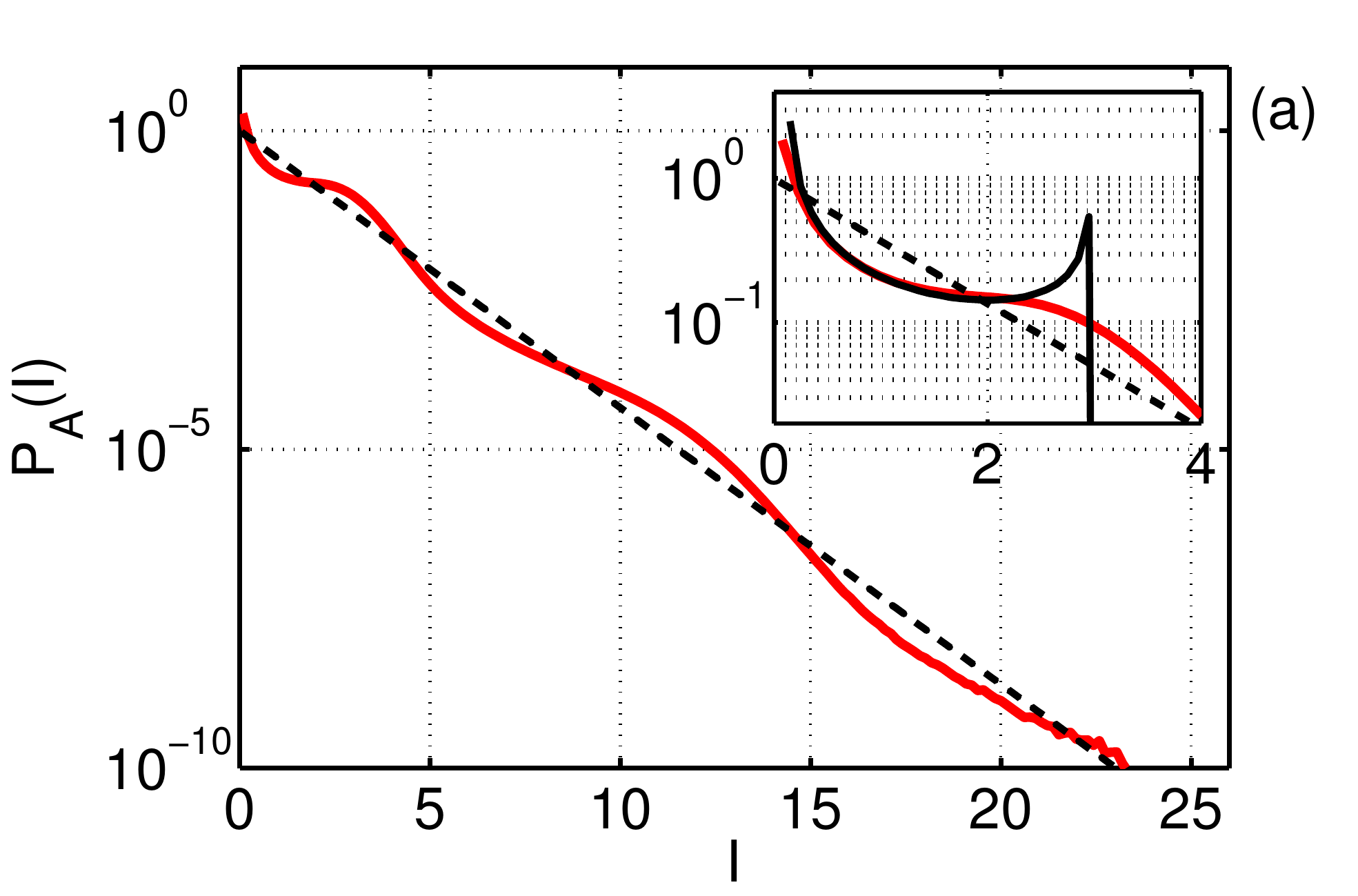}
\includegraphics[width=8.0cm]{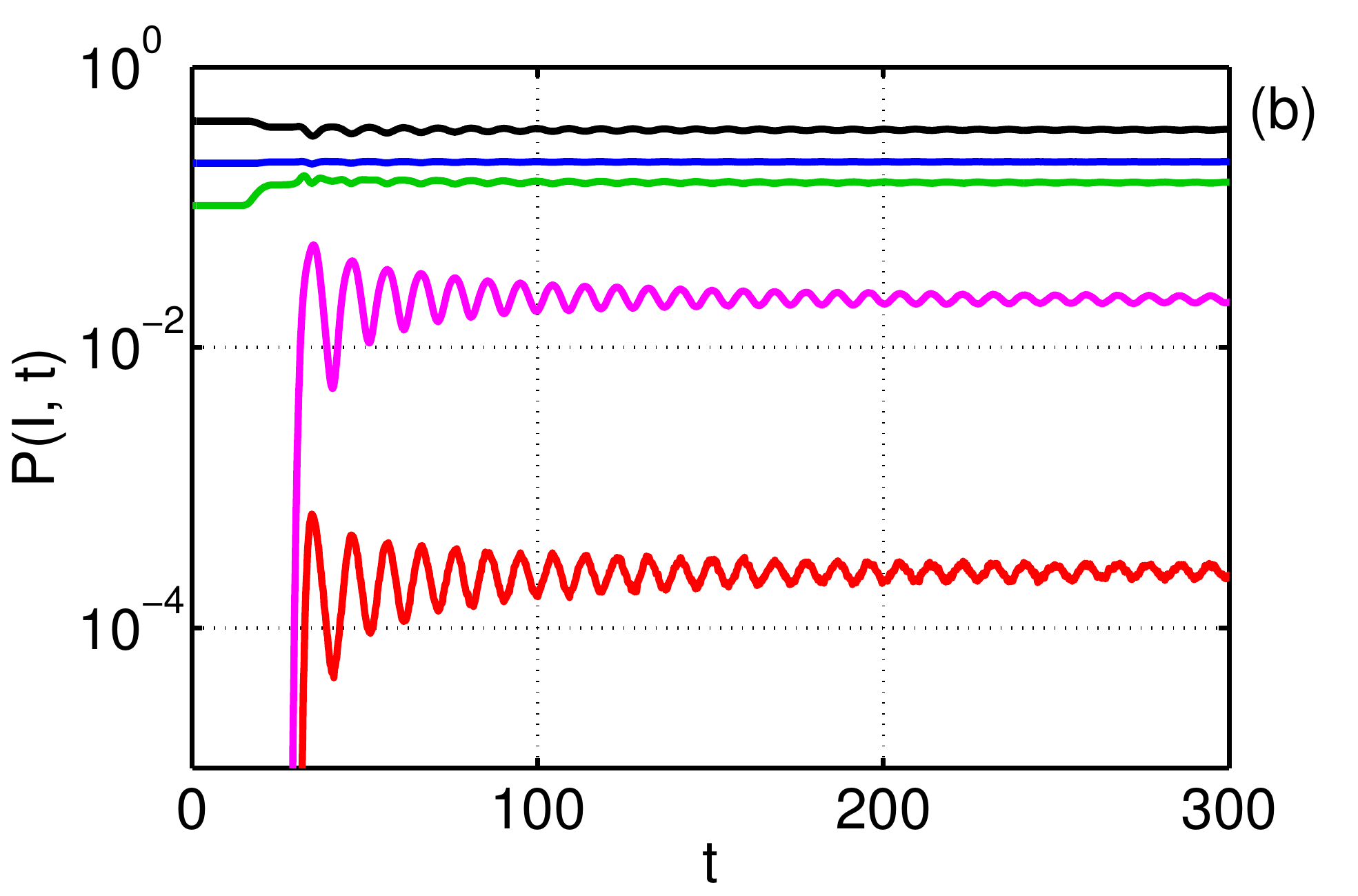}

\caption{\small {\it (Color on-line)} 
(a) Asymptotic PDF $\mathcal{P}_{A}(I)$ (thick red) and exponential PDF~(\ref{Rayleigh}) (dashed black). The inset shows the same PDFs and also the initial PDF $\mathcal{P}(I,t)$ at $t=0$ (black solid line).
(b) Time dependence of the PDF $\mathcal{P}(I,t)$ at different relative intensities $I=0.5$ (black), $I=1$ (blue), $I=2$ (green), $I=4$ (pink), $I=8$ (red).
}
\label{fig:PDFa}
\end{figure}

As shown in Fig.~\ref{fig:moments_asymptotic}, the asymptotic moments $M^{(n)}_{A}$ differ from exponential moments~(\ref{MnR}). This means that the PDF $\mathcal{P}_{A}(I)$ in the asymptotic state must differ from exponential PDF~(\ref{Rayleigh}). This is indeed the case, as shown in Fig.~\ref{fig:PDFa}(a). The asymptotic PDF exceeds the exponential PDF for relative intensities $I<0.22$, $I\in[1.9, 4.2]$ and $I\in[8.7, 14.4]$. According to rogue waves criterion $|\Psi|^{2}>8\langle |\Psi|^{2}\rangle$ (see e.g.~\cite{kharif2003physical, dysthe2008oceanic, agafontsev2014integrable}), or $I>8$, only the last region contains rogue waves. In this region $I\in[8.7, 14.4]$ the asymptotic PDF exceeds the exponential PDF by about 2.5 times maximum at $I=12$. Note, that in the region $I\in[0,2]$ the asymptotic PDF turns out to be very close to the initial PDF $\mathcal{P}(I,t)$ at $t=0$, as shown in the inset of Fig.~\ref{fig:PDFa}(a).

\begin{figure}[t] \centering
\includegraphics[width=8.0cm]{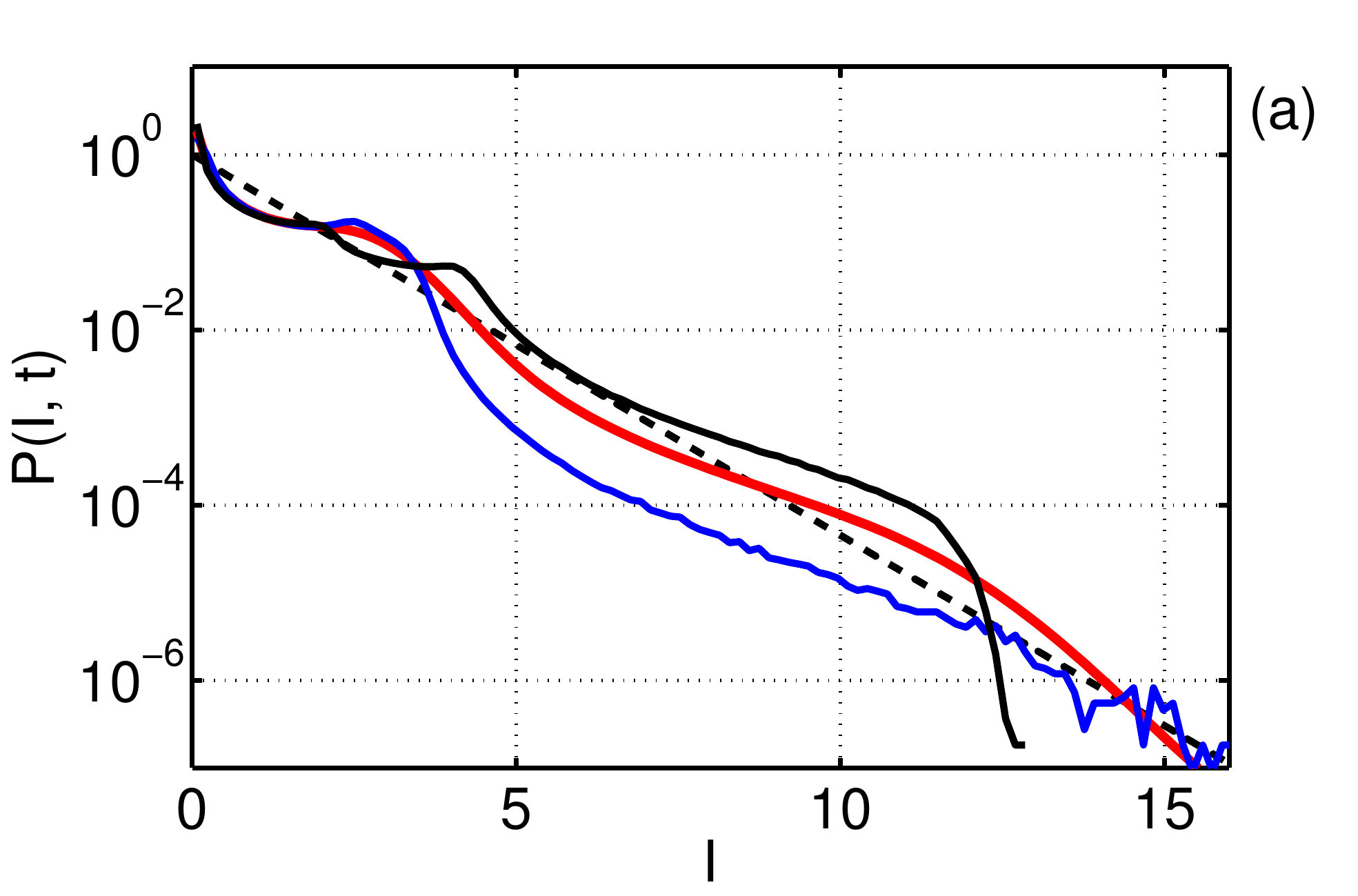}
\includegraphics[width=8.0cm]{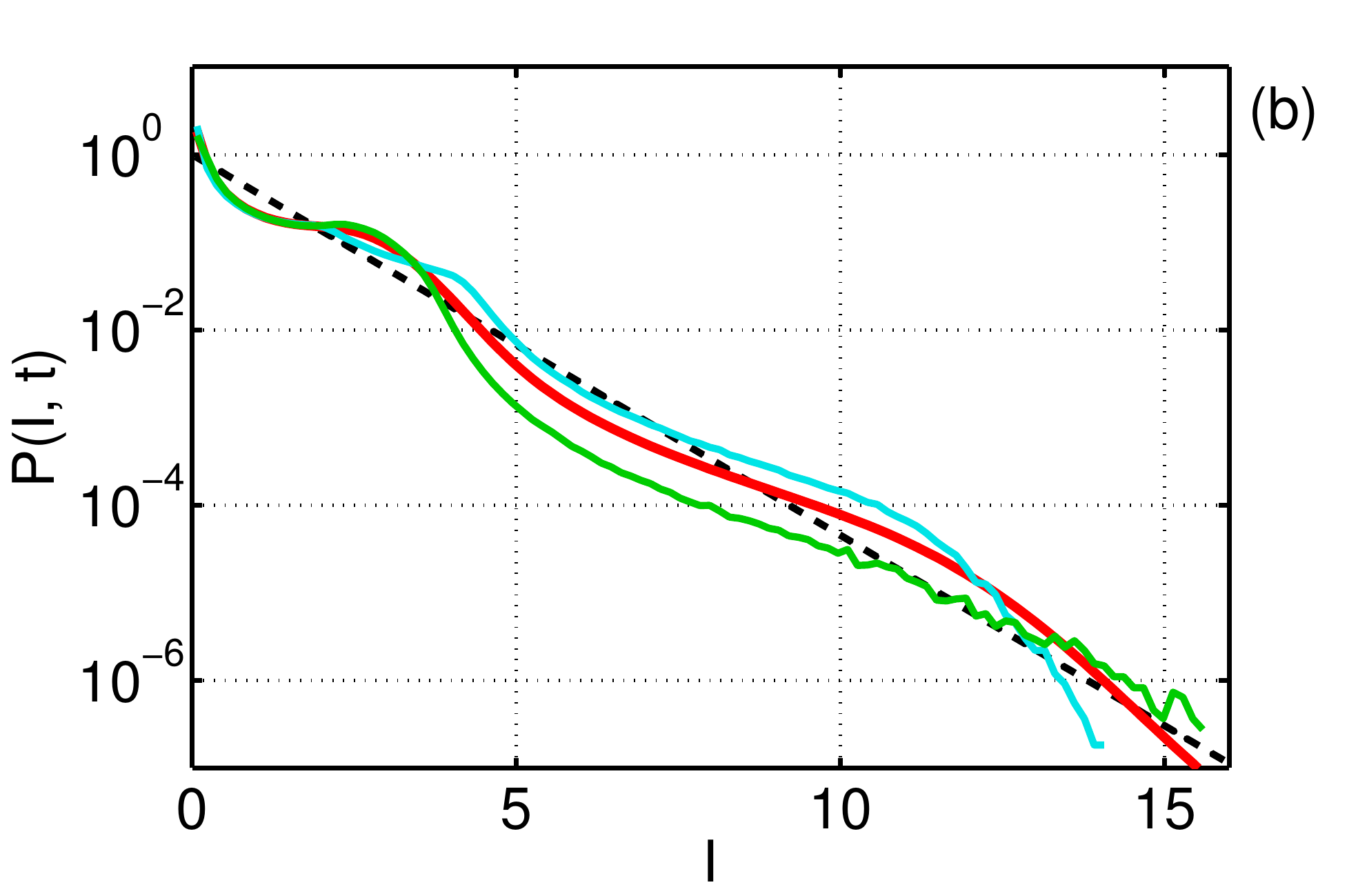}

\caption{\small {\it (Color on-line)} PDF $\mathcal{P}(I,t)$ at different times corresponding to the first several potential energy modulus $|\langle H_{4}(t)\rangle|$ extremums: (a) at $t=34.8$ (black, first local maximum of $|\langle H_{4}(t)\rangle|$), $t=40.6$ (blue, first local minimum) and (b) at $t=46.2$ (cyan, second local maximum), $t=51.4$ (green, second local minimum). Thick red line shows the asymptotic PDF $\mathcal{P}_{A}(I)$, dashed black line is exponential PDF~(\ref{Rayleigh}).}
\label{fig:PDF12}
\end{figure}

After development of the MI, the PDF $\mathcal{P}(I,t)$ evolves with time in oscillatory way approaching to the asymptotic PDF at late times, as shown in Fig.~\ref{fig:PDFa}(b),~\ref{fig:PDF12}(a),(b). This evolution is similar to that for wave-action spectrum and spatial correlation function, with the same ``turning points'' coinciding with local maximums and minimums of $|\langle H_{4}(t)\rangle|$. At the local maximums of $|\langle H_{4}(t)\rangle|$ and for sufficiently large intensities $4\lesssim I\lesssim 12$, the PDF $\mathcal{P}(I,t)$ takes (locally in time) maximal values, Fig.~\ref{fig:PDF12}(a),(b). At the local minimums of $|\langle H_{4}(t)\rangle|$ the PDF takes (locally in time) minimal values for the same region of intensities. The maximal excess of the PDF $\mathcal{P}(I,t)$ over exponential PDF~(\ref{Rayleigh}) by about 6 times is observed at the first local maximum of $|\langle H_{4}(t)\rangle|$ at $t=34.8$ for relative intensity $I=11.5$.

\begin{figure}[t] \centering
\includegraphics[width=8.0cm]{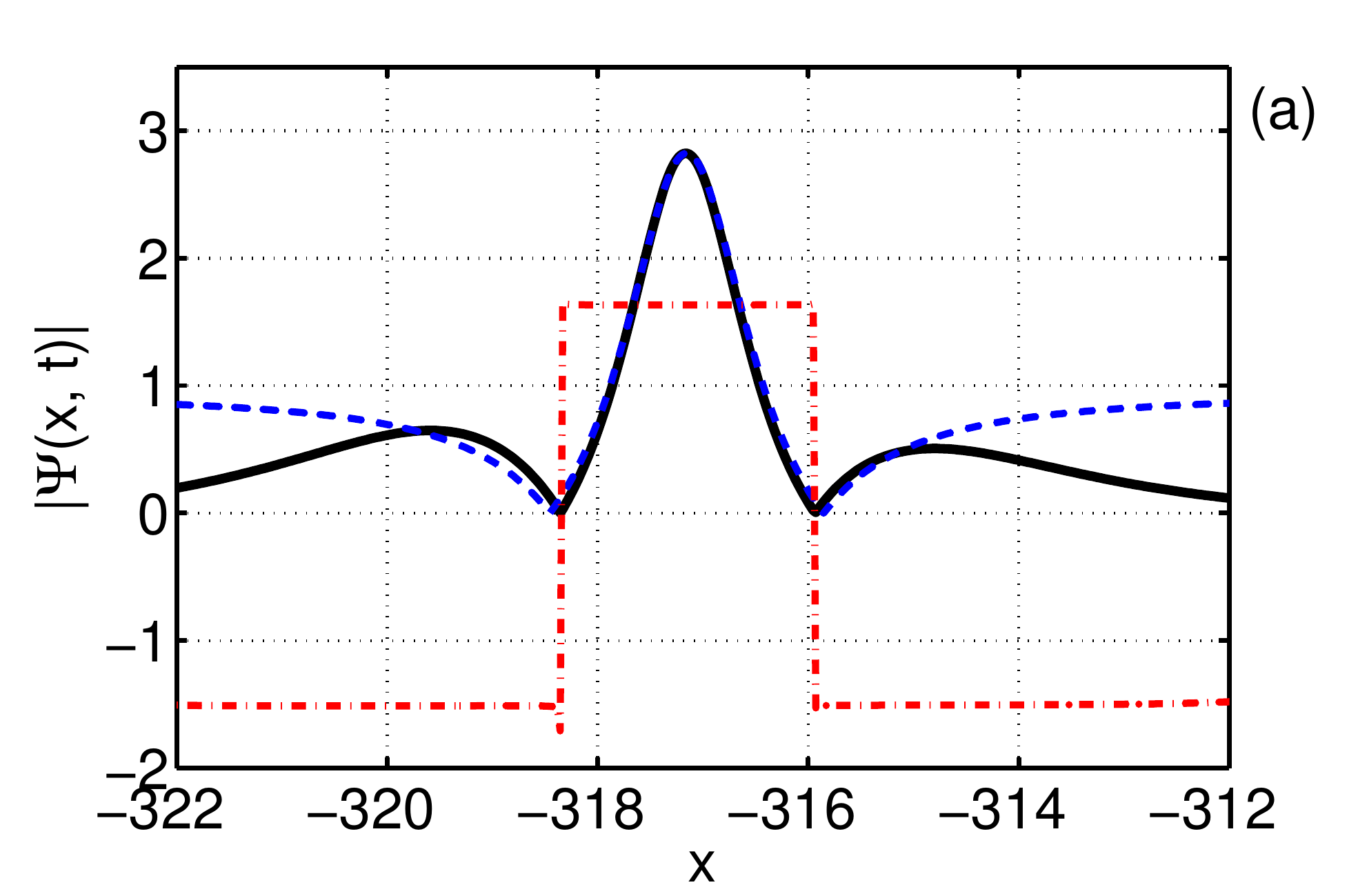}
\includegraphics[width=8.0cm]{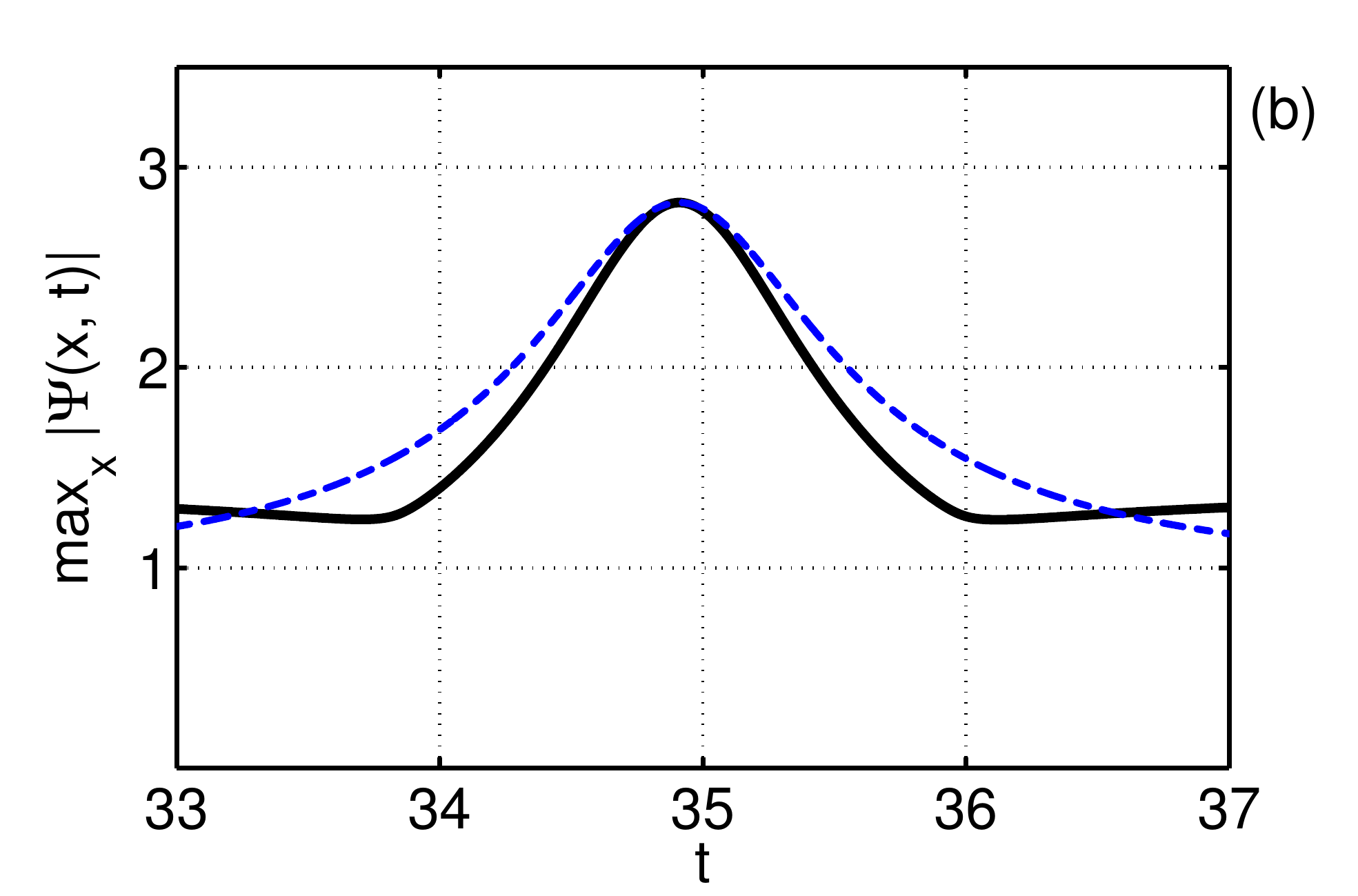}
\includegraphics[width=8.0cm]{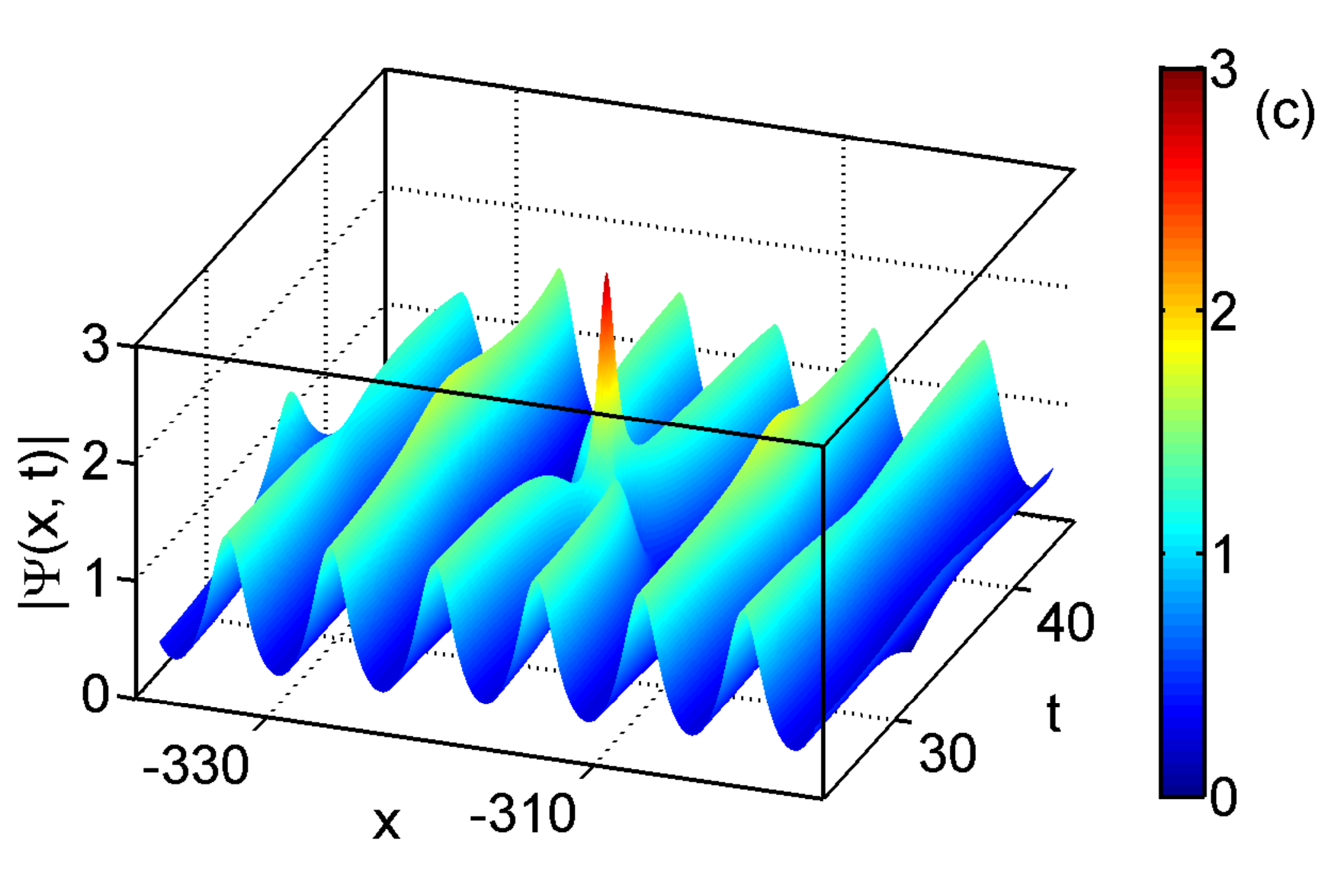}

\caption{\small {\it (Color on-line)} 
(a) Space distribution of amplitude $|\Psi(x,t_{0})|$ (solid black) and phase $\arg\,\Psi(x,t_{0})$ (dash-dot red) for a typical rogue wave at the time $t_{0}=38.9$ of its maximal elevation. Dashed blue line is fit by Peregrine solution~(\ref{Peregrine_x}) with $A=-0.94$, $x_{0}=-317.2$.
(b) Time evolution of maximal amplitude -- for rogue wave (solid black) and Peregrine solution~(\ref{Peregrine_t}) (dashed blue). 
(c) Space-time representation of amplitude $|\Psi(x,t)|$ near the rogue wave event.
}
\label{fig:RWmax}
\end{figure}

\begin{figure}[t] \centering
\includegraphics[width=8.0cm]{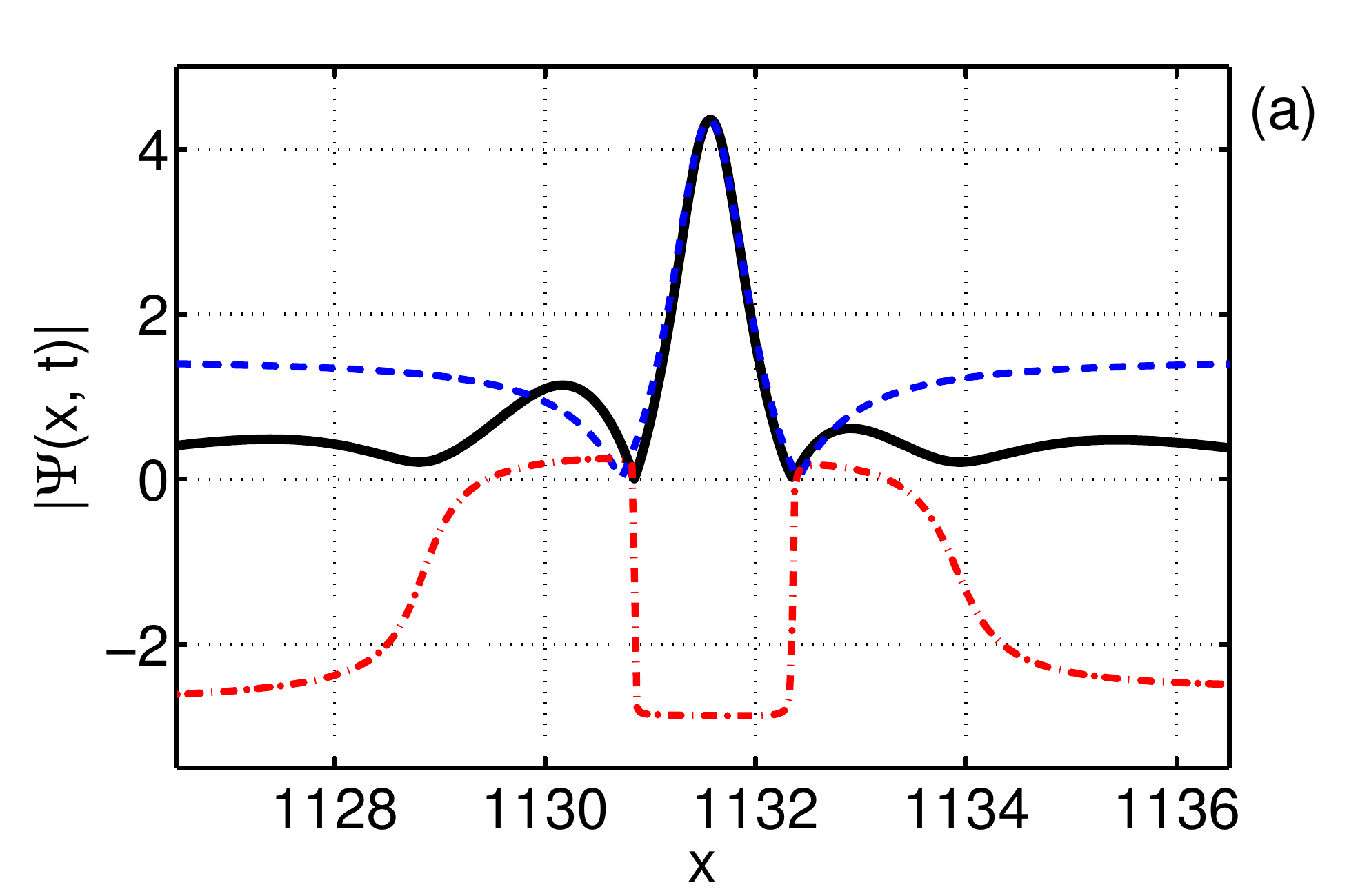}
\includegraphics[width=8.0cm]{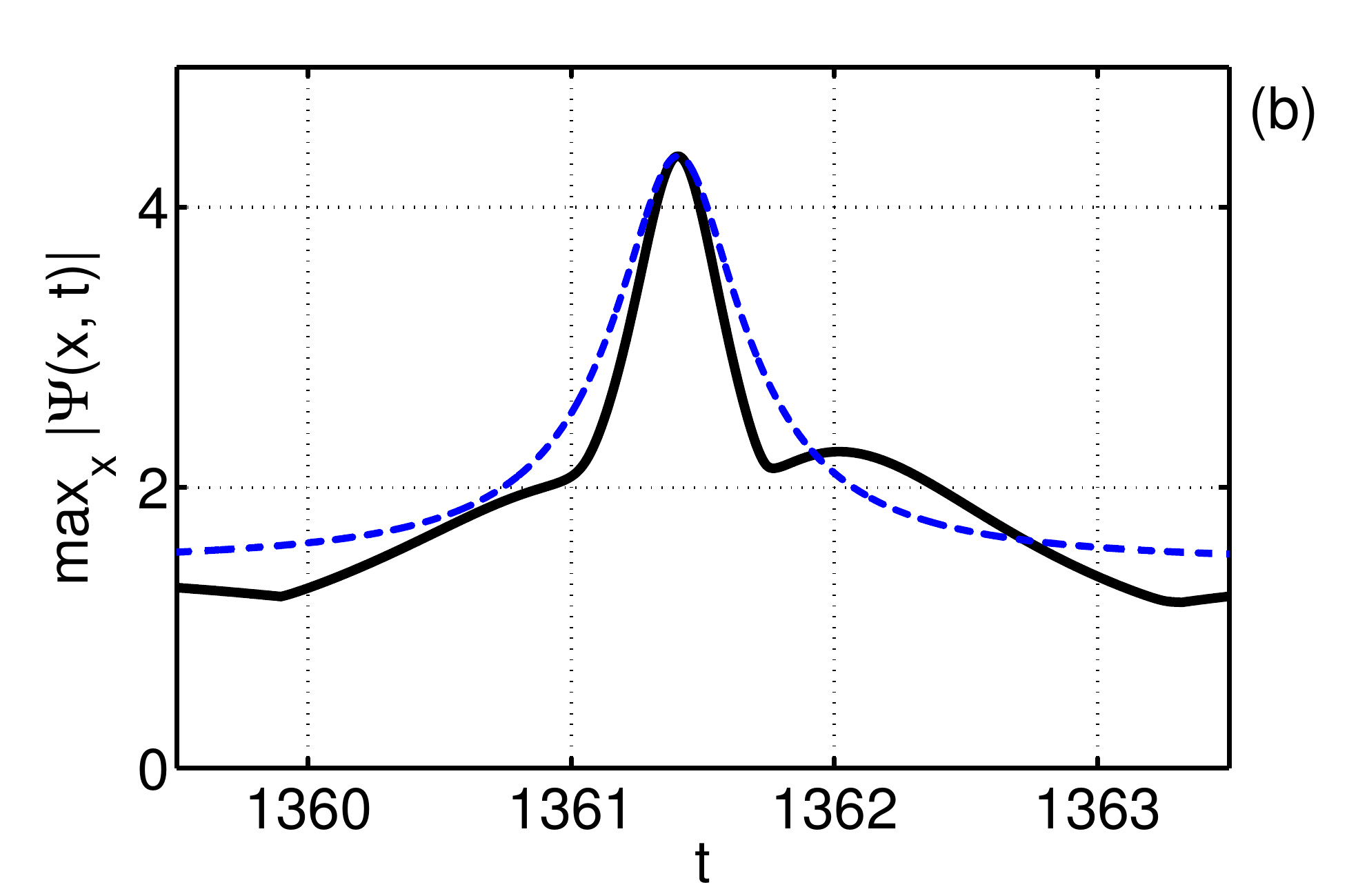}
\includegraphics[width=8.0cm]{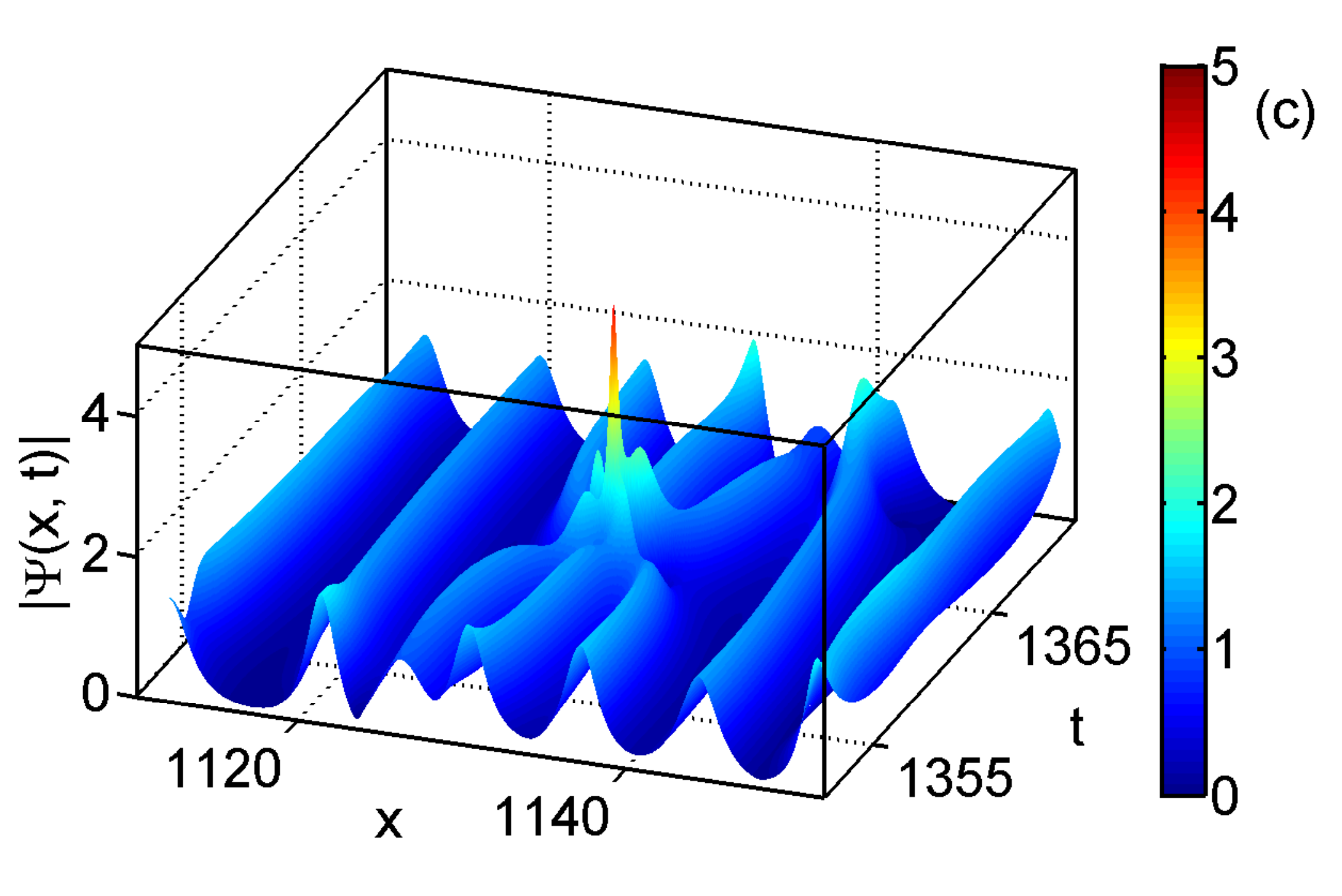}

\caption{\small {\it (Color on-line)} 
(a) Space distribution of amplitude $|\Psi(x,t_{0})|$ (solid black) and phase $\arg\,\Psi(x,t_{0})$ (dash-dot red) for the largest detected rogue wave at the time $t_{0}=1361.4$ of its maximal elevation. Dashed blue line is fit by Peregrine solution~(\ref{Peregrine_x}) with $A=-1.46$, $x_{0}=1131.6$.
(b) Time evolution of maximal amplitude -- for rogue wave (solid black) and Peregrine solution~(\ref{Peregrine_t}) (dashed blue).
(c) Space-time representation of amplitude $|\Psi(x,t)|$ near the rogue wave event.
}
\label{fig:RW}
\end{figure}

Fig.~\ref{fig:RWmax},~\ref{fig:RW} show two rogue wave events -- a typical one and the largest detected one respectively. Rogue wave in Fig.~\ref{fig:RWmax} had duration in time $\Delta T\sim 1$ and reached maximal amplitude $\max|\Psi|=2.8$ at $t_{0}=34.9$, that is close to the first local maximum of $|\langle H_{4}(t)\rangle|$ at $t=34.8$. With the average square amplitude $\langle |\Psi|^{2}\rangle = 0.66$, the crest of this wave corresponds to relative intensity $I=12$. The largest detected wave shown in Fig.~\ref{fig:RW} had duration in time $\Delta T\sim 0.5$ and reached maximal amplitude $\max|\Psi|=4.4$ at $t=1361.4$, that is sufficiently close to the asymptotic stationary state. The crest of this wave corresponds to relative intensity $I=29$.

We examined several hundreds of rogue waves detected in our experiments, and found that all of them at the time of their maximal elevation have quasi-rational profile similar to that of the Peregrine solution~\cite{peregrine1983water,kibler2010peregrine} (see the similar results in~\cite{suret2016direct} for incoherent wave initial conditions and in~\cite{agafontsev2014intermittency} for generalized NLS equation accounting for six-wave interactions, pumping and dumping terms). The Peregrine solution is localized in space and time algebraic solution of the NLS equation~(\ref{nlse}),
\begin{equation}\label{Peregrine_fit}
\psi_{P}(x,t)=e^{i t}\bigg[1-\frac{4(1+2 i t)}{1+2 x^{2}+4 t^{2}}\bigg].
\end{equation}
It is easy to see that $A\,\psi_{P}(X,T)$, where $X=|A|(x-x_{0})$ and $T=|A|^{2}(t-t_{0})$, is also a solution of the NLS equation, which becomes maximal at $x=x_{0}$ and $t=t_{0}$. 
Fig.~\ref{fig:RWmax}(a),~\ref{fig:RW}(a) show the similarity of rogue waves profile (at the time of their maximal elevation $t=t_{0}$) with that of the Peregrine solution, 
\begin{equation}\label{Peregrine_x}
|\Psi(x,t_{0})|\approx A\,\psi_{P}(X,0) = A\bigg[1-\frac{4}{1+2|A|^{2} (x-x_{0})^{2}}\bigg].
\end{equation}
Note, that the phase of rogue waves $\mathrm{arg}\,\Psi(x,t_{0})$ is almost constant near the amplitude maximum, as is the case for Peregrine solution at the time of its maximal elevation. 
Moreover, for one realization from ensemble of initial conditions we checked all waves that exceed maximal amplitude of the original cnoidal wave by 1.5 times or more. At the time of their maximal elevation, all such waves are well approximated in space by ansatz~(\ref{Peregrine_x}). However, time evolution of maximal amplitude $\max_{x}|\Psi(x,t)|$ for large waves is different from that, 
\begin{equation}\label{Peregrine_t}
A\,|\psi_{P}(0,T)| = A\bigg|1-4\frac{1+2i |A|^{2}(t-t_{0})}{1+4|A|^{4} (t-t_{0})^{2}}\bigg|,
\end{equation}
for the Peregrine solution, see examples in Fig.~\ref{fig:RWmax}(b),~\ref{fig:RW}(b).

The phase of those rogue waves, that appear near the first several local maximums of $|\langle H_{4}(t)\rangle|$, is very close to $\mathrm{arg}\,\Psi\approx \pi/2+\pi (m-1)$, where $m$ is the local maximum index number (i.e., $\mathrm{arg}\,\Psi\approx \pi/2$ for the first local maximum at $t= 34.8$, $\mathrm{arg}\,\Psi\approx 3\pi/2$ for the second local maximum at $t=46.2$, etc.). 
The phase of those rogue waves, that appear near the first several local minimums of $|\langle H_{4}(t)\rangle|$, is very close to $\mathrm{arg}\,\Psi\approx \pi+\pi (m-1)$, where $m$ is the local minimum index number. We observe such behavior for about ten first local maximums and minimums of potential energy modulus $|\langle H_{4}(t)\rangle|$. We checked this fact by direct observation (see e.g. Fig.~\ref{fig:RWmax}(a)) and also by measuring PDFs for real and imaginary parts of wave field $\Psi$. The same ``rotation of phase'' is present for the condensate case~\cite{agafontsev2014integrable} too.

As shown in Fig.~\ref{fig:RWmax}(c),~\ref{fig:RW}(c), rogue waves in these figures look like collisions of two and three pulses respectively. In terms of the recent study of rogue waves on cnoidal wave background~\cite{kedziora2014rogue}, wave in Fig.~\ref{fig:RWmax} might be the ``concentrated'' cnoidal rogue wave, while wave in Fig.~\ref{fig:RW} -- the ``fused'' second-order cnoidal rogue wave. 
There is also another possibility, that waves in Fig.~\ref{fig:RWmax},~\ref{fig:RW} are the collisions of breathers that decompose from ``superregular'' solitonic solutions on the cnoidal wave background~\cite{gelash2015private}. 
The similar solutions on the condensate background were recently found theoretically~\cite{gelash2014superregular} and observed experimentally~\cite{gelash2015superregular}. 
Such solutions on the cnoidal wave background should decompose to breathers, and subsequent collisions of these breathers may lead to almost algebraic behavior of the resulting pulse at the time of its maximal elevation (see e.g.~\cite{akhmediev2009extreme} of how this may happen for Akhmediev breathers). 
We plan to study the question of rogue waves origin in more details in another publication.


\section{Dependence on cnoidal wave parameters}
\label{Sec:Dependence}

In this section we describe the dependence of integrable turbulence on cnoidal wave parameters, namely the imaginary half-period $\omega_{1}$, which determines the ``overlapping'' between the solitons within the cnoidal wave. 
For this purpose we repeated numerical experiment of Section~\ref{Sec:Evolution} for another 9 cnoidal waves~(\ref{cnoidal1_stationary}) with imaginary half-periods from $\omega_{1}=0.8$ to $\omega_{1}=5$, which have maximal increments of the MI~(\ref{cnoidal1_increment}) $\gamma_{\max}$ from $0.065$ to $0.5$ respectively. These experiments were carried out in the box $L=256\pi$ up to final time from $t=200$ (for $\omega_{1}=5$) to $t=1000$ (for $\omega_{1}=0.8$). The properties of the generated integrable turbulence turned out to be qualitatively similar to that discussed in Section~\ref{Sec:Evolution}. Therefore, below we will focus mainly on the distinctions in the turbulence properties for different initial cnoidal waves.

\begin{figure}[t] \centering
\includegraphics[width=8.0cm]{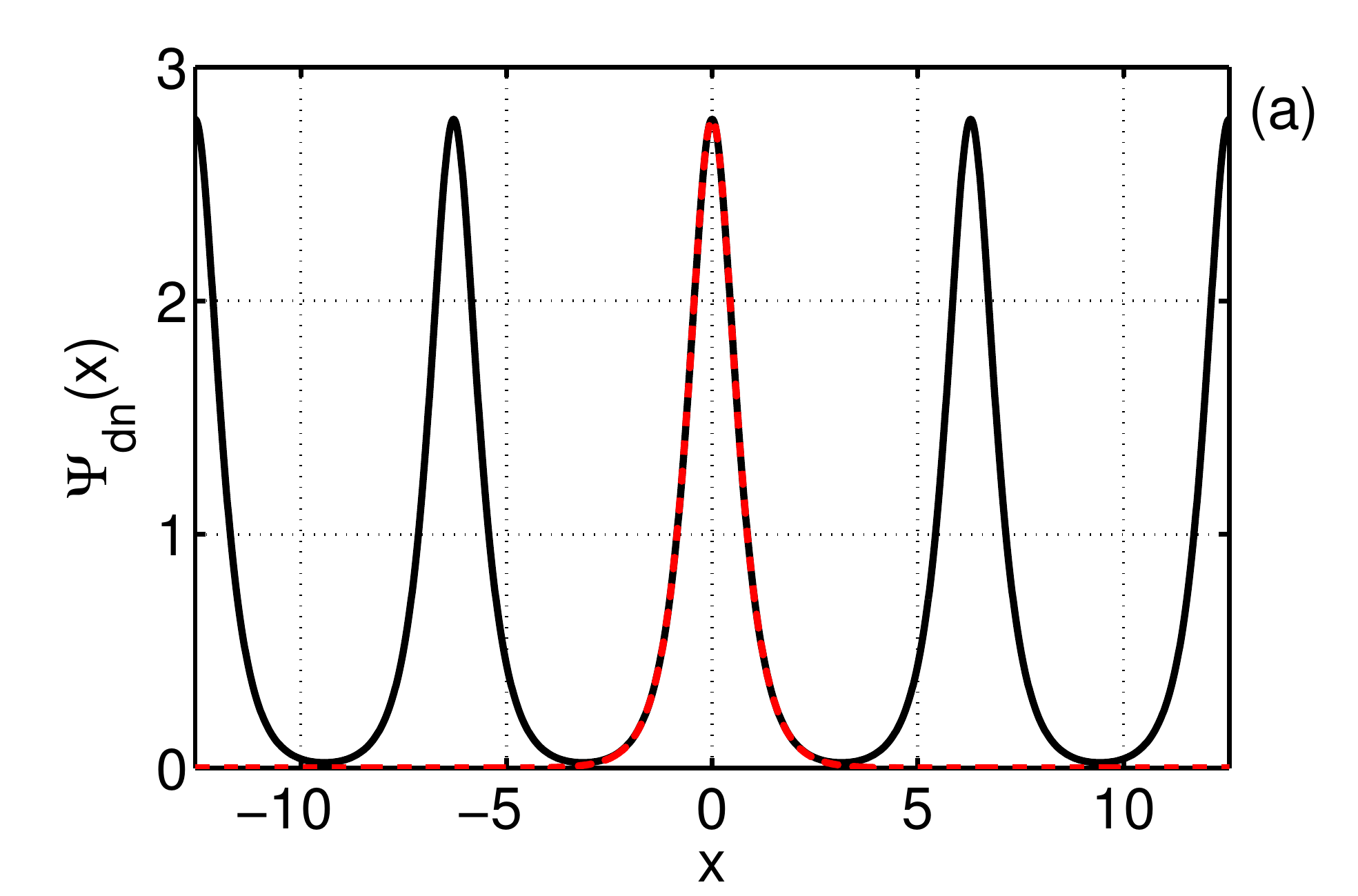}
\includegraphics[width=8.0cm]{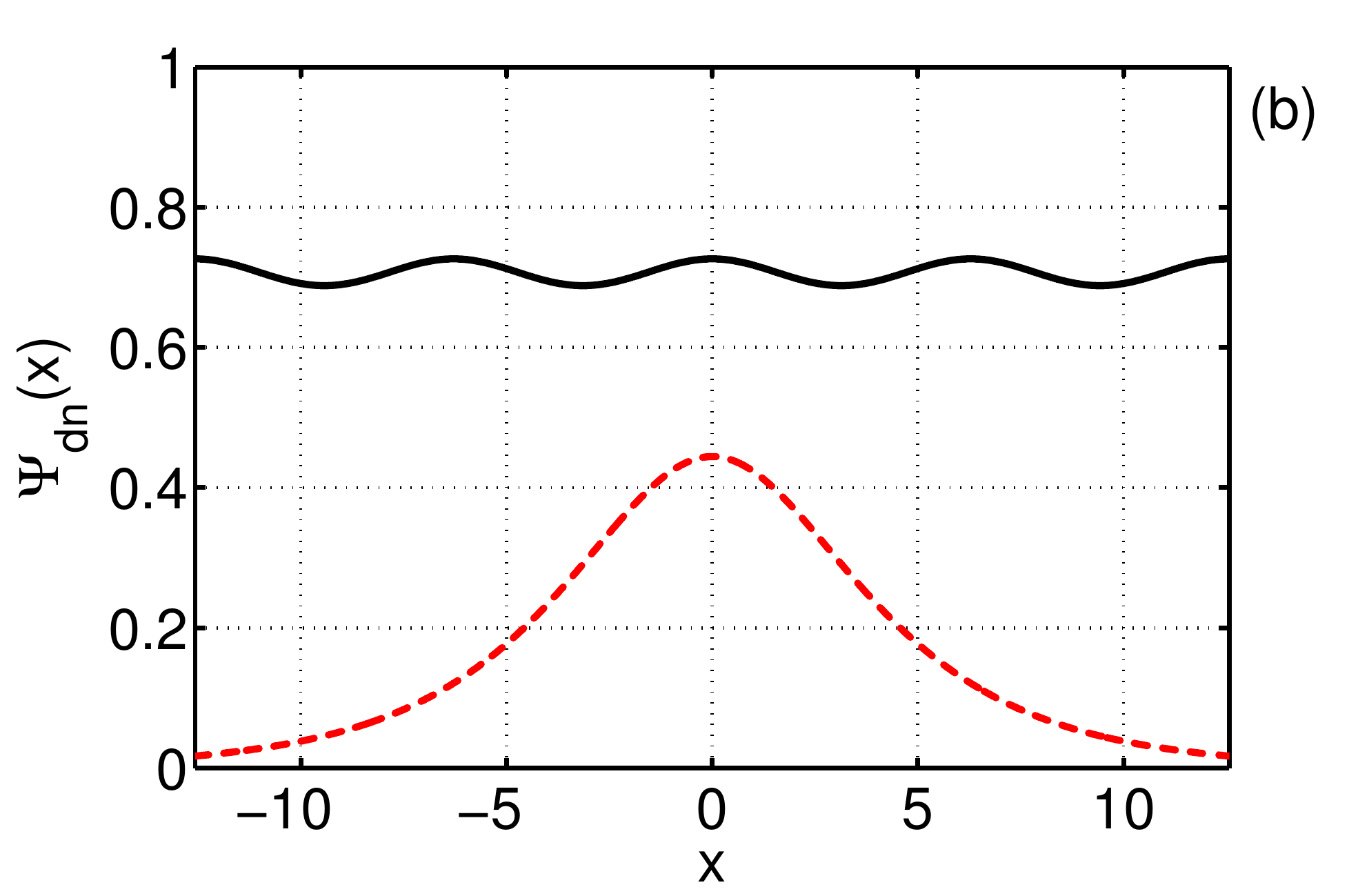}

\caption{\small {\it (Color on-line)} Cnoidal waves~(\ref{cnoidal1_stationary}) with $\omega_{1}=0.8$ (a) and $\omega_{1}=5$ (b) at $t=0$. Dashed red lines show solitons~(\ref{nlse_soliton}) with $\lambda=\pi/2\omega_{1}$. }
\label{fig:cnoidal_wave_B112}
\end{figure}

\begin{figure}[t] \centering
\includegraphics[width=8.0cm]{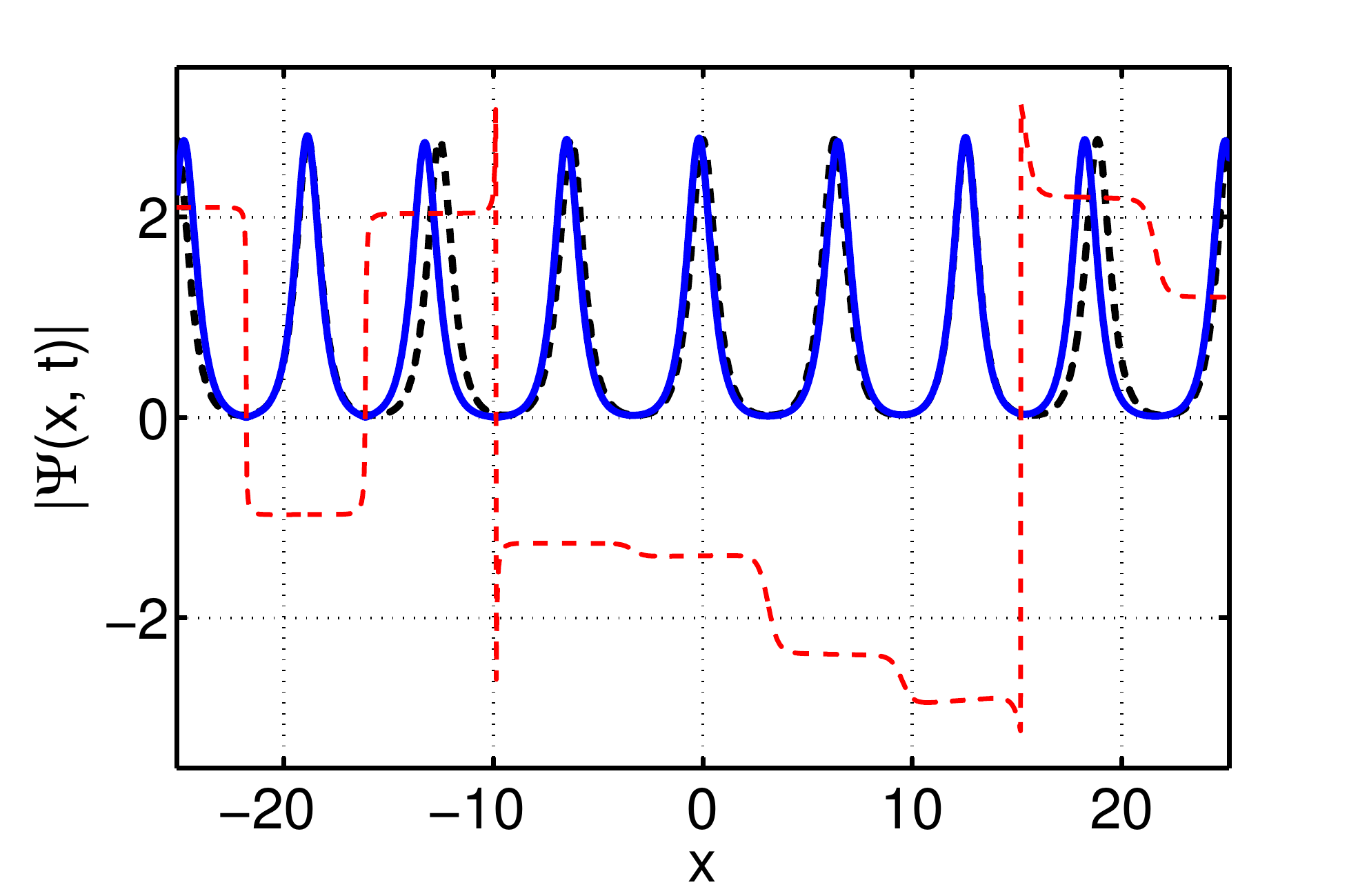}

\caption{\small {\it (Color on-line)} Amplitude $|\Psi(x,t)|$ for one of the realizations of initial conditions for cnoidal wave~(\ref{cnoidal1_stationary}) with $\omega_{1}=0.8$: at $t=0$ (dashed black) and close to the asymptotic state at $t=1000$ (solid blue). Dashed red line shows phase $\mathrm{arg}\,\Psi(x,t)$ at $t=1000$.}
\label{fig:cnoidal_wave_B1_final}
\end{figure}

\begin{figure}[t] \centering
\includegraphics[width=8.0cm]{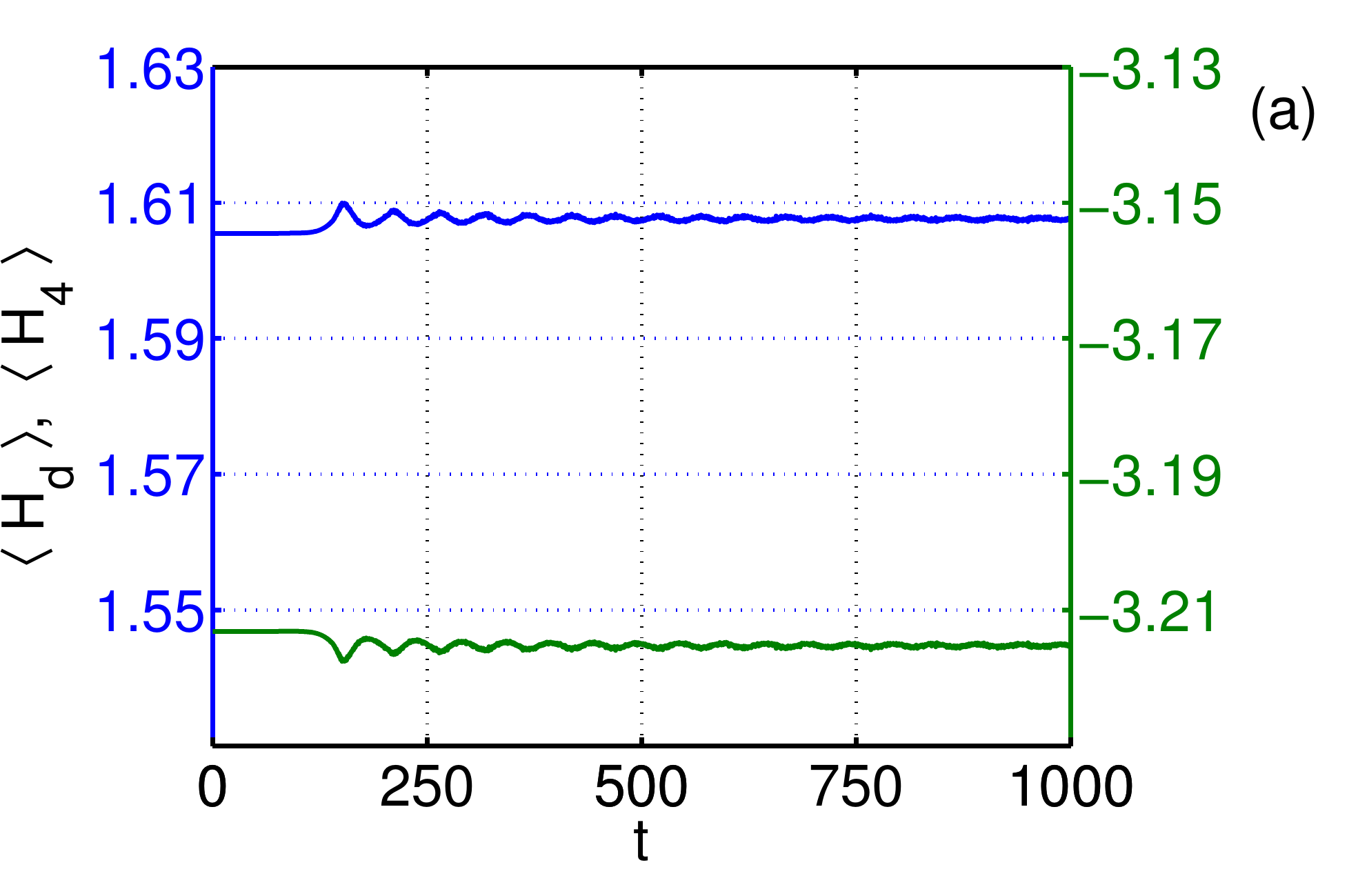}
\includegraphics[width=8.0cm]{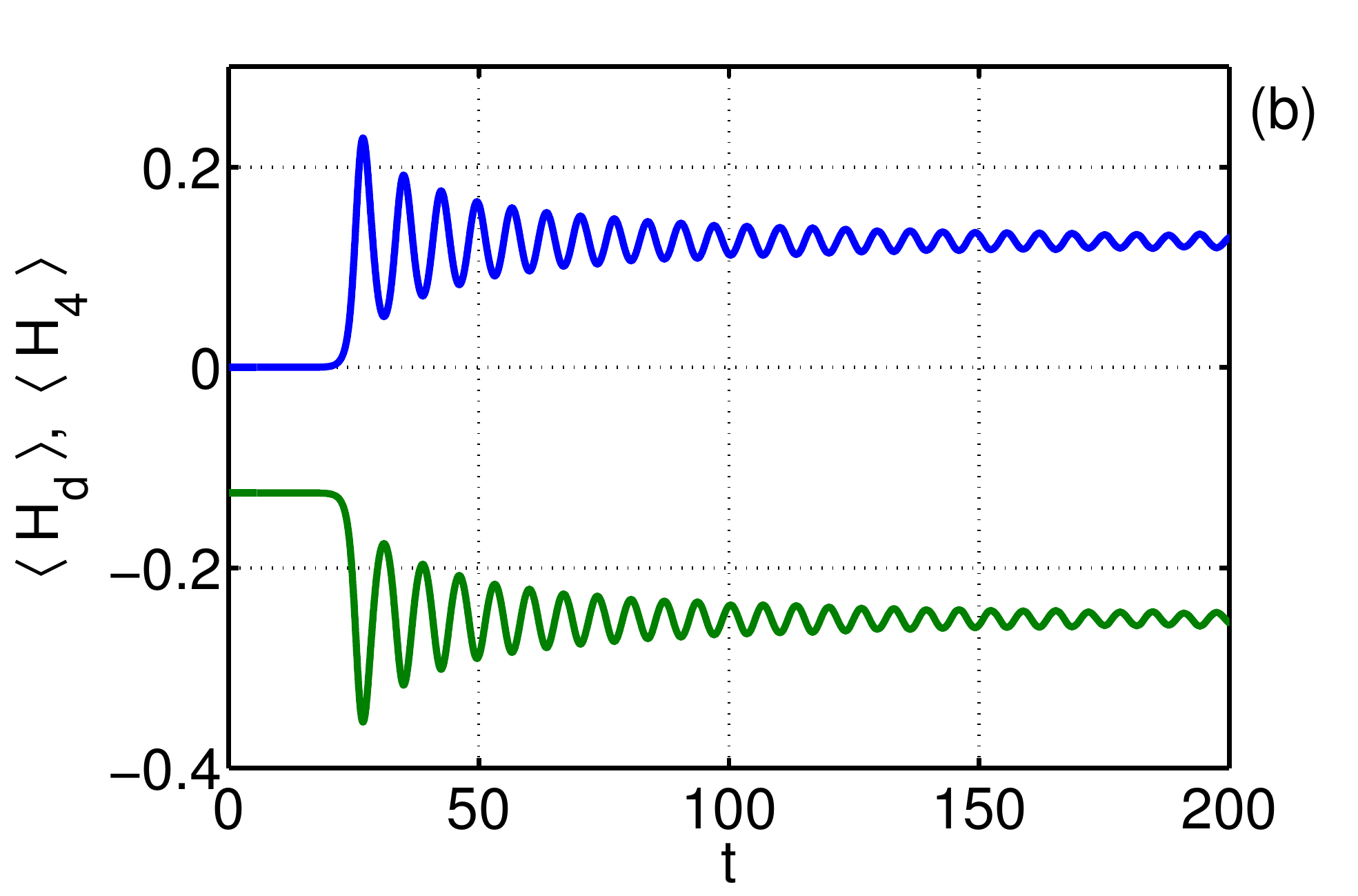}

\caption{\small {\it (Color on-line)} Evolution of ensemble average kinetic $\langle H_{d}(t)\rangle$ (blue) and potential $\langle H_{4}(t)\rangle$ (green) energies for cnoidal waves with $\omega_{1}=0.8$ (a) and $\omega_{1}=5$ (b). Note the difference in scales between graphs (a) and (b), and also different OY-axis (left for $\langle H_{d}(t)\rangle$ and right for $\langle H_{4}(t)\rangle$) in graph (a). In the asymptotic stationary state the energy ratio is equal to $Q_{A} = \langle H_{4}\rangle/\langle H_{d}\rangle = -2$ for both cnoidal waves. }
\label{fig:evolutionE12}
\end{figure}

For small $\omega_{1}$, cnoidal waves are very close to arithmetic sum of singular solitons~(\ref{cnoidal_limit_w0}), as demonstrated by the example in Fig.~\ref{fig:cnoidal_wave_B112}(a) for $\omega_{1}=0.8$. 
After development of the MI from such waves (see Fig.~\ref{fig:CW1_MI} in Appendix~\ref{Sec:Annex-B}), wave field remains close to a composition of singular solitons~(\ref{nlse_soliton}) with different phases and positions, even after a very long time when the system is close to the asymptotic stationary state, Fig.~\ref{fig:cnoidal_wave_B1_final}. Moreover, the positions of these singular solitons remain generally very close to the positions of ``solitons'' of the original cnoidal wave. 
Note, that the phase $\mathrm{arg}\,\Psi(x,t)$ stays almost constant on the solitons and randomly jumps between them.
Therefore, turbulence generated from cnoidal waves with small $\omega_{1}$ transforms into soliton turbulence in integrable system. The potential to kinetic energy ratio $Q(t)$ for such turbulence at all times remains very close to -2, Fig.~\ref{fig:evolutionE12}(a), the same as for a singular NLS soliton~(\ref{nlse_soliton}).

\begin{figure}[t] \centering
\includegraphics[width=8.0cm]{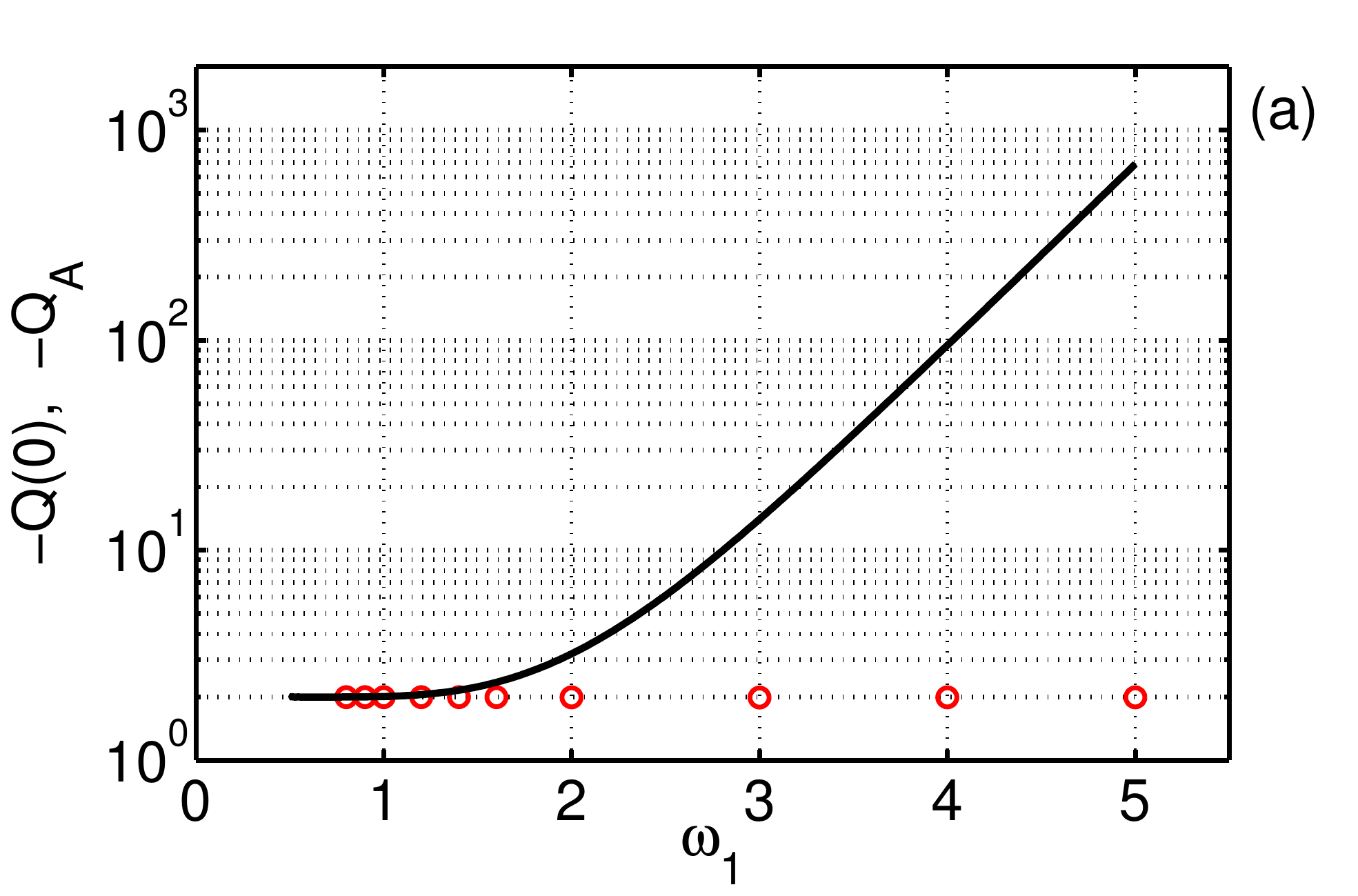}
\includegraphics[width=8.0cm]{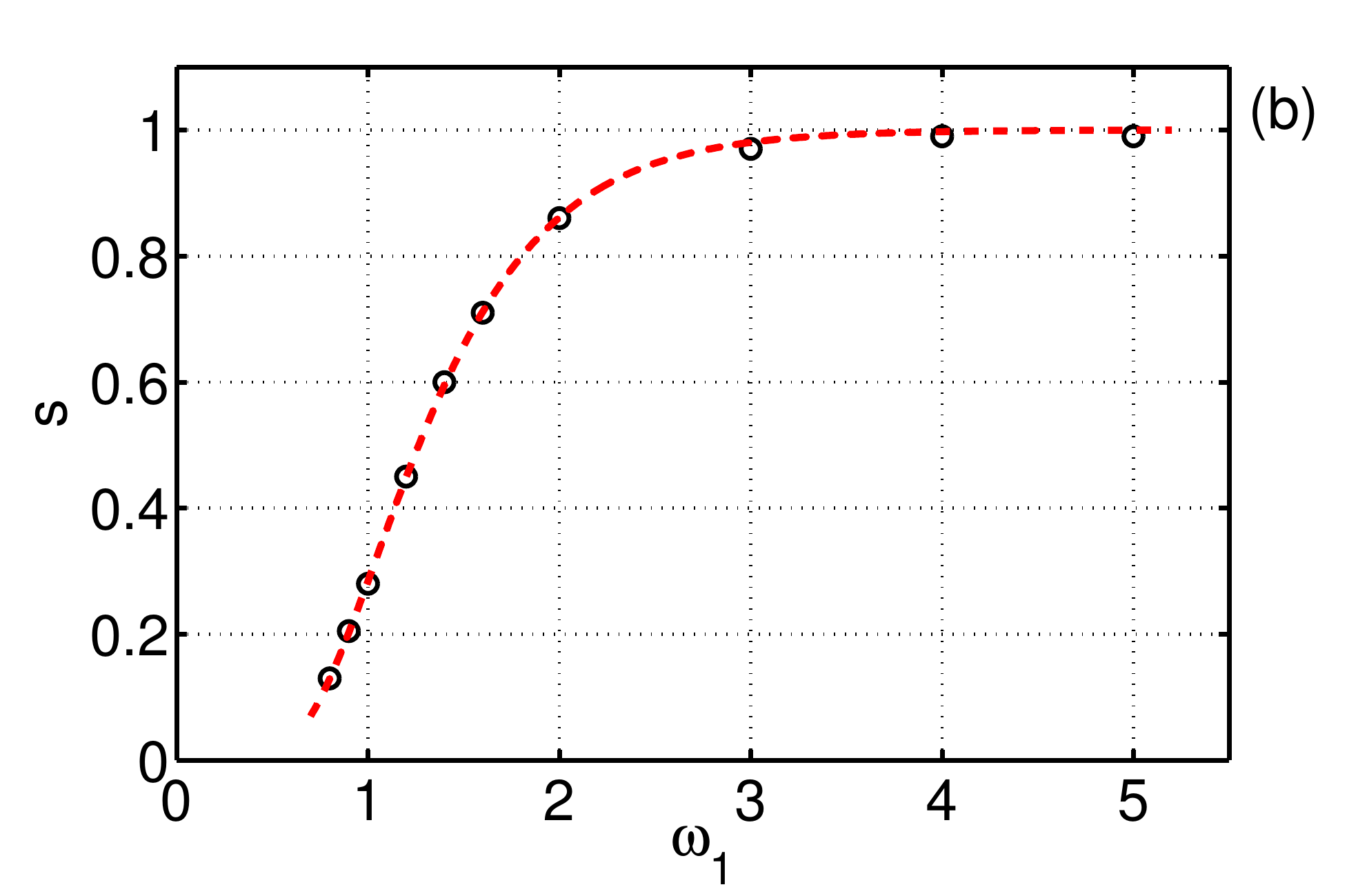}

\caption{\small {\it (Color on-line)} 
(a) Initial $-Q(0)$ (solid black line) and asymptotic $-Q_{A}$ (red circles) potential to kinetic energy ratio, versus $\omega_{1}$.
(b) Frequency of the oscillations $s$ for different cnoidal waves (black circles), versus $\omega_{1}$. Dashed red line shows the double maximal growth rate of the MI $2\gamma_{max}$, see Eq.~(\ref{cnoidal1_increment}), for these cnoidal waves.}
\label{fig:s_gamma}
\end{figure}

Development of the MI from cnoidal waves with large $\omega_{1}$ is similar to that for the condensate (see Fig.~\ref{fig:CW3_MI} in Appendix~\ref{Sec:Annex-B}), and these waves are themselves close to condensate~(\ref{cnoidal_limit_w1}), Fig.~\ref{fig:cnoidal_wave_B112}(b). The initial energy ratio for such cnoidal waves is large, $-Q(0)\gg 1$, see Fig.~\ref{fig:s_gamma}(a), and the asymptotic ratio $Q_{A}=-2$ is the same as for the condensate case~\cite{agafontsev2014integrable}, see Fig.~\ref{fig:evolutionE12}(b). Moreover, the asymptotic energy ratio is equal to $Q_{A}=-2$ for all cnoidal waves of dn-branch that we studied, though the nature of this beautiful relation remains unclear for us so far, Fig.~\ref{fig:s_gamma}(a).

In the nonlinear stage of the MI, kinetic $\langle H_{d}(t)\rangle$ and potential $\langle H_{4}(t)\rangle$ energies, as well as the moments $M^{(n)}(t)$, approach in oscillatory way to their asymptotic values. These oscillations are very well approximated by functions~(\ref{oscillations}) for all cnoidal waves that we studied. However, for $\omega_{1}\le 1$ we were able to check this only for the first four moments, since oscillations for such cnoidal waves are very small, compare Fig.~\ref{fig:evolutionE12}(a) and (b). 
Exponent $\alpha$ for power-law decay $\propto t^{-\alpha}$ of the amplitude of these oscillations is different for different cnoidal waves and moments $M^{(n)}(t)$, and stays in the range $1<\alpha< 1.5$. It turns out that the frequency of the oscillations is equal to the double maximal growth rate of the MI~(\ref{cnoidal1_increment}), $s=2\gamma_{\max}$, Fig.~\ref{fig:s_gamma}(b). For small $\omega_{1}$ the maximal growth rate~(\ref{cnoidal1_increment}) is exponentially small~\cite{kuznetsov1999modulation}, 
\begin{equation}\label{cnoidal1_increment_w1_small}
\gamma_{\max}\to 8\bigg(\frac{\pi}{\omega_{1}}\bigg)^{2}\exp\bigg(-\frac{\pi\omega_{0}}{\omega_{1}}\bigg),
\end{equation}
and for large $\omega_{1}$ it approaches to that for the condensate~(\ref{cnoidal_limit_w1}) with amplitude $1/\sqrt{2}$,
\begin{equation}\label{cnoidal1_increment_w1_large}
\gamma_{\max}\to 1/2.
\end{equation}
The same relation between the oscillations' frequency and the maximal growth rate of the MI is valid for the condensate case~\cite{agafontsev2014integrable} too. More study is necessary to clarify the nature of this beautiful phenomenon.

\begin{figure}[t] \centering
\includegraphics[width=8.0cm]{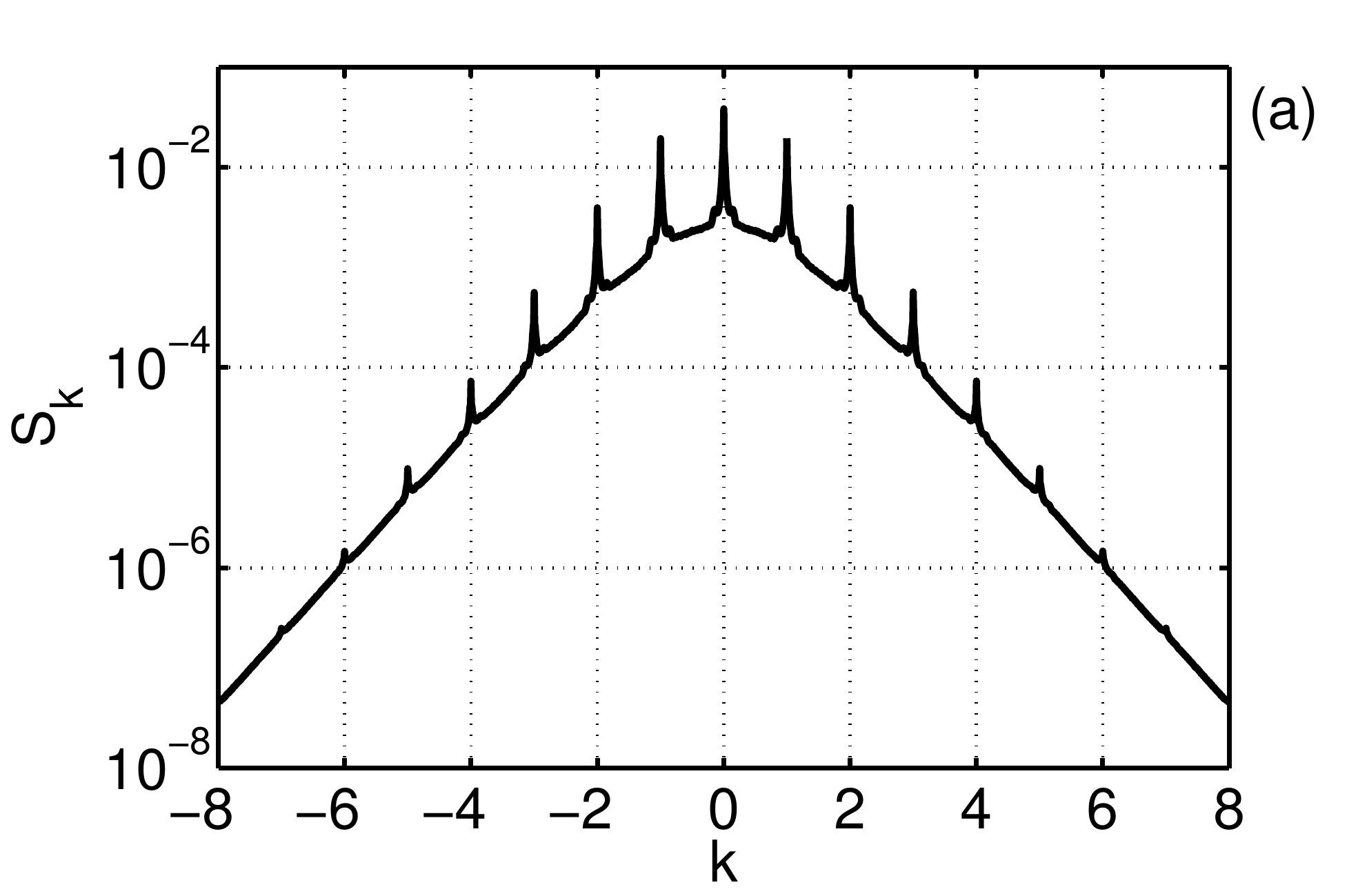}
\includegraphics[width=8.0cm]{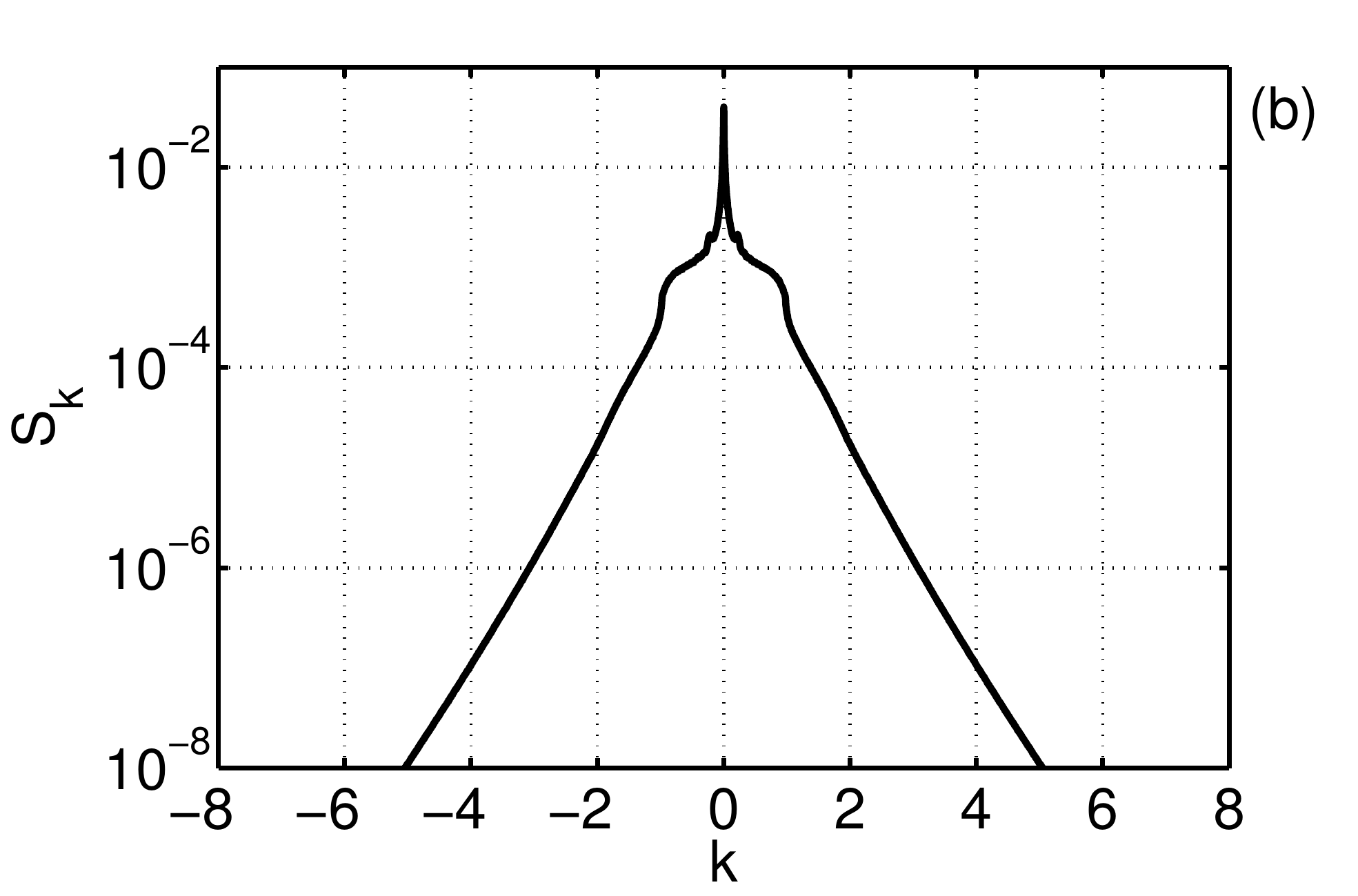}

\caption{\small Asymptotic wave-action spectrum $S_{k}$ for cnoidal waves with $\omega_{1}=0.8$ (a) and $\omega_{1}=5$ (b). Small irregularities at the bases of the peaks disappear at later times. At large $k$ the spectrum decays exponentially $S_{k}\propto e^{-\rho |k|}$ with $\rho= 0.81$ for $\omega_{1}=0.8$ and $\rho= 1.42$ for $\omega_{1}=5$.}
\label{fig:spectra_B1_w12}
\end{figure}

Asymptotic wave-action spectrum decays exponentially at large wavenumbers, $S_{k}\propto e^{-\rho |k|}$. Exponent $\rho$ increases with $\omega_{1}$, so that for larger $\omega_{1}$ the spectrum is narrower, Fig.~\ref{fig:spectra_B1_w12}(a),(b). 
For $\omega_{1}=5$ we obtain $\rho=1.42$, that after the scaling transformations quantitatively corresponds to the condensate case~\cite{agafontsev2014integrable}. 
Only one peak at zeroth harmonic ``survives'' in the asymptotic spectrum for cnoidal waves with large $\omega_{1}$, while for small $\omega_{1}$ many of such peaks at integer wavenumbers $k_{0}$ remain. 
The spectrum in these peaks behaves by power law, $S_{k}\propto |k-k_{0}|^{-\beta}$, with different exponents $\beta>0$ for different cnoidal waves and peaks $|k_{0}|$. The peaks accumulate about 40\% of all wave action $\langle N\rangle$, for all cnoidal waves that we studied. For large $\omega_{1}$ this fraction of wave action is concentrated in quasi-condensate modes $|k|\le \delta k$, $\delta k\sim 0.1$, only. For 
small $\omega_{1}$ the peak at zeroth harmonic becomes narrower and the other peaks become wider, so that their widths and concentrated in them fractions of wave action become comparable. Thus, quasi-condensate is replaced by ``quasi-cnoidal wave'' -- a collection of power-law peaks at the same positions where the peaks of the original cnoidal wave were situated.

\begin{figure}[t] \centering
\includegraphics[width=8.0cm]{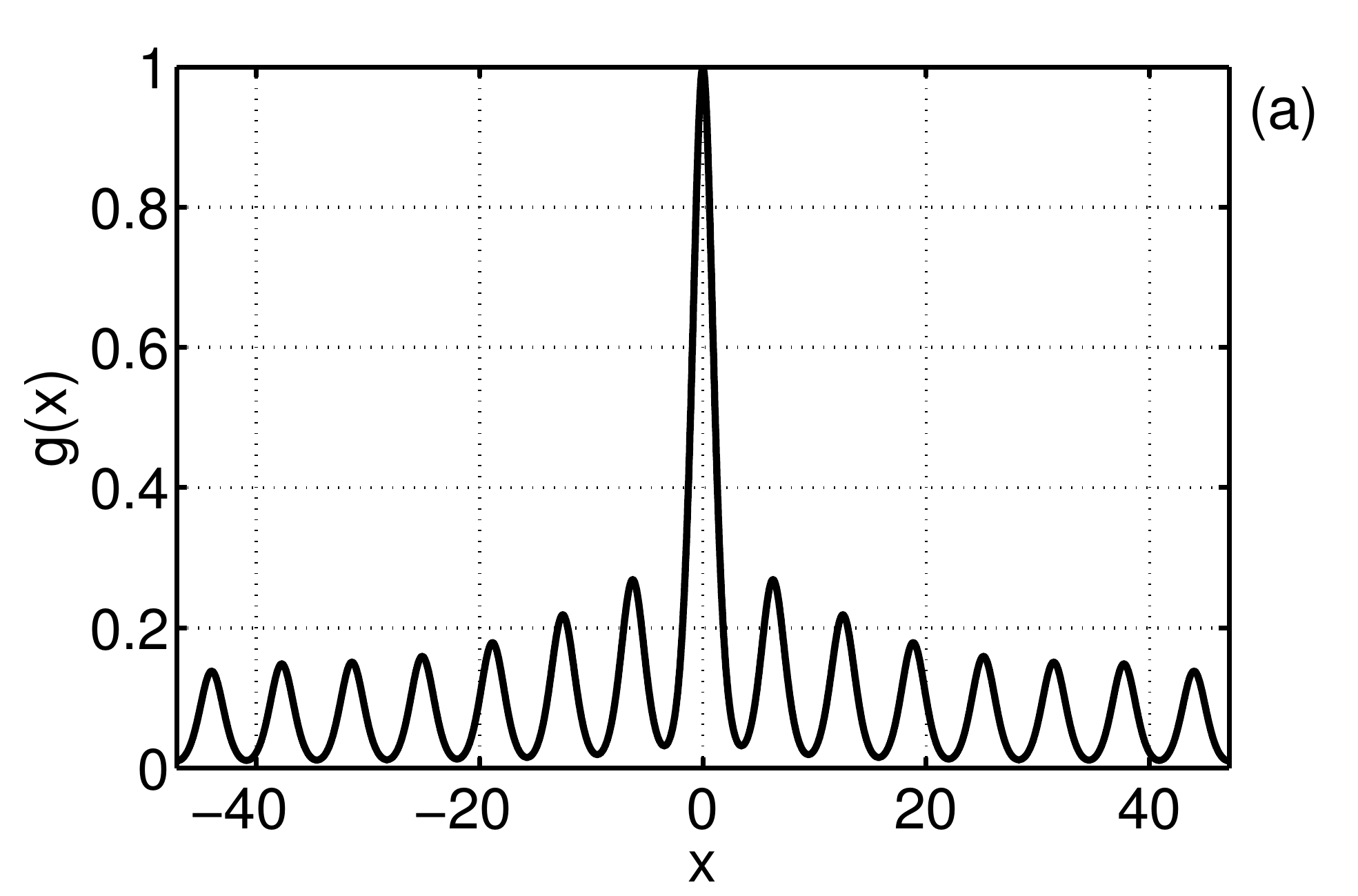}
\includegraphics[width=8.0cm]{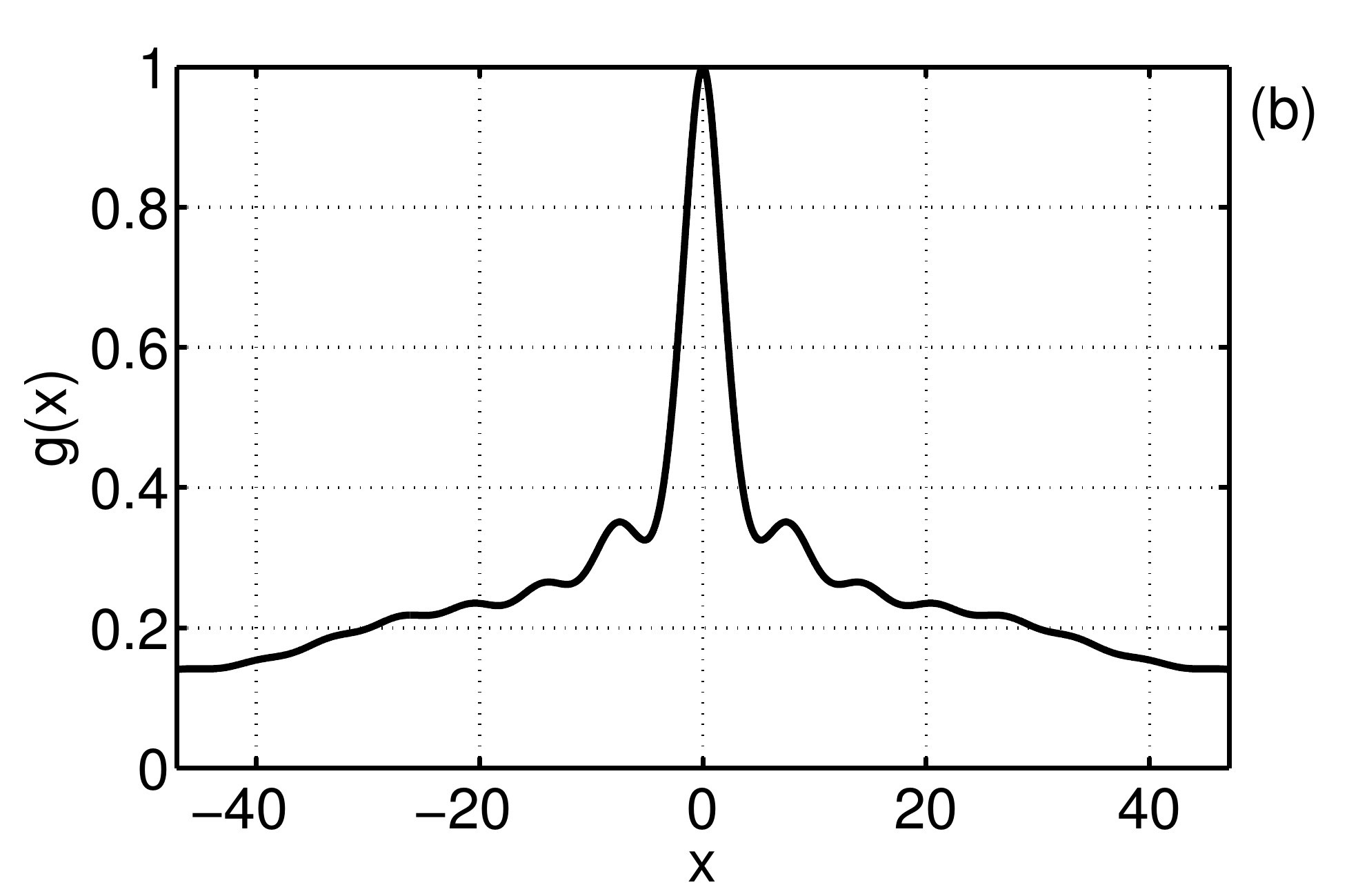}

\caption{\small Asymptotic spatial correlation function $g(x)$ for cnoidal waves with $\omega_{1}=0.8$ (a) and $\omega_{1}=5$ (b). In graph (a) the period of the oscillations is equal to $2\pi$.}
\label{fig:corr_x0_B1_w12}
\end{figure}

At small lengths $|x|<x_{corr}/2$ the asymptotic spatial correlation function is well approximated by Gaussian~(\ref{corr_Gaussian}). It's full width at half maximum increases with $\omega_{1}$; for $\omega_{1}=0.8$ we measure $x_{corr}=2.2$ and for $\omega_{1}=5$ we find $x_{corr}=5.6$. After the scaling transformations, the latter value almost coincides with that for the condensate case~\cite{agafontsev2014integrable}. For small $\omega_{1}$ the asymptotic correlation function decays with $|x|$ in oscillatory way; the period of these oscillations is equal to $2\pi$, Fig.~\ref{fig:corr_x0_B1_w12}(a). The oscillations become more pronounced for cnoidal waves with smaller $\omega_{1}$. We think that their amplitude decays with $|x|$ exponentially, similarly to Fig.~\ref{fig:corr_x0}(b) for $\omega_{1}=1.6$. However, for cnoidal waves with sufficiently small $\omega_{1}\le 1.2$ we cannot check this hypothesis with our computational resources. 
These oscillations with period $2\pi$ are connected with the peaks at nonzero integer wavenumbers $|k_{0}|>0$ in the asymptotic spectrum, since the oscillations change drastically if we ``erase'' the spectrum near the corresponding modes. 
For large $\omega_{1}$ both the peaks in the asymptotic spectrum at $|k_{0}|>0$ and the oscillations of the asymptotic spatial correlation function disappear, Fig.~\ref{fig:spectra_B1_w12}(b),~\ref{fig:corr_x0_B1_w12}(b), and the correlation function decays inverse-proportionally with $|x|$, $g(x)\propto |x|^{-1}$, as for the condensate case~\cite{agafontsev2014integrable}.

\begin{figure}[t] \centering
\includegraphics[width=8.0cm]{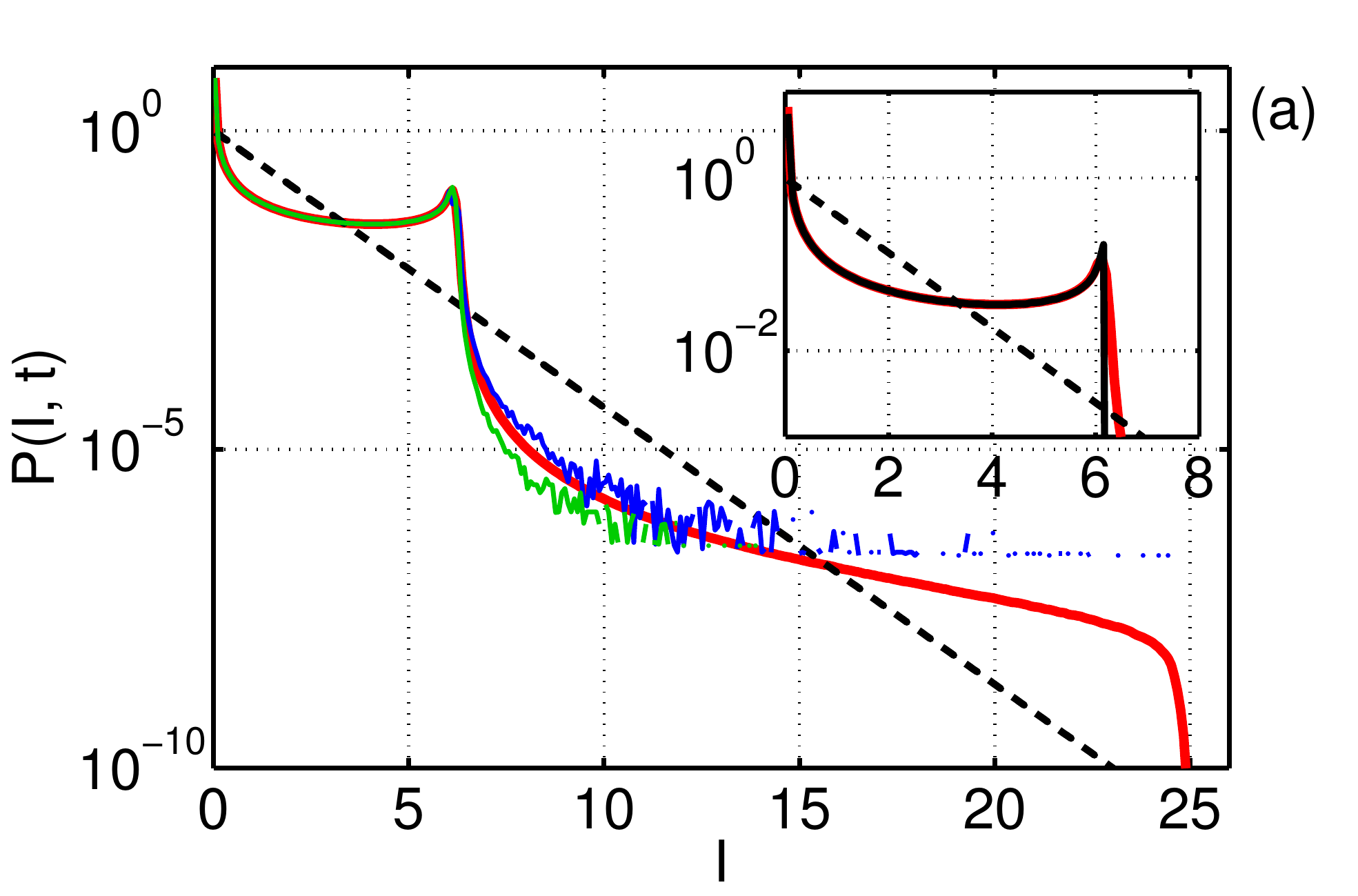}
\includegraphics[width=8.0cm]{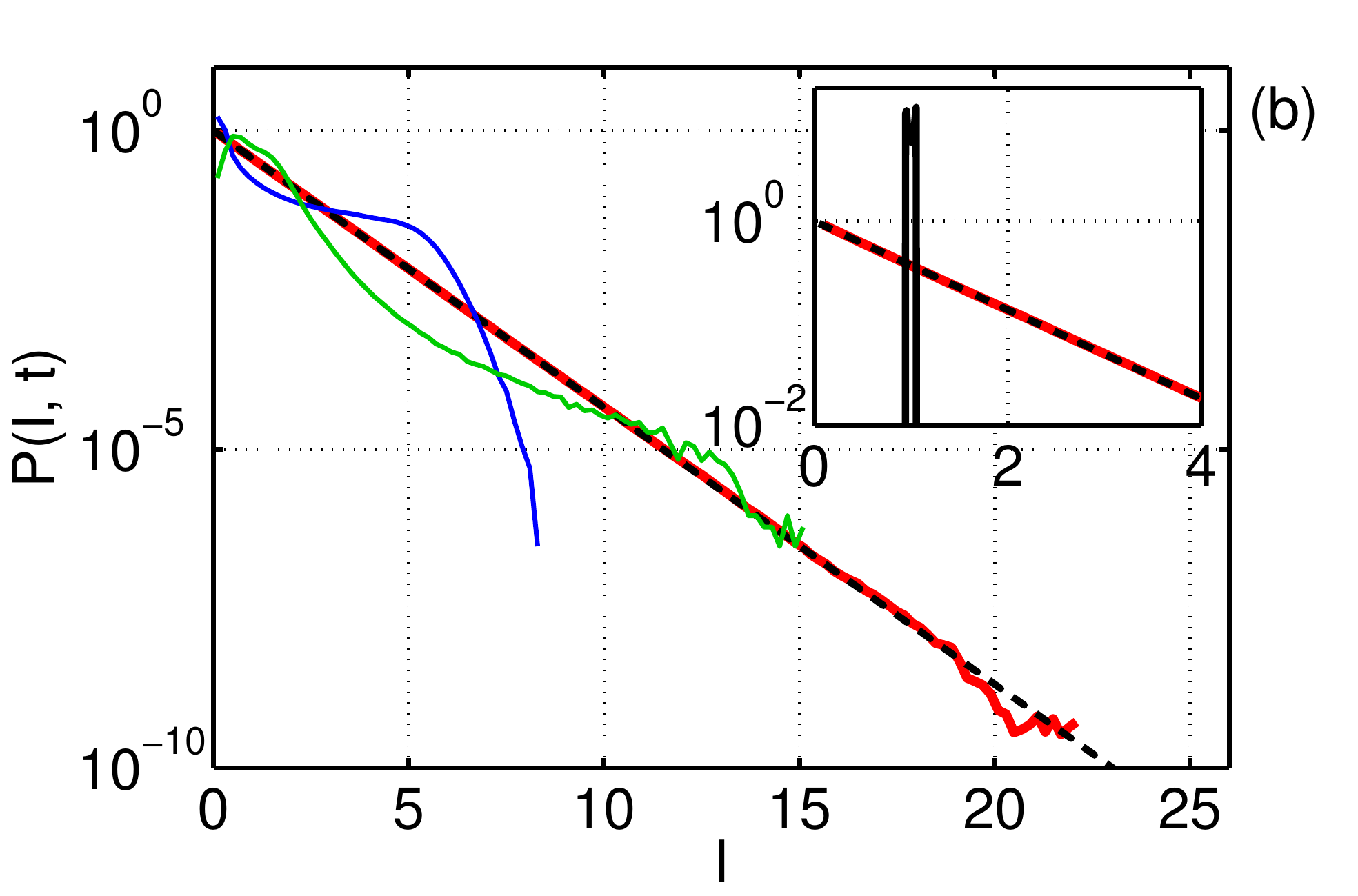}

\caption{\small {\it (Color on-line)} Asymptotic PDF $\mathcal{P}_{A}(I)$ (thick red) and exponential PDF~(\ref{Rayleigh}) (dashed black) for cnoidal waves with $\omega_{1}=0.8$ (a) and $\omega_{1}=5$ (b). Solid black lines in the insets show the corresponding initial PDFs $\mathcal{P}(I,t)$ at $t=0$. Blue lines are the PDFs at the first local maximum of potential energy modulus $|\langle H_{4}(t)\rangle|$ at $t=151.4$ (a) and $t=26.85$ (b), green lines are the PDFs at the first local minimum of $|\langle H_{4}(t)\rangle|$ at $t=179.8$ (a) and $t=31.05$ (b). Note, that the asymptotic PDFs $\mathcal{P}_{A}(I)$ are additionally averaged over time close to the asymptotic stationary state, which allows us to measure these PDFs for larger relative intensities $I$ than the PDFs at specific times $\mathcal{P}(I,t)$.}
\label{fig:PDFa12}
\end{figure}

\begin{figure}[t] \centering
\includegraphics[width=8.0cm]{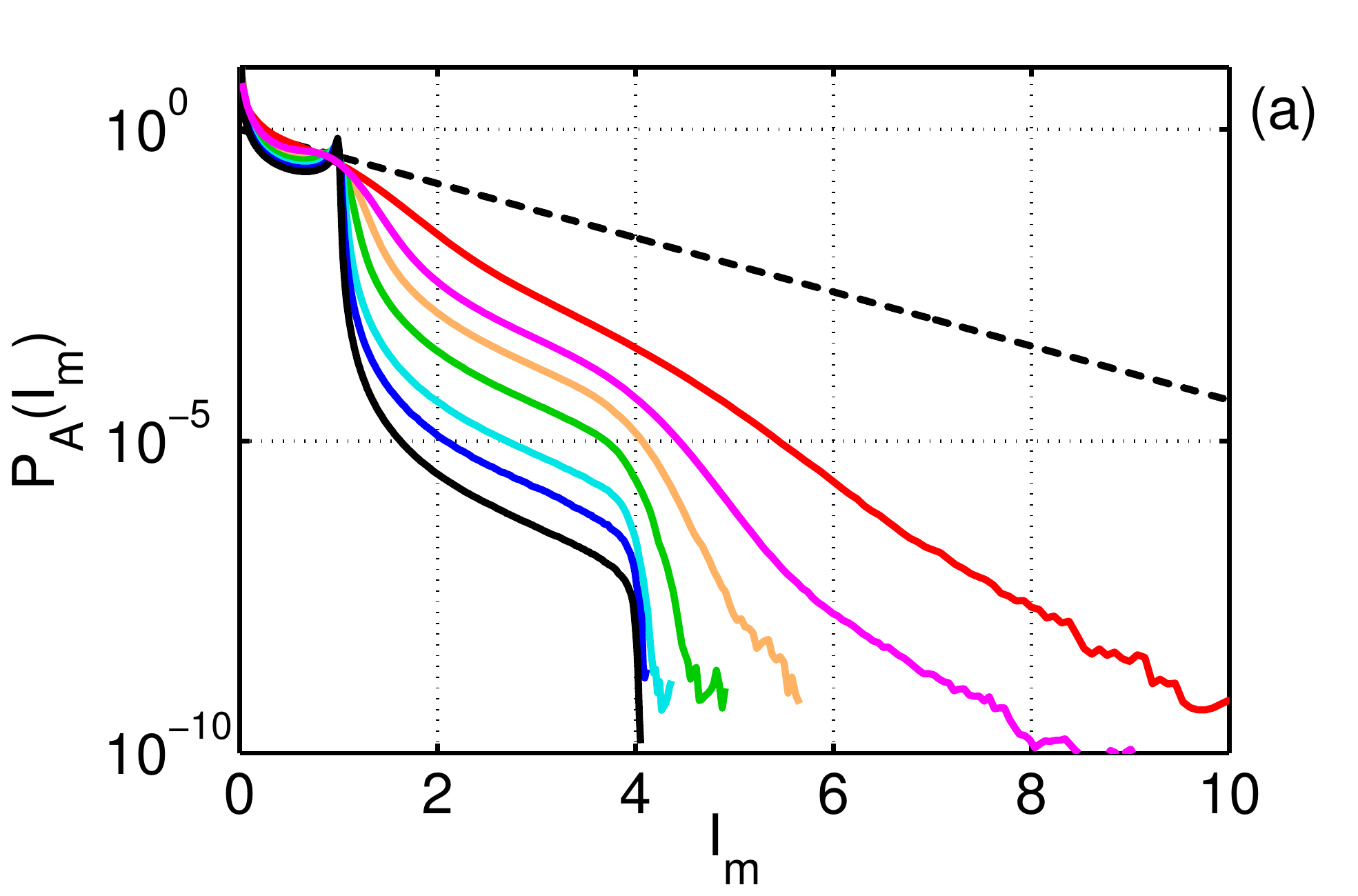}
\includegraphics[width=8.0cm]{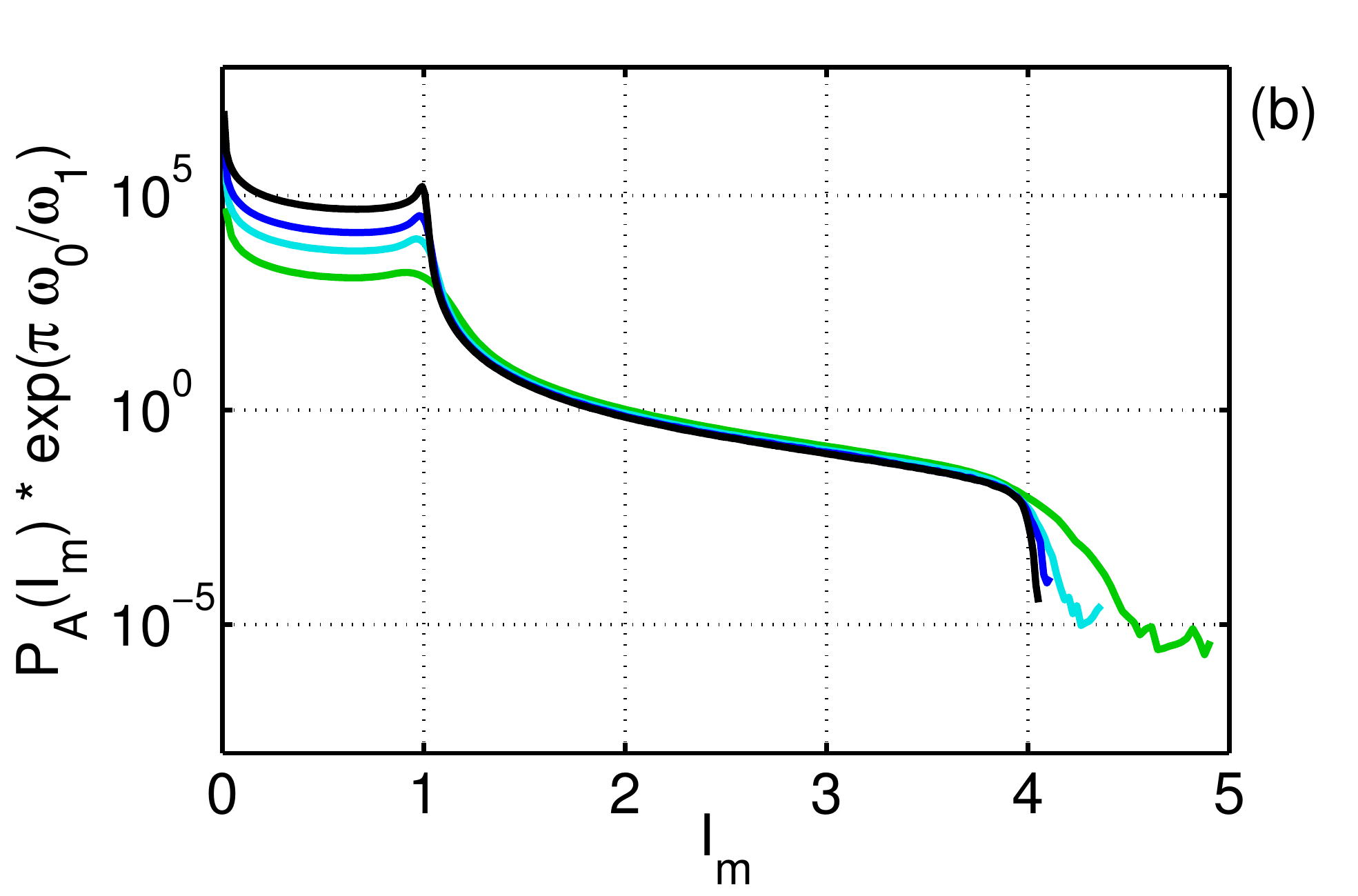}

\caption{\small {\it (Color on-line)} Asymptotic PDF $\mathcal{P}_{A}(I_{m})$ (a) and renormalized asymptotic PDF $\mathcal{P}_{A}(I_{m})\times e^{\pi\omega_{0}/\omega_{1}}$ (b) for renormalized square amplitude $I_{m}=|\Psi|^{2}/\max|\Psi_{dn}|^{2}$: for $\omega_{1}=0.8$ (black), $\omega_{1}=0.9$ (blue), $\omega_{1}=1$ (cyan), $\omega_{1}=1.2$ (green), $\omega_{1}=1.4$ (yellow), $\omega_{1}=1.6$ (pink) and $\omega_{1}=2$ (red). Here $\max|\Psi_{dn}|$ is the maximal amplitude of the original cnoidal wave.}
\label{fig:PDFa_w1all}
\end{figure}

For small $\omega_{1}$ the asymptotic PDF is significantly non-exponential, Fig.~\ref{fig:PDFa12}(a). 
In particular, for $\omega_{1}=0.8$ the maximal deviation from exponential PDF~(\ref{Rayleigh}) is observed at $I=24.2$, where the asymptotic PDF $\mathcal{P}_{A}(I)\approx 7.8\times 10^{-9}$ exceeds the exponential PDF by about 250 times. However, for these $\omega_{1}$ the typical deviation of square amplitude $|\Psi|^{2}$ is significantly larger than its mean value $\langle|\Psi|^{2}\rangle$, see e.g. Fig.~\ref{fig:cnoidal_wave_B1_final}. 
Therefore, it is also instructive to measure the PDF $\mathcal{P}(I_{m},t)$ for square amplitude $I_{m}=|\Psi|^{2}/\max|\Psi_{dn}|^{2}$, renormalized to the maximal amplitude of the original cnoidal wave $\max|\Psi_{dn}|$. 
In this case, $I_{m}=1$ corresponds to the maximal amplitude $|\Psi|=\max|\Psi_{dn}|$ of the initial cnoidal wave, $I_{m}=4$ corresponds to two-fold increase in amplitude $|\Psi|=2\max|\Psi_{dn}|$, and so on. 
As shown in Fig.~\ref{fig:PDFa_w1all}(a), for small $\omega_{1}$ the asymptotic PDF $\mathcal{P}_{A}(I_{m})$ decreases sharply at $I_{m}=1$ and $I_{m}=4$, and ends almost exactly at $I_{m}=4$. Fig.~\ref{fig:PDFa12}(a) demonstrates also that the PDF almost does not change with time at $I_{m}\in[0,1]$ (which corresponds to $I\in[0,6]$ in the figure), and the asymptotic PDF almost coincides in this region with the the initial PDF at $t=0$. 
These facts corroborate our observation, that after development of the MI from cnoidal waves with small $\omega_{1}$, wave field at all times remains close to a composition of singular solitons~(\ref{nlse_soliton}). Then, the first part of the PDF $I_{m}\in[0,1]$ in Fig.~\ref{fig:PDFa_w1all}(a) represents this composition of solitons, and the second part $I_{m}\in[1,4]$ corresponds to very rare two-soliton collisions. During these collisions wave amplitude exceeds maximal amplitude of the original cnoidal wave by two times maximum. 

The rate of such collisions should be proportional to the maximal growth rate of the MI $\gamma_{\max}$, which is exponentially small~(\ref{cnoidal1_increment_w1_small}) for small $\omega_{1}$. As shown in Fig.~\ref{fig:PDFa_w1all}(b), in the region of two-soliton collisions $I_{m}\in[1,4]$ the renormalized asymptotic PDFs $\mathcal{P}_{A}(I_{m})\times e^{\pi\omega_{0}/\omega_{1}}$ coincide almost exactly for sufficiently small $\omega_{1}$ (this coincidence is worse if we compare $\mathcal{P}_{A}(I_{m})/\gamma_{\max}$). 
Thus, for small $\omega_{1}$ the rate of two-soliton collisions and the probability of occurrence of waves twice larger than the original cnoidal wave are proportional to $e^{-\pi\omega_{0}/\omega_{1}}$, which is exponentially small value. 
We think that three-soliton collisions should also be present on the PDF, but they are extremely rare and we do not detect them in our experiments. 
We observe the first rogue waves that look like three-pulses collisions starting from cnoidal wave with $\omega_{1}=1.6$, Fig.~\ref{fig:RW}(c). 

For sufficiently large $\omega_{1}$ the PDF in the asymptotic stationary state coincides with exponential PDF~(\ref{Rayleigh}), Fig.~\ref{fig:PDFa12}(b). This coincidence is almost exact already from $\omega_{1}=3$. In the nonlinear stage of the MI the PDF may significantly deviate from the exponential PDF. If we limit ourselves with rogue waves $I>8$ only, then for cnoidal wave with $\omega_{1}=5$ the maximal excess of the PDF $\mathcal{P}(I,t)$ over the exponential PDF of about 2.5 times is achieved at the first local minimum of potential energy modulus $|\langle H_{4}(t)\rangle|$ at $t=31.05$ and for relative intensity $I=12.5$. These results coincide with that for the condensate case~\cite{agafontsev2014integrable}. 

Thus, for cnoidal waves with $\omega_{1}\gtrsim 1.5$ the probability of rogue waves occurrence is not significantly larger than for a random wave field governed by linear equations. 
For $\omega_{1}\lesssim 1.5$ wave field is close to a collection of singular solitons rarely interacting with each over. 
The amplitude of these solitons is already significantly larger than the mean amplitude $\langle|\Psi|^{2}\rangle^{1/2}$. 
Therefore, it is no surprise that rogue waves of intensities much larger than the mean one may appear in this system with probabilities exceeding exponential PDF~(\ref{Rayleigh}) by orders of magnitude. However, the amplitude of these rogue waves almost never exceeds the double maximal amplitude of the original cnoidal wave.

For each of the ten studied cnoidal waves $\omega_{1}$, we examined several of the largest rogue waves detected in our experiments. 
For $\omega_{1}\ge 1.6$ many of these rogue waves look like three-pulses collisions, similar to that in Fig.~\ref{fig:RW}(c), while for $\omega_{1}< 1.6$ all of these rogue waves look like two-pulses collisions similar to that in Fig.~\ref{fig:RWmax}(c). 
For the first several extremums of potential energy modulus, the phases of the rogue waves are close to $\mathrm{arg}\, \Psi\approx \pi/2+\pi(m-1)$ for local maximums and $\mathrm{arg}\, \Psi\approx \pi+\pi(m-1)$ for local minimums of $|\langle H_{4}(t)\rangle|$, where $m$ is the local maximum or local minimum index number respectively.
Also, we studied evolution of one realization from ensemble of initial conditions for each of the ten cnoidal waves $\omega_{1}$, examining all waves that at any time exceed the maximal amplitude of the original cnoidal wave by 1.5 times or more. 
For $\omega_{1}=0.8$ we found 14 such events in the time interval $t\in[0,1000]$, and for $\omega_{1}=5$ -- more than 1500 events in the interval $t\in[0,200]$. 
All these waves, and also all of the largest rogue waves generated from the entire ensembles of initial conditions, at the time of their maximal elevation have quasi-rational profile similar to that of the Peregrine solution~(\ref{Peregrine_x}); see examples in Fig.~\ref{fig:RWmax}(a),~\ref{fig:RW}(a). Time evolution of maximal amplitude for these waves differs from that for the Peregrine solution~(\ref{Peregrine_t}), see examples in Fig.~\ref{fig:RWmax}(b),~\ref{fig:RW}(b).


\section{Conclusions}
\label{Sec:Conclusions}

In this paper we studied integrable turbulence generated from MI of dn-branch of cnoidal waves. The corresponding problem of MI essentially depends on one free parameter, and the ratio $\omega_{1}/\omega_{0}$ between the imaginary and real half-periods of the cnoidal wave can be used as such. Using the scaling transformations, we fixed $\omega_{0}=\pi$ and studied dependence of integrable turbulence on the imaginary half-period $\omega_{1}$. 

We found that properties of the integrable turbulence change gradually with $\omega_{1}$, so that cnoidal waves with ``intermediate'' $\omega_{1}$ lead to turbulence with ``intermediate'' properties between the two limits $\omega_{1}\to 0$ and $\omega_{1}\to +\infty$. Our results show, that in the nonlinear stage of the MI, statistical characteristics of the turbulence evolve with time in oscillatory way, approaching to their asymptotics at late times. This means that the system asymptotically approaches in oscillatory way toward the stationary state of the integrable turbulence. This state depends on cnoidal wave parameters and is defined by infinite series of invariants~(\ref{all_integrals1}),~(\ref{all_integrals2}). 

During the evolution toward the asymptotic state, kinetic $\langle H_{d}(t)\rangle$ and potential $\langle H_{4}(t)\rangle$ energies, and also the moments $M^{(n)}(t)$, oscillate around their asymptotic values according to ansatz~(\ref{oscillations}). The amplitudes of these oscillations decay with time as $t^{-\alpha}$,  with different exponents $1<\alpha< 1.5$ for different cnoidal waves and moments $M^{(n)}(t)$. The oscillations are very small for cnoidal waves with small $\omega_{1}$, and pronounced for cnoidal waves with large $\omega_{1}$. 
The phases of the oscillations contain nonlinear phase shift decaying with time as $t^{-1/2}$, and the frequency is equal to the double maximal growth rate of the MI, $s=2\gamma_{\max}$. Thus, for cnoidal waves with small $\omega_{1}$ the frequency of the oscillations $s$ is exponentially small and the oscillations themselves are very small too, and for large $\omega_{1}$ the oscillations are pronounced and their frequency approaches to $s=1$. 
The ratio of potential to kinetic energy in the asymptotic state is equal to $Q_{A}=\langle H_{4}\rangle/\langle H_{d}\rangle=-2$ for all cnoidal waves of dn-branch, while the initial energy ratio is different $Q(0)\in(-\infty, -2)$ for different cnoidal waves.
The other characteristics of the turbulence -- i.e. wave-action spectrum, spatial correlation function and the PDF -- evolve with time in oscillatory way coherently with kinetic and potential energies. We describe their evolution using points in time when potential energy modulus $|\langle H_{4}(t)\rangle|$ takes maximal and minimal values; at these points the evolution of the spectrum, the correlation function and the PDF turns to roughly the opposite.

For unperturbed cnoidal wave~(\ref{cnoidal1_stationary}) with $\omega_{0}=\pi$, wave-action spectrum represents a collection of peaks at integer wavenumbers $k_{0}\in\mathbb{Z}$, and spatial correlation function is periodic with period $2\pi$. In the linear stage of the MI, wave-action spectrum starts to rise most notably near half-integer wavenumbers $k_{0}+1/2$, while spatial correlation function does not change visibly. 
In the nonlinear stage and at the local maximums of $|\langle H_{4}(t)\rangle|$, the peaks in the spectrum at $k_{0}$ are the smallest and the rest of the spectrum is maximally excited, while the correlation function takes (locally in time) minimal values at $|x|>0$. 
At the local minimums of $|\langle H_{4}(t)\rangle|$, the peaks in the spectrum are the largest and the rest of the spectrum is minimally excited, while the correlation function takes (locally in time) maximal values at $|x|>0$. Thus, during the evolution toward the asymptotic state, wave action is being ``pumped'' in oscillatory way between the peaks at integer wavenumbers and the rest of the spectrum, while spatial correlation function ``forms'' its tails at large lengths $x$.

The asymptotic wave-action spectrum decays exponentially $S_{k}\propto e^{-\rho|k|}$ at large $k$. Exponent $\rho$ increases with $\omega_{1}$, so that for larger $\omega_{1}$ the spectrum is narrower. For cnoidal waves with large $\omega_{1}$ only the peak at zeroth harmonic ``survives'' in the asymptotic spectrum, while for small $\omega_{1}$ many of the peaks at integer wavenumbers remain. Contrary to the original cnoidal wave, these peaks in the asymptotic spectrum occupy not single harmonics $k_{0}$ only, but small regions of modes around $k_{0}$ where the spectrum behaves by power law $S_{k}\propto |k-k_{0}|^{-\beta}$, $\beta>0$, with different exponents $\beta$ for different cnoidal waves and peaks $|k_{0}|$. 
These power-law peaks contain about 40\% of all wave action $\langle N\rangle$, for all cnoidal waves that we studied. For sufficiently large $\omega_{1}$ most of this wave action is concentrated in quasi-condensate modes $|k|\le \delta k$, $\delta k\sim 0.1$, which have extremely large scales $\ell\gg 2\pi$ in the physical space. For small $\omega_{1}$ quasi-condensate is replaced by ``quasi-cnoidal wave'' -- a collection of power-law peaks at the same positions where the peaks of the original cnoidal wave were situated.

The asymptotic spatial correlation function is close to Gaussian~(\ref{corr_Gaussian}) at small lengths $|x|<x_{corr}/2$; its full width at half maximum $x_{corr}$ increases with $\omega_{1}$ so that for larger $\omega_{1}$ the correlation function is wider. For sufficiently small $\omega_{1}$ the asymptotic correlation function decays at large $|x|$ in oscillatory way; these oscillations with period $2\pi$ are connected with the peaks at nonzero integer wavenumbers $|k_{0}|>0$ in the asymptotic spectrum. We think that the amplitude of these oscillations decays with $|x|$ exponentially as in Fig.~\ref{fig:corr_x0}(b). For large $\omega_{1}$ both the peaks in the asymptotic spectrum and the oscillations of the asymptotic correlation function disappear, and the correlation function decays inverse-proportionally $g(x)\propto |x|^{-1}$ with $|x|$, as for the condensate case~\cite{agafontsev2014integrable}.

After development of the MI from cnoidal waves with small $\omega_{1}$, wave field at all times remains close to a collection of singular solitons~(\ref{nlse_soliton}) with different phases and positions. Moreover, the positions of these solitons remain generally very close to the positions of ``solitons'' of the original cnoidal wave. Thus, integrable turbulence transforms into integrable soliton turbulence of very thin and high solitons~(\ref{nlse_soliton}). The amplitude of these solitons is already significantly larger than the mean amplitude $\langle|\Psi|^{2}\rangle^{1/2}$. 
In the asymptotic stationary state of this turbulence, the PDF of wave intensity is significantly non-exponential, and dynamics of the system reduces to two-soliton collisions. These collisions provide up to two-fold increase in amplitude compared with the original cnoidal wave and occur with exponentially small rate $\propto e^{-\pi\omega_{0}/\omega_{1}}$. Still, the probability of occurrence of large waves during these collisions is much larger than would be for exponential PDF~(\ref{Rayleigh}) with the same mean square amplitude $\sigma^{2}=\langle|\Psi|^{2}\rangle$. The potential to kinetic energy ratio $Q(t)$ for such turbulence at all times remains very close to -2, as for the singular soliton~(\ref{nlse_soliton}).

Integrable turbulence generated from cnoidal waves with sufficiently large $\omega_{1}$ is qualitatively and quantitatively very similar to that for the condensate case~\cite{agafontsev2014integrable}. The PDF in the asymptotic state of this turbulence is exponential. During the evolution toward the asymptotic state, the PDF may significantly deviate from the exponential PDF, however, at all times it does not exceed the exponential PDF by more than several times. 

Overall, for $\omega_{1}\gtrsim 1.5$ the probability of rogue waves appearance does not exceed significantly that for the exponential PDF. For $\omega_{1}\lesssim 1.5$ rogue waves may appear much more frequently than predicted by the exponential PDF, however the amplitude of these waves almost never exceeds the double maximal amplitude of the original cnoidal wave.

According to our observations, all sufficiently large waves, that appear after development of the MI, at the time of their maximal elevation have quasi-rational profile similar to that of the Peregrine solution of the NLS equation~(\ref{Peregrine_x}). 
We would like to stress that this similarity isn't a sign that Peregrine solution emerges in the problem of MI of cnoidal waves, but merely a characteristic of rogue waves spatial profile. 
In terminology of~\cite{kedziora2014rogue}, rogue waves that we observe could be the ``concentrated'' cnoidal rogue wave and the ``fused'' second-order cnoidal rogue wave. However, there is also another possibility that the observed rogue waves were formed as collisions of breathers, which decomposed from ``superregular'' solitonic solutions on the background of cnoidal wave~\cite{gelash2015private}, similar to that in the condensate case~\cite{gelash2014superregular, gelash2015superregular}. For sufficiently small $\omega_{1}$, solutions that describe rogue waves formation should transform into collisions of singular solitons~(\ref{nlse_soliton}).
It is interesting, that for the first several extremums of potential energy modulus, the phases of the rogue waves are close to $\mathrm{arg}\, \Psi\approx \pi/2+\pi(m-1)$ for local maximums and $\mathrm{arg}\, \Psi\approx \pi+\pi(m-1)$ for local minimums of $|\langle H_{4}(t)\rangle|$, where $m$ is the local maximum or local minimum index number respectively. 
We plan to examine the question of rogue waves origin in more details in another publication.

MI of cn-branch of cnoidal waves~(\ref{cnoidal2}) should lead to integrable turbulence with many similar properties. 
In particular, we expect the similar oscillatory evolution for characteristics of the turbulence toward their asymptotics at late times. It is possible though, that oscillations of the moments, and also kinetic and potential energies, are described not by ansatz~(\ref{oscillations}) exactly. 
We think that in the limit $\omega_{1}\to 0$ both branches of cnoidal waves should lead to quantitatively similar stationary states, since in this limit integrable turbulence should transform into integrable soliton turbulence of very thin and high solitons~(\ref{nlse_soliton}). 

However, for cn-branch some properties of the integrable turbulence should be different. 
For instance, we expect that cnoidal waves~(\ref{cnoidal2}) with large $\omega_{1}$ will lead to almost linear integrable turbulence with very small asymptotic potential to kinetic energy ratio $|Q_{A}|\ll 1$, since these cnoidal waves are close to sinusoidal wave with exponentially small amplitude~(\ref{cnoidal2_limit_w1}).
Therefore, for cn-branch the asymptotic energy ratio $Q_{A}$ should not be fixed to -2, but should instead vary from $Q_{A}\to -2$ for $\omega_{1}\to 0$ to $Q_{A}\to 0$ for $\omega_{1}\to +\infty$. Our preliminary experiments confirm these suggestions. Note, that the initial energy ratio for cn-branch of cnoidal waves is also limited in the same region $Q(0)\in(-2,0)$. 
According to our observations, the study of MI for cn-branch of cnoidal waves is significantly more difficult than for dn-branch, since development of the MI takes much more time and the subsequent oscillatory evolution toward the asymptotic stationary state is much slower too.

Combined with the studies~\cite{agafontsev2014integrable,walczak2015optical,suret2016direct} for the condensate and incoherent wave initial conditions, this publication poses a series of questions concerning the general properties of integrable turbulence. Our results demonstrate that turbulence generated from MI of dn-branch of cnoidal waves is quite similar to its limiting case $\omega_{1}\to +\infty$, studied in~\cite{agafontsev2014integrable} as the condensate initial conditions. In both cases, the turbulence approaches toward its asymptotic stationary state in oscillatory way for a very long time. 
For incoherent wave initial conditions the system reaches its stationary state very quickly~\cite{walczak2015optical,suret2016direct}. 
Then, what specifically in the initial conditions leads to such dramatic difference in the time of arrival to the stationary state?

The oscillatory evolution of integrable turbulence for cnoidal wave initial conditions might be connected with excitation of some multi-phase solutions of the NLS equation during the linear stage of MI. This would explain why the frequency of the oscillations is equal to the double maximal growth rate of MI. If true, then what are these multi-phase solutions? During our studies we also observe another interesting phenomenon, that potential to kinetic energy ratio in the asymptotic stationary state is limited as $Q_{A}\in[-2,0]$ (this ratio is equal to -2 for dn-branch, and should be between -2 and 0 for cn-branch of cnoidal waves). This raises a question, whether asymptotic potential energy may exceed the kinetic one by more than two times, and if not, then what is the nature of this constrain? We plan to continue our studies of integrable turbulence in the future publications.

\section*{Acknowledgements}

The authors thank E. Kuznetsov for valuable discussions concerning this publication, M. Fedoruk for access to and V. Kalyuzhny for assistance with Novosibirsk Supercomputer Center. 
Development of the numerical code and simulations were supported by the Russian Science Foundation grant No. 14-22-00174 "Wave turbulence: theory, numerical simulation, experiment", with the simulations performed at the Novosibirsk Supercomputer Center (NSU). 
Analysis of the results was supported by the Russian Science Foundation grant No. 14-50-00095 "The World Ocean in the XXI Century: Climate, Ecosystems, Resources, Natural Disasters".

\appendix
\section{Cnoidal waves}
\label{Sec:Annex-A}

Cnoidal waves of real $\omega_{0}$ and imaginary $\omega_{1}$ half-periods have general form~\cite{kuznetsov1999modulation}
\begin{eqnarray}\label{cnoidal_common}
\psi(x,t) = \sqrt{2}\,\frac{\sigma(x+i\omega_{1}+a)}{\sigma(x+i\omega_{1})\sigma(a)}\,e^{i\Omega t-\zeta(a)x-a\zeta(i\omega_{1})},
\end{eqnarray}
and are exact periodic solutions of the NLS equation~(\ref{nlse}). 
Here $\Omega=3\,\wp(a)$, where $a$ corresponds to one of the two possible branches of cnoidal waves $a=\omega_{0}$ (dn-branch) or $a=\omega_{0}+i\omega_{1}$ (cn-branch), $\wp(z)$ is Weierstrass elliptic function defined on complex plane $z\in \mathbb{C}$ and periodic along both real and imaginary axes with periods $2\omega_{0}$ and $2\omega_{1}$ respectively, while $\sigma(z)$ and $\zeta(z)$ are auxiliary Weierstrass sigma- and zeta-functions defined as $\zeta'(z)=-\wp(z)$ and $\sigma'(z)/\sigma(z)=\zeta(z)$.
The dn- and cn-branches of cnoidal waves can be rewritten as Eq.~(\ref{cnoidal1}) and Eq.~(\ref{cnoidal2}) respectively. In these equations $\nu = (e_{1}-e_{3})^{1/2}$, $s=(e_{2}-e_{3})^{1/2}/\nu$, and $e_{1}>e_{2}>e_{3}$ are values of $\wp(z)$ at $z=\omega_{0}$, $z=\omega_{0}+i\omega_{1}$ and $z=i\omega_{1}$ respectively. 

Square amplitude for both branches of cnoidal waves $a=\omega_{0}$ and $a=\omega_{0}+i\omega_{1}$ is equal to~\cite{kuznetsov1999modulation} 
\begin{eqnarray}\label{cnoidal_common2}
|\psi(x,t)|^{2} = 2\wp(a)-2\wp(x+i\omega_{1}).
\end{eqnarray}
Weierstrass function $\wp(z)$ has a representation in the form of infinite series of KdV-solitons~\cite{kuznetsov1974stability}
\begin{eqnarray}\label{cnoidal_soliton_lattice}
-2\,\wp(x+i\omega_{1}) = \frac{\zeta(i\omega_{1})}{i\omega_{1}} + \sum_{m=-\infty}^{+\infty}\frac{2\lambda^{2}}{\cosh^{2}\,\lambda(x-2m\omega_{0})},\quad \lambda = \pi/2\omega_{1}.
\end{eqnarray}
Note, that the square root of KdV-soliton gives soliton solution~(\ref{nlse_soliton}) for the NLS equation~(\ref{nlse}). 
Thus, one can say that cnoidal waves represent infinite lattices of overlapping solitons, with the width of the solitons proportional to $\omega_{1}$ and the distance between them equal to $2\omega_{0}$.

\section{Development of MI on the cnoidal wave background}
\label{Sec:Annex-B}

In this Appendix we demonstrate how MI develops on the background of three cnoidal waves with $\omega_{1}=0.8$ (weak overlapping), $\omega_{1}=1.6$ (moderate overlapping) and $\omega_{1}=5$ (strong overlapping). We consider one realization of initial noise for each of these cnoidal waves. 

\begin{figure}[H] \centering
\includegraphics[width=14cm]{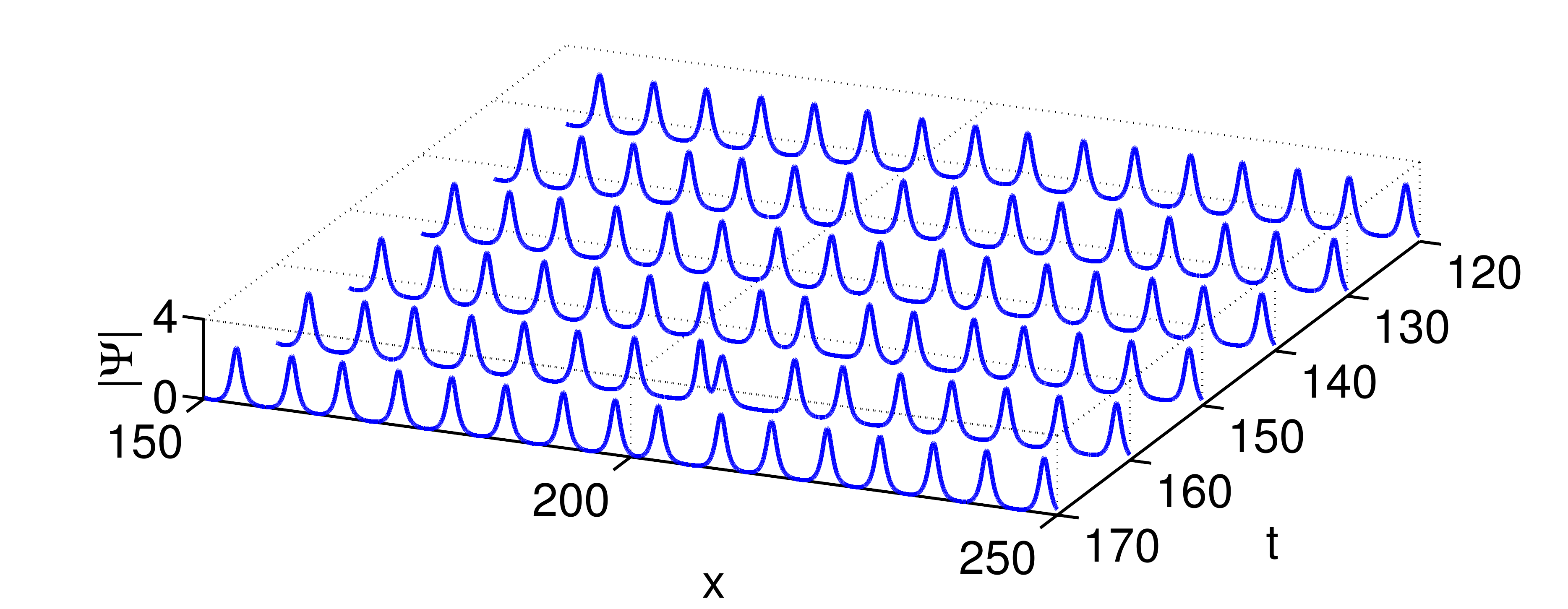}

\caption{\small 
Development of MI on the background of cnoidal wave with $\omega_{0}=\pi$, $\omega_{1}=0.8$. Significant perturbations in amplitude $|\Psi|$ become visible starting from $t=140$; compare with Fig.~\ref{fig:evolutionE12}(a).
}
\label{fig:CW1_MI}
\end{figure}

\begin{figure}[H] \centering
\includegraphics[width=14cm]{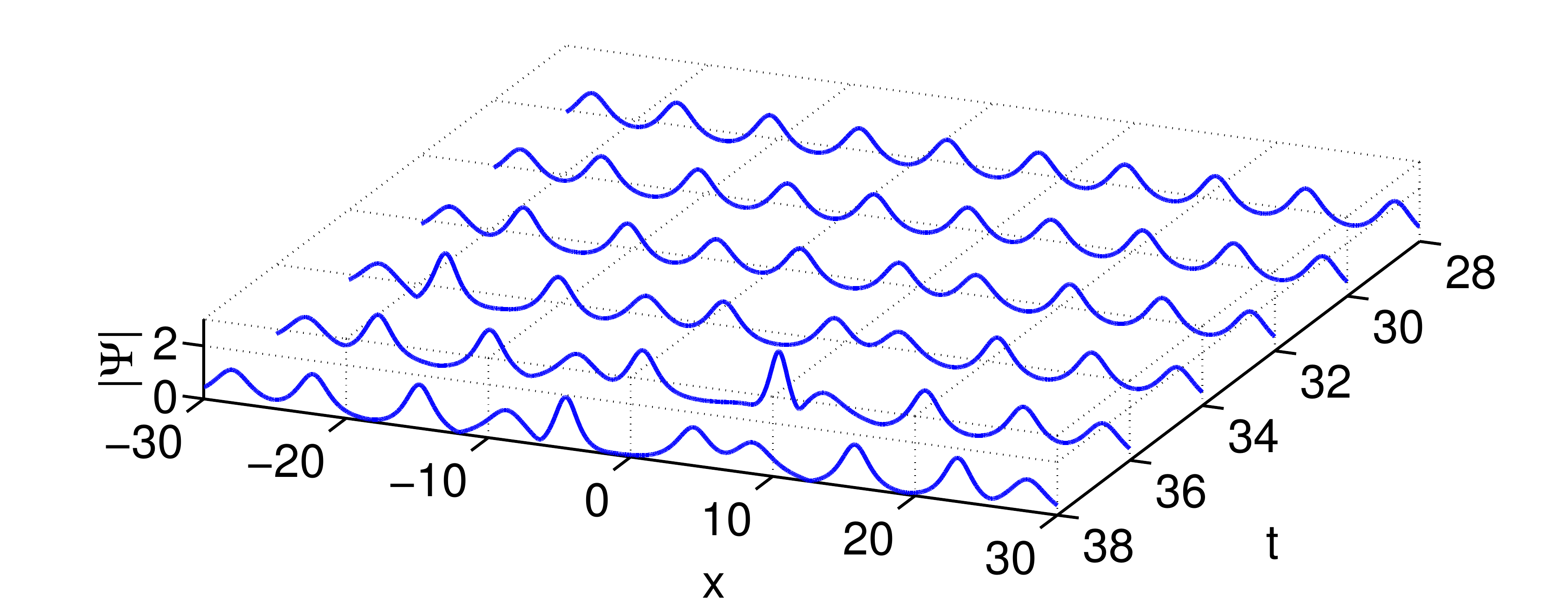}

\caption{\small 
Development of MI on the background of cnoidal wave with $\omega_{0}=\pi$, $\omega_{1}=1.6$. Significant perturbations in amplitude $|\Psi|$ become visible starting from $t=32$; compare with Fig.~\ref{fig:evolutionEM}(a).
}
\label{fig:CW2_MI}
\end{figure}

\begin{figure}[H] \centering
\includegraphics[width=14cm]{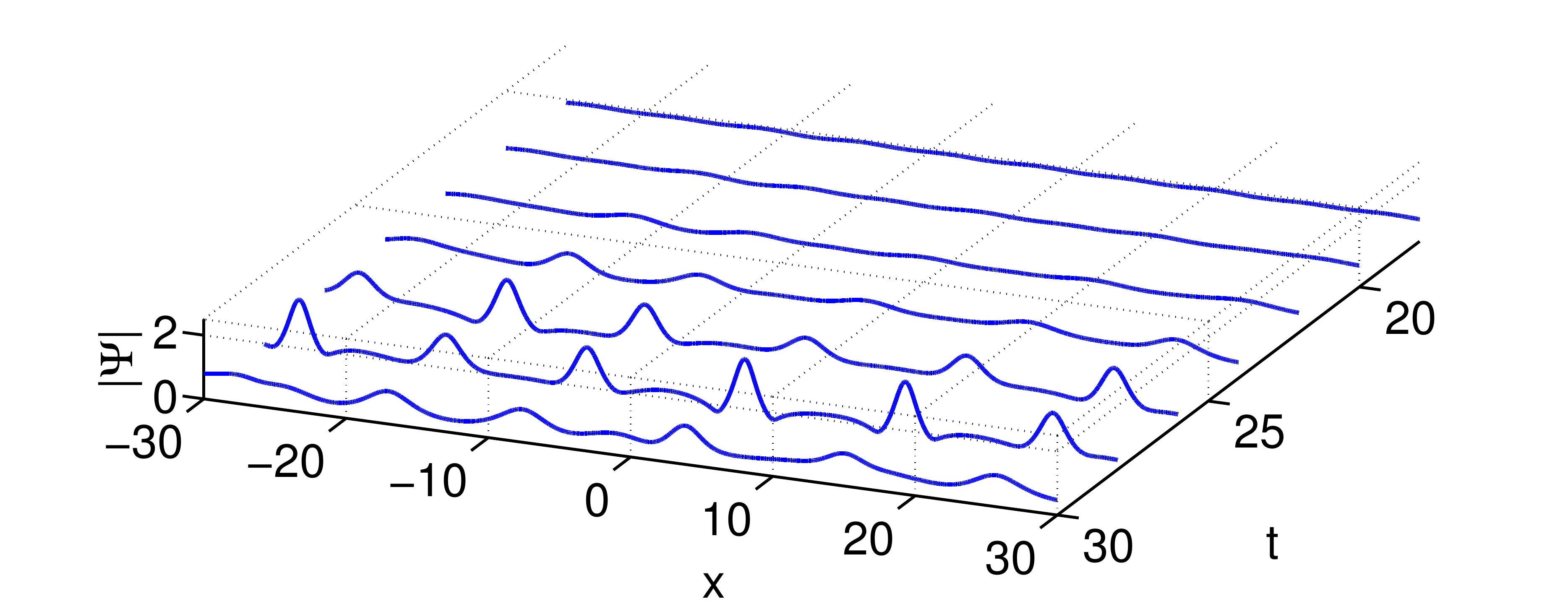}

\caption{\small 
Development of MI on the background of cnoidal wave with $\omega_{0}=\pi$, $\omega_{1}=5$. Significant perturbations in amplitude $|\Psi|$ become visible starting from $t=22$; compare with Fig.~\ref{fig:evolutionE12}(b).
}
\label{fig:CW3_MI}
\end{figure}

\end{document}